\documentclass[fleqn,usenatbib]{mnras}

\usepackage{newtxtext,newtxmath}
\usepackage{xcolor}

\usepackage[T1]{fontenc}
\usepackage{scalerel}

\DeclareRobustCommand{\VAN}[3]{#2}
\let\VANthebibliography\thebibliography
\def\thebibliography{\DeclareRobustCommand{\VAN}[3]{##3}\VANthebibliography}

\usepackage{graphicx}	
\usepackage{amsmath}	
\usepackage{indentfirst}
\usepackage{orcidlink}
\usepackage{fontawesome5}
\usepackage{array}
\usepackage{multirow}
\usepackage{booktabs}
\usepackage{threeparttable}

\newcommand{\kB}{\ensuremath{\, {k_{\rm B}}}}  

\newcommand{\Rtwohc}{R_{\rm 200c}}

\newcommand{\Mtwohc}{M_{\rm 200c}}

\newcommand{\Rfivehc}{R_{\rm 500c}}
\newcommand{\Mfivehc}{M_{\rm 500c}}

\newcommand{\eg}{{\sl e.g.},}

\newcommand{\mfiveh}{M_{\rm 500c}}             
\newcommand{\LX}{L_{\rm X}}
\newcommand{\LXce}{L_{\rm X, ce}}

\newcommand{\LXsoft}{L_{\rm X, soft}}
\newcommand{\Mgas}{M_{\rm gas}}
\newcommand{\fgas}{f_{\rm gas}}

\newcommand{\Tsl}{T_{\rm sl}}
\newcommand{\Tmw}{T_{\rm mw}}

\newcommand{\Tspec}{T_{\rm spec}}
\newcommand{\YX}{Y_{\rm X}}
\newcommand{\TX}{T_{\rm X}}
\newcommand{\TXce}{T_{\rm X, ce}}
\newcommand{\YSZ}{Y_{\rm SZ}}

\newcommand{\mgas}{M_{\rm gas}}

\newcommand{\mdelta}{M_{\Delta}} 
\newcommand{\rdelta}{R_{\Delta}}

\newcommand{\rhocrit}{\rho_{\rm crit }}
\newcommand{\Mfid}{M_{\rm fid}}

\newcommand{\varmus}{\ensuremath{\sigma^2_{\mu \vert \mathbf{s} }}}
\newcommand{\sigmamus}{\ensuremath{\sigma_{\mu \vert s }}}

\newcommand{\DeltaSS}{\Delta_{{\rm SS}, a}}

\newcommand{\hinv}{\ensuremath{\, h^{-1}}}                       
\newcommand{\msol}{\ensuremath{\, {\rm M}_\odot}}    
\newcommand{\msun}{\ensuremath{\, {\rm M}_\odot}} 
\newcommand{\kpc}{\ensuremath{\, {\rm kpc}}}         
\newcommand{\mpc}{\ensuremath{\, {\rm Mpc}}}         
\newcommand{\gpc}{\ensuremath{\, {\rm Gpc}}}

\newcommand{\myr}{\ensuremath{\, {\rm Myr}}}   

\newcommand{\kev}{\ensuremath{\, {\rm keV}}}

\newcommand{\lcdm}{$\Lambda$CDM}
\newcommand{\sigmaeight}{\ensuremath{\sigma_8}}
\newcommand{\ns}{\ensuremath{n_{\rm s}}}
\newcommand{\omegam}{\ensuremath{\Omega_{\rm m}}}
\newcommand{\omegab}{\ensuremath{\Omega_{\rm b}}}

\newcommand{\omegaLambda}{\ensuremath{\Omega_{\Lambda}}}

\newcommand{\TNGcombined}{\rm TNG300 + TNG-{\textsc{cluster}}}
\newcommand{\Flam}{\rm {\textsc{flamingo}}-L1\_m8}
\defcitealias{Planck2014}{Planck Collab. (2014)}

\newcommand{\github}[1]{%
   \href{#1}{\faGithub}%
}

\definecolor{bleudefrance}{rgb}{0.19, 0.55, 0.91}
\definecolor{forestgreen}{rgb}{0.13, 0.55, 0.13}
\definecolor{violet}{RGB}{128, 0, 128}%

\title[Mass Proxy Quality of Gas Properties]{Mass Proxy Quality of Massive Halo Properties in the {\textsc{illustris}}TNG and {\textsc{flamingo}} Simulations: I. Hot Gas}

\author[Eddie Aljamal et al.]{
Eddie Aljamal$^{\orcidlink{0009-0002-2253-4583}}$, $^{1}$\thanks{E-mail: ealjamal@umich.edu}
August E. Evrard$^{\orcidlink{0000-0002-4876-956X}}$,$^{1}$
Arya Farahi$^{\orcidlink{0000-0003-0777-4618}}$,$^{2,3}$
Annalisa Pillepich$^{\orcidlink{0000-0003-1065-9274}}$,$^{4}$
Dylan Nelson$^{\orcidlink{0000-0001-8421-5890}}$,$^{5}$
\newauthor
\hspace{4pt}Joop Schaye$^{\orcidlink{0000-0002-0668-5560}}$$^{6}$
Matthieu Schaller$^{\orcidlink{0000-0002-0668-5560}}$,$^{6,7}$
Joey Braspenning$^{4}$
\\ \\
$^{1}$Department of Physics and Leinweber Center for Theoretical Physics, University of Michigan\\
$^{2}$Departments of Statistics and Data Sciences, University of Texas at Austin, Austin, TX 78712, USA\\
$^{3}$The NSF-Simons AI Institute for Cosmic Origins, University of Texas at Austin, Austin, TX 78712, USA\\
$^{4}$Max-Planck-Insitut für Astronomie, Königstuhl 17, 69117 Heidelberg, Germany\\
$^{5}$Universität Heidelberg, Institut für Theoretische Astrophysik, ZAH, Albert-Ueberle-Str. 2, 69120 Heidelberg, Germany\\
$^{6}$Leiden Observatory, Leiden University, PO Box 9513, 2300 RA Leiden, the Netherlands\\
$^{7}$Lorentz Institute for Theoretical Physics, Leiden University, PO box 9506, 2300 RA Leiden, the Netherlands
}

\date{Accepted XXX. Received YYY; in original form ZZZ}

\pubyear{2025}

\begin{document}
\label{firstpage}
\pagerange{\pageref{firstpage}--\pageref{lastpage}}
\maketitle

\begin{abstract}
We examine scale and redshift dependence of mass-property relations (MPRs) for five hot gas properties of two large group- and cluster-scale halo samples realized by the {\textsc{illustris}}TNG, TNG-{\textsc{cluster}} and {\textsc{flamingo}} cosmological hydrodynamical simulations. For intrinsic properties of i) hot gas mass ($\Mgas$), ii) spectroscopic-like temperature ($\Tsl$), iii) soft-band X-ray luminosity ($\LX$), and iv) X-ray ($\YX$) and v) Sunyaev-Zel'dovich ($\YSZ$) thermal energies, we use MPR parameters to infer mass proxy quality (MPQ) --- the implied scatter in total halo mass conditioned on a property --- for halos with $\Mfivehc \geq 10^{13}\msol$ at redshifts, $z \in \{0, 0.5, 1, 2\}$. 
We find: (1) in general, scaling relation slopes and covariance display moderate to strong dependence on halo mass, with redshift dependence secondary;
(2) for halos with $\Mfivehc > 10^{14}\msol$, scalings of $\Mgas$ and $\YSZ$ simplify toward self-similar slope and constant intrinsic scatter (5 and 10\%, respectively) nearly independent of scale, making both measures ideal for cluster finding and characterization to $z=2$; (3) halo mass-conditioned likelihoods of hot gas mass and thermal energy closely follow a log-normal form; 
(4) despite normalization differences up to $0.4$ dex between the two simulations, higher-order scaling features such as slopes and property covariance show much better agreement. Slopes show appreciable redshift dependence at the group scale, while redshift dependence of the scatter is exhibited by low mass {\textsc{flamingo}} halos only; (5) property correlations are largely consistent between the simulations, with values that mainly agree with existing empirical measurements.  
We close with a literature survey placing our MPR slopes and intrinsic scatter estimates into community context.
\end{abstract}



\section{Introduction}
Dark matter (DM) halos are the nurseries wherein galaxies first form \citep[\emph{e.g.},][]{OstrikerPeeblesYahil1974, WhiteRees1978, SilkMamon2012}, then cluster into larger structures such as groups and clusters of galaxies, whose abundance across cosmic time informs studies of cosmological parameters \citep[\emph{e.g.},][]{White1993a, Voit2005, Allen2011, Kravtsov2012, Weinberg2013, Norton2024}.  At the population level, halos exhibit a rich variety of bulk observable properties with magnitudes driven primarily by scale (total system mass), and secondarily by astrophysical processes integrated over their hierarchical merger histories, large-scale environmental effects and other factors.  At any redshift, the measurable characteristics of stellar and hot gas contents of halos will, in the mean, scale in some manner with total mass, and variations in formation history and astrophysical evolution will drive intrinsic covariance in such properties for halos of fixed total mass. 

The differential comoving space density of halos as a function of their total mass -- the Halo Mass Function (HMF) -- is a sensitive probe of structure growth and cosmology that has been studied with cluster samples selected by the optical-IR galaxy content of massive halos \citep{Gladders2007RCSclustercosmo, Rozo2010SDSSclustercosmo, Rykoff2014redMaPPerI,  Gonzales2019WISEclusters, Abdullah2020SDSSclustercosmo, Costanzi2020DESY1, Miyatake2021HSCcosmo, Aguena2021WAZP, Wen2022DES+unWISE, Maturi2023JPASamicoClusters} as well as the thermal Sunyaev-Zel'dovich (tSZ) effect on the cosmic microwave background \citep{Sehgal2011ACTclustercosmo, deHaan2016SPTclustercosmo, PlanckXXIV2016clustercosmo, Bocquet2019SPTclustercosmo, SPT2024} and the extended X-ray emission \citep{Bohringer2007REXCESS,  Vikhlinin2009, Mantz2010cosmo, Mehrtens2012XCS, Mantz2015WtGcosmo, Pierre2016XXLintro, Pacaud2018XXLcosmo, Chitham2020CODEXcosmo, Chiu2023eFEDScosmo} that arises from their hot intracluster plasma. Since the true masses of DM halos in the sky are not directly measurable, empirical determination of the HMF requires observable proxies, defined by scaling relations \citep[\eg][]{Giodini2013scalingRelationReview}, to statistically map true halo masses to observable features of clusters used by the above surveys.  

Considerable effort has therefore been directed to understanding scaling relations for properties such as brightest central galaxy magnitude, count of galaxies above a certain size threshold, hot gas mass, temperature, X-ray luminosity, and integrated electron pressure that manifests the thermal SZ effect.  Results from a number of such studies are presented below (\S\ref{sec:literature-comparison}). Cluster sample sizes with high quality observations are typically dozens to hundreds, and sample selection effects must be treated carefully to ensure unbiased results \citep{Mantz2019, Mulroy2019}.  Large volume simulations can generate 10,000 or more massive halos (though many studies are smaller), which enables precise scaling relation analysis for samples complete above some minimum total mass.  The upper mass scale is set by the simulation's volume, and to avoid excessive computational expense many studies selectively perform full physics treatment within specific sub-volumes enclosing only the most massive halos \citep[\eg][]{Barnes2017a, Nelson2024}. 

We employ the term \emph{mass-property relation} (MPR) to refer to the subset of scaling relations that relate true mass, defined with a spherical overdensity condition, to intrinsic properties of halos measured within the same spherical region.  MPRs for massive halos have been studied using cosmological simulations beginning with the first generation of available codes \citep[\eg][]{EMN1996, BryanNorman1998}.  
Modern simulations that include complex feedback from stars and active galactic nuclei (AGN) show that non-gravitational processes drive systematic shifts in scaling relations away from simple self-similar expectations founded on purely gravitational evolution \citep{Bhattacharya2008, Puchwein2008, Fabjan2010, McCarthy2010}. 

In particular, MPR slopes for hot gas mass tend to be steeper than self-similar because the efficiency of converting gas into stars decreases as halo mass increases \citep{Kravtsov2006, Nagai2006, Nagai2007, Planelles2014, Biffi2014, Truong2018}.  The action of AGN feedback also steepens slopes by preferentially expelling gas from the shallow potential wells of lower mass systems \citep{Planelles2014, LeBrun2017, Barnes2017a, Barnes2017b, Farahi2018, Truong2018, Henden2019, Pop2022}. Despite consensus on steeper than self-similar slopes, there remains disagreement about the magnitude of the slope and the level of intrinsic scatter around the mean. This may partially be due to disparate mass ranges used in each analysis, the choice of which affects MPR parameters when significant mass-dependence exists \citep{LeBrun2017, Farahi2018, Pop2022}.

When population samples are large, thousands or more, and span a wide dynamic range in halo mass, single power law MPRs are likely to be inadequate. To tease out scale-dependent features in scaling relations, Kernel-Localized Linear Regression \citep[KLLR][]{Farahi2022KLLR} is a powerful approach. The method introduces a single parameter to regression analysis, the width of the kernel applied to the primary scale variable.  In the limit of infinite sample size this scale shrinks toward zero and differential forms for conditional probabilities are recovered (see Appendix~\ref{sec:KLLR}). The method was first applied by \citet{Farahi2018} to hot gas and stellar masses in large halo samples from the BAHAMAS \citep{McCarthy2010} and MACSIS \citep{Barnes2017a} simulations. Non-monotonic behaviors in the slope and scatter of both properties is found, with scatter in gas mass reaching a low of $\sim 5\%$ for the largest halos. 

The work we present here is an extension of the aforementioned studies to incorporate a broader set of intrinsic halo properties derived from two large-volume simulation campaigns employing different galaxy formation physics treatments with independent modeling approaches: {\textsc{illustris}}TNG \citep{Weinberger2017TNGmethods, Pillepich2018TNGmethods} and {\textsc{flamingo}} \citep{Kugel2023, Schaye2023Flamingo}.  We report MPRs for five hot gas properties of simulated samples containing many thousands of halos above $10^{13} \msol$ at redshift zero.  Subsequently, we report stellar property MPRs (Aljamal \textit{et al.}, in prep., Paper II). 

The local intrinsic scatter and slope of a property's scaling with true mass determines its value as a proxy for the latter.  Steeper slopes map a fixed property range onto a narrower range of mass, and the limit of zero intrinsic scatter maps a property uniquely to mass.  We introduce the term \emph{mass-proxy quality} (MPQ) to refer to the \emph{precision} with which true halo mass can be estimated from the value of an intrinsic property. The standard deviation of the conditional likelihood, $\Pr(M|S,z)$, of true halo mass, $M$, conditioned on a chosen property, $S$, at redshift, $z$ is a natural measure of that property's quality as a mass proxy.  We employ instead shape parameters from the complementary MPR likelihood, $\Pr(S|M,z)$, to define the MPQ.  When likelihoods are log-normal, the population model of \citet[][hereafter E14]{E14} expresses the second moment of $\Pr(M|S,z)$ in terms of MPR and HMF parameters.  We test the performance of this model in \S\ref{sec:directMassScatter} below.

Note that this definition of proxy quality ignores the issue of \emph{accuracy} when inverting a scaling relation. For example, previous work used a simulation-based prior on the mean gas mass fraction in clusters to derive total masses from hot gas mass measurements \citep{Mantz2010XrayScaling}.  If this prior value for gas fraction were systematically biased, so too would be the implied mass values.  Indeed, MPR normalizations differ between the two simulations we study, as discussed in \S\ref{sec:normalization-redshift-evolution} below.  

The MPQ concept is not new. Using mock X-ray images of $16$ simulated clusters spanning roughly a decade in final mass. 
\citet{Kravtsov2006} and \citet{Nagai2007} highlighted gas thermal energy as superior to either gas mass or temperature in terms of mass proxy quality.  Even earlier, \citet{EMN1996} noted that masses derived from the virial scaling of X-ray temperature alone produced mass estimates with smaller scatter compared to a hydrostatic mass treatment, because the latter included extra variance induced by the outer gas profile slope. More recently, deep neural networks trained on mock observations of hydrodynamical simulations have shown promise in improving the accuracy and reducing the scatter in cluster mass estimation using X-ray \citep{Ntampaka2019MLXraymasses, GreenMLXraymasses2019, YanMLXraymasses2020, Ho2023MLmasses, Krippendorf2023ERositaMLmasses}, microwave \citep{CohnBattaglia2020MLmasses, WadekarMLSZmasses2023, WadekarMLSymbSZmasses2023, deAndres2022DLYSZ}, and optical data \citep{Ntampaka2016MLdynamicalMass, Ho2019, HoMLComa2022, KodiRamanahDynmasses2020, KodiRamnahMLDyn2021}.

Our use of the term MPR intentionally avoids confusion with \emph{mass-observable relations} (MORs) that aim to study directly observable features of clusters, such as the number of red galaxies defined to lie in some cluster-centric sky region and photometric color window.  Such measures require integration along the line of sight and potentially observer-specific treatments. The hot gas properties we study here are certainly ``observable'', but extracting unbiased measurements of intrinsic values\footnote{The term \emph{intrinsic} here refers to spherically integrated values given in Table~\ref{tab:prop-definitions}.} from actual observations entails a variety of data analysis methods that depend on the telescope/instrument combination, observing conditions, and other factors that lie beyond the scope of this work.

This paper is organized as follows. In \S\ref{sec:methods-sims} we define the gas properties used in this paper, review self-similar MPR scalings and relevant aspects of the E14 model, and describe the simulation samples used in this work.  We introduce our results for the mass and redshift dependence of the MPQs of five hot gas properties in \S\ref{sec:MPQ}, while \S\ref{sec:MPR-parameters} presents the MPR parameters (slope, scatter, normalization) that underlie the MPQs at redshifts of 0, $0.5$, 1 and 2. In \S\ref{sec:correlations}, we discuss halo mass-conditioned correlations of the five gas properties. In \S\ref{sec:kernel-shapes}, we examine the MPR likelihood shapes. We then compare our results to the literature in \S\ref{sec:literature-comparison}, followed by a brief summary in \S\ref{sec:conclusion}.  

In terms of primary scale variable, we employ a mass ($\mfiveh$) and radius ($\Rfivehc$) convention defined by an interior spherical density contrast of 500 times the critical density, $\rhocrit(z)$.


\section{Methods and Simulations}\label{sec:methods-sims}

In this section, we introduce the halo properties being used (\S\ref{sec:defintions-of-properties}), 
review the self-similar scaling model (\S\ref{sec:self-similar-theory}), present the framework for mass variance estimation (\S\ref{sec:E14-model-MPQ}), and finish with a description of the simulated samples (\S\ref{sec:simulations}).  Readers familiar with the basics may choose to review Tables~\ref{tab:prop-definitions}, \ref{tab:sim-constants}
and \ref{tab:counts} and examine our MPQ definition in \S\ref{sec:E14-model-MPQ} before moving to the next section.

\subsection{Halo properties}\label{sec:defintions-of-properties}

As stated above, all properties are intrinsic to a halo, meaning they involve only material within a spherical region defined by $\Rfivehc$.  

\textbf{True halo mass.} We adopt the common spherical overdensity true mass convention of $M_{\Delta}$, for which the summed mass of all components within a sphere of radius $R_{\Delta}$ encloses a mean density of $\Delta$ times the critical density, $\rhocrit(z) = 3H(z)^2/(8\pi G)$, thus
\begin{equation}
    \mdelta = \frac{4\pi}{3}\Delta \rhocrit (z) \rdelta^3. \label{eq:SO-true-mass}
\end{equation}
We use a value $\Delta = 500$, a scale commonly used by the X-ray and SZ community. For the statistics we study, results using $\Delta = 200$ are qualitatively and quantitatively similar.  

\textbf{Hot gas mass.} 
We define the hot gas mass, $\Mgas$, to be the sum of all gas elements having temperature $\geq 10^5$ K within $\Rfivehc$. This component represents the majority of baryons at cluster scales at $z = 0$, though we will show in paper II that stars are an important component for low-mass groups.  Low temperature $T < 10^{5}$~K gas constitutes approximately $4\%$ of the total gas phase within $\Rfivehc$ at the group scale, declining to $0.1\%$ for high-mass clusters in TNG at $z = 0$. These fractions increase with redshift at fixed halo mass \citep{Rohr2024, Staffehl2025}. For {\textsc{flamingo}}, these fractions are $6\%$ and $1\%$, respectively.

\textbf{Hot gas temperature.} 
While the thermal structure of the ICM is not isothermal \citep{Chatzigiannakis2025}, 
there are several common aggregate measures for gas temperature in use \citep{Lee2022, Kay2024}.  We primarily employ a core-excised spectroscopic-like temperature, $\Tsl$, defined by a density weighting 
\begin{equation}
    \Tsl = \frac{\int n_{\rm e}^2 T^{1 - \alpha}dV}{\int n_{\rm e}^2 T^{-\alpha}dV} \approx \frac{\sum_i n_{{\rm e}, i}m_i T_i^{1 - \alpha}}{\sum_i n_{{\rm e}, i}m_i T_i^{-\alpha}}. \label{eq:spect-like-temp}
\end{equation}
where the sum is over all gas cells within $0.15\Rfivehc < r < \Rfivehc$ and with temperature $\kB T_i \geq 0.1 \kev$. In the above, $\alpha = 3/4$ and $m_i$ and $n_{{\rm e}, i}$ are the gas mass and electron density of the fluid element.  We employ a temperature threshold of $0.1\kev$ to better align the comparison with observations from X-ray telescopes, such as \textit{XMM} and \textit{Chandra}, which are typically insensitive to photons below $\sim 0.1\kev$ \citep{Weisskopf2000, Jansen2001}. 

Secondarily, we employ a mass-weighted average temperature, 
\begin{equation}
    \Tmw = \frac{\sum_i m_i T_i}{\sum_i m_i} \label{eq:mass-weighted-avg-temp},\\
\end{equation}
where the sum is over the gas elements of mass $m_i$ and temperature $T_i$ that contribute to $\Mgas$.

The X-ray spectra of hot gas in clusters and groups of galaxies are usually fit with a single or multitemperature thermal model \citep{MazzottaOct2004, Vikhlinin2006} to arrive at a spectroscopic temperature, $\Tspec$. Previous work on simulated clusters with mass $\Mfivehc > 10^{14} \hinv \msol$ found that the spectroscopic-like temperature reproduces $\Tspec$ to within a few percent \citep{Rasia2005, MazzottaDec2004, Rasia2014}. We therefore consider $\Tsl$ as the simulation analogue of the observable X-ray temperature $\Tspec$. Caution must be exercised in comparing $\Tsl$ to $\Tspec$ for clusters with $\Mfivehc < 10^{14} \hinv \msol$ or $\Tspec < 3$\kev. To extend the temperature range of a comparison between $\Tsl$ and $\Tspec$, a more complicated weighting scheme for $\Tsl$ is required \citep{Vikhlinin2006}.

We note that the density weighting aspect of $\Tsl$ makes it sensitive to recently heated material near a supermassive black hole (SMBH). The hot gas properties reported by \citet{Braspenning2024} for {\textsc{flamingo}} halos omit contributions from such recently heated particles. Specifically, for hot gas properties in {\textsc{flamingo}}
we exclude cells which have been heated in the past $15\myr$ and that have temperature in the range $10^{0.1}\Delta T_{\rm AGN} \leq T_i \leq 10^{0.3}\Delta T_{\rm AGN}$ where $\Delta T_{\rm AGN} = 10^{7.78}$ K is the temperature change of the cell after an AGN heating event in the {\textsc{flamingo}} simulation.
The amount of material removed is small.

\textbf{Soft-band X-ray Luminosity.} 
The X-ray luminosity, $\LX$, in the observed soft band $[0.5-2.0]\kev$ is summed for each gas element within $\Rfivehc$. Following the approach of {\textsc{flamingo}} \citep{Braspenning2024}, the contribution of each gas element is calculated by interpolating density, temperature, individual metal abundances of nine elements (H, He, C, N, O, Ne, Mg, Si, Fe), and redshift using the Cloudy \citep{Ferland2017} photoionization spectral synthesis code. We use luminosities measured in the observer frame and therefore different parts of the rest-frame X-ray spectra will contribute to the $[0.5-2.0]\kev$ band at different redshifts. For {\textsc{flamingo}}, these luminosities were included in the halo catalog, derived from previous calculations using the Spherical Overdensity and Aperture Processor \citep[SOAP][]{McGibbonSOAP2025}. For TNG, we use the density, temperature, and metallicity of each gas cell to calculate each cell's luminosity by interpolating the {\textsc{flamingo}} X-ray tables. Finally, we sum the contributions of all gas cells within a three-dimensional aperture of $\Rfivehc$. We compared the results of these calculations to the halo luminosities reported by \citet{Nelson2024} for TNG300 and TNG-{\textsc{cluster}} at $z = 0$, finding good agreement with CLOUDY-based measurements being, on average, lower by $0.1$ dex at the group scale and $0.05$ dex at the cluster scale. The X-ray luminosity-halo mass scaling relation in {\textsc{flamingo}} has been shown to be in agreement with a variety of observational measurements \citep{Braspenning2024}. \citet{Nelson2024} demonstrate similar agreement for TNG-{\textsc{cluster}}, with a caveat discussed below. 

\textbf{Hot gas thermal energy.} 
The thermal Sunyaev–Zel’dovich effect is a distortion of the CMB spectrum that arises from inverse Compton scattering of CMB photons off free electrons in the ICM \citep{Sunyaev1972}. The total tSZ flux density from a cluster is given by the angular integral of the tSZ sky brightness \citep{Silva2004, Pop2022}.  

For a spherical halo isolated on the sky with no gas beyond $\Rfivehc$, the angular integral of the tSZ signal is equivalent to a volume integral within the halo, yielding a measure of intrinsic gas thermal energy
\begin{equation}
   Y = \frac{\kB \sigma_{\rm T}}{m_{\rm e}c^2}  \int T_{\rm e} n_{\rm e} dV \ \rightarrow \ \frac{k_\text{B} \sigma_\text{T}}{m_{\text{e}}c^2} \sum_i n_{\text{e}, i} T_i \frac{m_i}{\rho_i},
\end{equation}
where $\sigma_{\rm T} = 6.65 \times 10^{-25}{\rm cm}^{2}$ is the Thomson cross section and $m_{\rm e}$ the electron mass. The discrete sum is over gas cells contributing to $\Mgas$ with electron density $n_{{\rm e}, i}$, temperature $T_i$, mass $m_i$, and gas density $\rho_i$. Because $n_{\text{e}, i} \propto \rho_i$, the gas density cancels up to a factor that is effectively constant for the ionization state in the ICM.  Thus, $Y \propto \sum_i m_iT_i$ and so we define the tSZ thermal energy as 
\begin{equation}\label{eq:YSZ-def}
    \YSZ = \Mgas \Tmw 
\end{equation}
and express this quantity in units of $\msol\,$K.

X-ray observations also produce estimates of this quantity using $\Tspec$ and gas masses derived from X-ray imaging  \citep{LeBrun2017, Nagai2007, Henden2019}.  
We therefore also employ the X-ray thermal energy 
\begin{equation}\label{eq:YX-def}
    \YX = \Mgas \Tsl.
\end{equation}

Note that in {\textsc{flamingo}} these gas properties are calculated using all gas cells within a 3-D aperture of $\Rfivehc$ with the exclusion of recently heated AGN cells. In {\textsc{illustris}}TNG, we use all particles in the same aperture, irrespective of temporal proximity to feedback.

Observational scaling relation studies sometimes use properties within a fixed metric aperture. KLLR parameters in that case would behave similarly to those within $\Rfivehc$ across only a limited mass range. Extending fixed aperture measurements to the full mass range considered here would result in a mixture of different dynamical regimes across the mass spectrum.  A choice of 1 Mpc, for example, would capture the virialized regions of the most massive clusters but extend into the infall regions of low-mass groups.  A fixed aperture approach would thus increase the complexity and limit the interpretability of the KLLR analysis.

\begin{table}
    \centering
    \caption{Primary gas properties measured within a 3-D aperture of $ \Rfivehc$. In the TNG simulations, we use all gas cells in this aperture, while in {\textsc{flamingo}} we exclude recently heated AGN gas cells as defined below.}
    \label{tab:prop-definitions}
    \begin{threeparttable}
        \begin{tabular}{p{1cm} p{6cm}}
             \hline
             Property & Definition\\
             \hline
             $\Mgas$ & Total hot ($T \geq 10^{5}$ K) gas mass \\
             $\Tsl$ & Core-excised, spectroscopic-like temperature, eq.~\eqref{eq:spect-like-temp}\\
             $\LX$ & Observer-frame, soft X-ray luminosity, $[0.5-2.0]\kev$  \\
             $\YX$ & X-ray gas thermal energy, $\YX = \Mgas \Tsl$\\
             $\YSZ$ & tSZ thermal energy, $\YSZ = \Mgas \Tmw$\\
             \hline
        \end{tabular}
    \end{threeparttable}
\end{table}

\subsection{Self-similar MPRs}\label{sec:self-similar-theory}
The self-similar model of structure formation \citep{Kaiser1986} assumes that halos: i) are in virial and hydrostatic equilibrium, ii) possess a common internal structure in scaled units, and iii) have a constant internal baryon fraction. While highly idealized, this model forms a baseline to which more sophisticated models can be compared.  Self-similar scaling expectations for a certain property have fixed slopes in mass and redshift, but the normalizations are arbitrary.  

The self-similar gas temperature will satisfy the virial theorem, $\kB T \propto \frac{G\Mfivehc}{\Rfivehc}$, which results in the following mass scaling and redshift evolution,
\begin{equation}
    T \propto E(z)^{2/3} \Mfivehc^{2/3},
\end{equation}
where we choose to quantify the redshift dependence using the evolution of the Hubble parameter $E(z) = \sqrt{\omegam (1 + z)^{3/2} + \omegaLambda}$ assuming a flat {\lcdm} cosmology.

For halos in a self-similar model where larger halos are simply scaled up from smaller halos, the total gas mass, $\Mgas$, is a constant fraction, $\fgas$, of the total halo mass $\Mfivehc$:
\begin{equation}
    \Mgas = \fgas \Mfivehc \propto E^0(z) \Mfivehc.
\end{equation}
By the definition of the tSZ and X-ray thermal energies above, the self-similar mass scaling and evolution predicts that,
\begin{align}
    &\YSZ \propto \YX \propto E^{2/3}(z) \Mfivehc^{5/3}.
\end{align}
The self-similar $\LXsoft-\Mfivehc$ relation is derived in \citep{Lovisari2022} and given by:
\begin{equation}
    \LXsoft \propto E^2(z) \Mfivehc
\end{equation}

Note that the redshift evolution parameterized by $E(z)$ is a consequence of the definition of halo mass in terms of the critical density.

\subsection{Mass Proxy Quality: E14 model and KLLR method}\label{sec:E14-model-MPQ}

The MPQ of a single property, gas mass or temperature, for example, is the scatter in true halo mass conditioned on the chosen property.  In general this will be both scale and redshift dependent.  The definition generalizes to the case of two or more properties.  

When multi-property statistics take a Gaussian form in log-space, 
a convolution of the mass-conditioned property likelihood with a quadratic HMF representation in log-mass yields closed form expressions for halo counts and other population statistics.  The compact forms presented by E14 are expanded in \citet{Norton2024}, including an explicit function for the halo space density as a function of observed properties. This expression makes clear the degeneracies between population MPR parameters (set largely by astrophysics) with parameters describing the HMF (set by cosmology). 

Due to the convolution with a steeply falling HMF needed to predict cluster counts, the mean mass selected by some chosen property is shifted low relative to the simple inverse of that property's MPR.  The shift in magnitude scales as the product of the local HMF slope magnitude and the true mass variance of the property.  A property that is a more effective mass proxy, one with lower halo mass variance, will thus have a smaller shift is selected mean mass.  

\subsubsection{MPQ definition }

Our measure of MPQ is based on the E14 model expressions for mass variance conditioned on a chosen property\footnote{Because the implied mass variance is small for all properties studies here, the HMF curvature term in the mass variance can be ignored. See equations (4) and (11) in E14.}. This approach maintains true halo mass as the primary scale variable and allows us to be volume complete above our minimum chosen mass. In \S\ref{sec:directMassScatter} we show that the MPQ of equation~\eqref{eq:mass-proxy-quality} matches direct estimates to better than $20\%$ for the three mass proxies with likelihoods closest to log-normal.

Using $\mu = \ln(\Mfivehc / \Mfid)$, where $\Mfid$ is a fiducial mass scale (\eg $10^{14} \msol$), the MPQ for a single observable property, denoted by the subscript $a$, is then the ratio 
\begin{equation}
    \sigma_{\mu \vert a}(\mu, z) = \frac{\sigma_a(\mu, z)}{|\alpha_a(\mu, z)|} , \label{eq:mass-proxy-quality}
\end{equation}
of the standard deviation in the property at fixed mass, $\sigma_a(\mu,z)$, to the magnitude of the local MPR slope, $\alpha_a(\mu, z)$.  

Better mass proxies have steeper slopes and/or smaller MPR scatter.  In \S\ref{sec:MPR-parameters} we demonstrate that the slope and scatter of many of hot gas MPRs generally are dependent on both scale (halo mass or property value) and redshift. 

A benefit of this approach is that the MPQ can be generalized to complex cases involving more than one property. For a set of properties, $\textbf{s}$, the mass variance about the selected mean is 
\begin{equation}\label{eq:implied-variance}
    \varmus (\mu, z) = \Big(\boldsymbol{\alpha}(\mu, z)^T\textbf{C}^{-1}(\mu, z)\boldsymbol{\alpha}(\mu, z)\Big)^{-1},
\end{equation}
where $\boldsymbol{\alpha}(\mu, z)$ is the vector of MPR slopes, formally defined in  equation~\eqref{eq:KLLR-scaling}, and $\textbf{C}(\mu, z)$ is the mass- and redshift-dependent log-property covariance maPlanck2014trix given in equation \eqref{eq:KLLR-covariance}.  The implied mass scatter is the square root of the variance, as usual. A smaller value of the mass scatter corresponds to better mass proxy quality.

\begin{table*}
\centering
    \caption{Cosmological and simulation parameters for the three simulations; the dimensionless Hubble constant $h \equiv H_0/100 \text{km s}^{-1} \text{Mpc}^{-1}$, the normalized density of matter $\omegam$, the normalized density of baryons $\omegab$, the normalized density of vacuum energy $\omegaLambda$, the normalization of the power spectrum $\sigmaeight$, the power-law index of the primordial matter power spectrum $\ns$, the box side-length $L$, the initial number of dark matter particles $N_{\rm DM}$, the dark matter particle mass $m_{\rm DM}$, the baryonic particle mass $m_{\rm baryon}$, and the gravitational softening length at $z = 0$ $\epsilon^{z = 0}$.}
    \label{tab:sim-constants}
\begin{threeparttable}
\begin{tabular}{cccccccccccc}
    \hline & \\[-9pt]
    Simulation & $h$ & $\omegam$ & $\omegab$ & $\omegaLambda$ & $\sigmaeight$ & $\ns$ & $L$ [\mpc/h] & $N_{\rm DM}$ & $m_{\rm DM}$ [$\msol$] &  
        $m_{\rm baryon}$ [$\msol$] & $\epsilon^{z = 0}$ [\kpc]\\ 
    \hline & \\[-9pt]
        \text{TNG300} & $0.6774$ & $0.3089$ & $0.0486$ & $0.6911$ & $0.8159$ & $0.9667$ & $205$ & $2500^3$ & $5.9 \times 10^7$ & $1.1 \times 10^7$ & $1.48$\\
        \text{TNG-{\textsc{cluster}}}\tnote{*} & $0.6774$ & $0.3089$ & $0.0486$ & $0.6911$ & $0.8159$ & $0.9667$ & $680$ & $8192^3$\tnote{$\dagger$} & $6.1 \times 10^7$ & $1.2 \times 10^7$ & $1.48$\\
        \text{FLAM-L1\_m8} & $0.681$ & $0.306$ & $0.0486$ & $0.694$ & $0.807$ & $0.967$ & $1000$ & $3600^3$ & $7.06 \times 10^8$ & $1.34 \times 10^8$ & $2.85$\\
    \hline
\end{tabular}
\begin{tablenotes}
    \item[*] Zoom simulation
    \item[$\dagger$] Effective resolution
\end{tablenotes}
\end{threeparttable}
\end{table*}

\subsubsection{KLLR: continous parameters from discrete populations }\label{sec:KLLRsubsec}
When sample sizes are small, under 100 or so, it is common practice to use simple linear regression of fixed slope to analyze cluster scaling relations. 
Such a fitting scheme does not allow for the possibility for a halo property to be a low quality mass proxy for estimating group masses and improve in quality at the cluster scale, or vice versa.  When sample sizes grow beyond a few thousand, a more flexible method can search for scale-dependent behavior. Our approach uses Kernel Localized Linear Regression \citep[KLLR][]{Farahi2022KLLR} to obtain scale-dependent MPQs and MPRs.  

The KLLR method has previously been used to validate the E14 model for gas properties in the BAHAMAS and MACSIS simulations \citep{Farahi2018}, to examine scale-dependent baryon back-reaction on dark matter scalings relations in the {\textsc{illustris}}TNG suite \citep{Anbajagane2022baryon}, to examine stellar properties \citep{Anbajagane2020} as well as galaxy velocity segregation \citep{Anbajagane2021velbias} in massive halo samples produced by multiple independent cosmological simulations. KLLR was also applied to various temperature measures in the {\textsc{flamingo}} simulations by \citet{Kay2024}.

In brief, the KLLR method extracts parameter estimates of a continuous statistical representation by applying a kernel density estimate approach to a discrete sample. Using total halo mass as the scale parameter, linear regression is performed --- assuming a lognormal distribution in each bin --- on property data weighted by a Gaussian in log mass centered at some scale. Regularly incrementing the center by shifts small compared with the widths allows near-continuous measures of normalization (Equation~\eqref{eq:KLLR-scaling}), slope (Equation~\eqref{eq:KLLR-scaling}), scatter (Equation ~\eqref{eq:KLLR-scatter}), and covariance (Equation ~\eqref{eq:KLLR-covariance}) of a set of properties by minimizing the weighted residual sum of square errors. 

In the limit of infinite sample size and zero KLLR kernel width, the method returns unbiased estimates of a model's continuous parameters, as long as the sample properties obey the underlying statistical form of TNG and {\textsc{flamingo}}.  With samples of roughly 2,500 and 94,000 at $z=0$, we must apply a finite width, and choose $0.2$ dex in true mass for most measurements.  We note that parameter estimates using a finite-width kernel  can be biased if the slope and scatter of MPRs have non-zero curvature in log-mass or if the population has a strong density gradient in mass.  As discussed below, we employ a narrower KLLR kernel of 0.1 dex when assessing the shape of the halo mass-conditioned property distributions in \S\ref{sec:kernel-shapes}. More details of the method can be found in Appendix \ref{sec:KLLR}.


\subsection{Simulations}\label{sec:simulations}

In this section, we describe the hydrodynamical simulations used in this work and the halo samples used in our analysis. 
A comparative summary of key simulation parameters is provided in Table~\ref{tab:sim-constants}.  Table~\ref{tab:counts} provides counts of halos above $10^{13} \msol$ for each simulation set at the redshifts of our study.  
 
\subsubsection{{\textsc{illustris}}TNG: TNG300 and TNG-{\textsc{cluster}}}\label{sec:TNG}

We use the {\textsc{illustris}}TNG cosmological hydrodynamic simulations \citep{Nelson2018, Marinacci2018, Pillepich2018, Springel2018, Naiman2018} which are run using the AREPO moving-mesh code with sub-grid prescriptions incorporating gas radiative mechanisms, metal-dependent radiative cooling and heating, multimode active galactic nucleus (AGN) feedback sourced by supermassive black holes, and more. We use the flagship $302.6 \mpc$ periodic box, TNG300, as well as the zoom-in $1003.8 \mpc$ simulation, TNG-{\textsc{cluster}} \citep{Nelson2024}, which adopt a flat {\lcdm} cosmology with characteristics summarized in Table \ref{tab:sim-constants}. As described in \cite{Pillepich2018}, the galaxy stellar mass function and cluster gas fraction produced in {\textsc{illustris}}TNG are in reasonable agreement with observational constraints, although \citet{Schaye2023Flamingo} present criticisms of the former.

Halos are found using a standard friends-of-friends (FoF) percolation algorithm with a linking length of $b = 0.2$. The smaller TNG300 box has a limited number of large mass halos; more specifically, $16$ halos with $10^{14.5} < \Mfivehc/\msol < 10^{15}$ and only $1$ halo with $\Mfivehc > 10^{15} \msol$ at $z = 0$. Our study focuses on halos in the mass range $10^{13} \lesssim \Mfivehc/\msol \lesssim 10^{15}$ and uses a localized scaling approach to the MPRs which requires us to have an appreciable number of halos outside this target range in order to avoid edge effects (Appendix~\ref{sec:KLLR}). For these reasons, we make use of the TNG-{\textsc{cluster}} ensemble \citep{Nelson2024} that performs "zoom" resimulations on $352$ cluster halos extracted from a $1003.8^3$ \gpc$^3$ N-body realization in order to boost the number statistics of massive clusters. Relative to TNG300, the cosmological parameters as well as the spatial resolution remain unchanged. This less computationally expensive simulation provides $204$ clusters with $10^{14.5} < \Mfivehc/\msol < 10^{15}$ and $31$ clusters with $\Mfivehc > 10^{15} \msol$ at $z = 0$. The TNG-{\textsc{cluster}} sample is volume complete above $\Mtwohc \ge 10^{15}\msol$ and randomly samples lower mass halos such that the number is uniform in log-mass in the combined TNG-300 and TNG-{\textsc{cluster}} samples.

The fact that the two TNG simulations were realized by the same code using the same mass resolution enables us to combine their halo samples for collective study of the super-sample. However, the runs differ in an important parameter, $\alpha$, controlling the star formation rate of cold gas. A change introduced for TNG50  \citep{Nelson2019} was retained for the TNG-{\textsc{cluster}} runs.   In Paper II, we discuss how this parameter boosts the total stellar mass of TNG-{\textsc{cluster}} halos by $\sim 30\%$ (a shift that is comparable to the impact of changing by one level of resolution in the TNG model \citep{Pillepich2018}) with respect to those in TNG300 halos at $z = 0$.  Other effects on stellar property covariance structure will be discussed there, but we also document effects on hot gas phase structure that affect $\Tsl$ below.

\subsubsection{{\textsc{flamingo}}: L1\_m8}\label{sec:FLAM}
We also study halos in the large-volume hydrodynamical simulations of the {\textsc{flamingo}} project \citep{Kugel2023, Schaye2023Flamingo} based on the SPHENIX smoothed particle hydrodynamics (SPH) scheme \citep{Borrow2022} using the SWIFT code \citep{Schaller2024}. The flagship simulations utilize sub-grid prescriptions calibrated to reproduce the galaxy stellar mass function and cluster gas fractions at $z = 0$. All simulations assume a spatially flat {\lcdm} cosmology with initial conditions created using MONOFONIC \citep{Hahn2021, Elbers2022} that include neutrinos. We employ the highest resolution simulation, {\textsc{flamingo}}-L1\_m8 (FLAM-L1\_m8 in Table \ref{tab:sim-constants}), which is run in a $1$ Gpc box with purely thermal AGN feedback. In addition to the flagship simulations, the suite provides variations in cosmologies, AGN feedback models, and gas fractions. Cosmic structure is identified using an updated version of the Hierarchical Bound Tracing algorithm \citep[HBT-HERONS][]{ForouharHBTHERONS} which uses a history-based approach to identify the subhalos using an iterative unbinding procedure on particles within FoF groups, and then tracks their evolution as they merge. 

Particular advantages of this simulation are: i) it supplies statistics for nearly 100,000 halos across our target mass range (Table \ref{tab:counts}), ii) It employs significantly different subgrid models, for example a single-mode thermal AGN feedback as opposed to {\textsc{illustris}}TNG's two-mode AGN feedback, as well as an SPH solver that is different from the quasi-Lagrangian mesh used in {\textsc{illustris}}TNG, offering an opportunity to assess the robustness of population statistics between two independent implementations and iii) {\textsc{flamingo}} reproduces observed stellar mass functions in the Galaxy and Mass Assembly (GAMA) survey \citep{Driver2022} as well as the cluster gas fractions in HSC-XXL \citep{Akino2022} thanks to careful calibration which is detailed in \citet{Kugel2023}.

\subsubsection{Halo sample sizes}\label{sec:counts}

\begin{table}
    \centering
    \caption{Sample sizes for halos with $\Mfivehc \geq 10^{13}\msol$ in the TNG300-1, TNG-{\textsc{cluster}} and {\Flam} (FLAM-L1\_m8) simulations.}
    \label{tab:counts}
    \begin{threeparttable}
        \begin{tabular}{p{1.5cm} p{1.5cm} p{1.5cm} p{1.75cm}}
             \hline
             Redshift & TNG300 & TNG-{\textsc{cluster}} & FLAM-L1\_m8 \\
             \hline
             $2$ & $299$ & $320$ & $11\;087$\\
             $1.0$ & $1\;290$ & $351$ & $47\;392$\\
             $0.5$ & $2\;015$ & $352$ & $74\;300$\\
             $0$ & $2\;548$ & $352$ & $93\;815$\\
             \hline
        \end{tabular}
    \end{threeparttable}
\end{table}

\begin{figure*}
    \includegraphics[width = 0.8\textwidth]{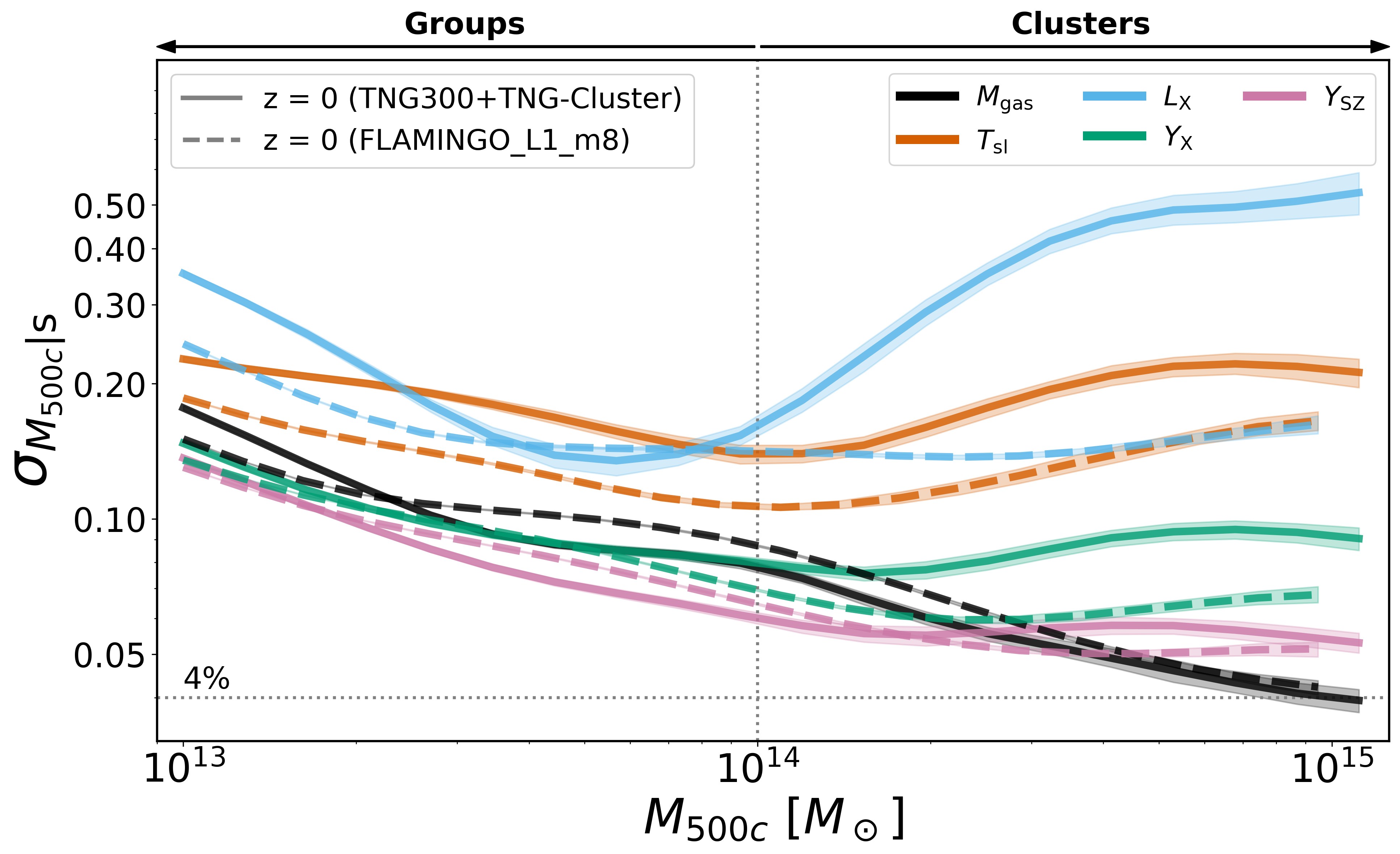}
    \caption{The mass proxy quality, quantified by the implied halo mass scatter, equation~\eqref{eq:mass-proxy-quality}, for gas properties within a 3-D aperture of $\Rfivehc$ listed in the legend derived 
    from TNG (solid lines) and {\textsc{flamingo}} (dashed) halo populations. Shaded regions are $1\sigma$ uncertainties based on $1000$ bootstrap samples.
    We indicate the low redshift mass regimes of groups and clusters as roughly divided by a halo mass of $10^{14}\msol$.  All MPQs show moderate to strong mass dependence.  The relative ordering of mass proxy quality is fairly consistent between the two simulation methods, with thermal energy ($\YSZ$, violet) being the best mass proxy below $10^{14.5} \msol$ and hot gas mass ($\Mgas$, black) best above this scale, reaching a minimum of $4\%$ halo mass scatter at $10^{15} \msol$. See text for further discussion. 
}
    \label{fig:MPQzero}
\end{figure*}

We employ samples of halos with $\Mfivehc \ge 10^{13} \msol$ at four specific redshifts. The mass range represents poor groups to the richest clusters at low redshift to massive protoclusters at $z=2$.  Counts for each simulation and redshift are given in Table~\ref{tab:counts}.  To avoid edge effects in KLLR statistics, we employ halos with $\Mfivehc \ge 10^{12.5} \msol$ in all  analysis.  

At the final epoch, there are a few thousand halos realized in the combined {\textsc{illustris}}TNG simulations and nearly one hundred thousand halos realized in {\textsc{flamingo}}.  The ratio of TNG300 to {\textsc{flamingo}}-L1\_m8 counts is fairly consistent with the volume ratio of the realizations (36), while the TNG-{\textsc{cluster}} sample, being biased to very high mass at $z=0$ provides nearly the same number of halos at $z=2$ (320) as $z=0$ (352). 

In the KLLR analysis of each sample, the upper bound on the center of the mass range is set by the 21st most massive halo. More massive halos contribute to the KLLR property estimates.  The similarity of the TNG-{\textsc{cluster}} and {\textsc{flamingo}} realization volumes means that the upper mass limits of the simulations are similar at all redshifts, lying just below ({\textsc{flamingo}}) or above (TNG) a value of $10^{15}\msun$ at redshift zero.  

From this point onwards, we will refer to the joint {\TNGcombined} as TNG and {\Flam} as {\textsc{flamingo}}.  Where appropriate we will explicitly distinguish between halos in TNG300 and TNG-{\textsc{cluster}} samples. 


\section{Mass Proxy Quality (MPQ) Assessment} \label{sec:MPQ}

Here we present the mass and redshift dependence of the MPQs (\S\ref{sec:E14-model-MPQ}) for the five primary properties listed in Table~\ref{tab:prop-definitions}.  
We first examine proxy quality at $z=0$, then explore evolution with redshift.  The underlying MPR behaviors for the five gas properties are presented in \S\ref{sec:MPR-parameters}, and their correlations in \S\ref{sec:correlations}.

The MPQs are derived by applying Equation~\eqref{eq:mass-proxy-quality} on the KLLR estimates of MPR scatter and slope. KLLR was performed using a kernel width of $0.2$ dex with centers beginning at $\Mfivehc = 10^{13} \msol$ and incrementing by $\sim 0.1$ dex up to the maximum value set by the 21st most massive halo at each epoch.  
Unless otherwise stated, all shaded regions in the figures below indicate $1\sigma$ estimated uncertainties using $1000$ bootstrap samples.

\subsection{Redshift Zero Behaviors}\label{sec:MPQzero}

Figure~\ref{fig:MPQzero} presents KLLR estimates of the logarithmic true halo mass scatter, $\sigmamus(\mu, z=0)$, for $\Mgas$ (black), $\Tsl$ (orange), $\LX$ (blue), $\YX$ (green), and $\YSZ$ (violet) in the halo samples of TNG (solid lines) and {\textsc{flamingo}} (dashed lines). 

The MPQs of all properties show a dependence on halo mass, with generally higher values of $\sigmamus$ seen at the group scale. Some MPQs decrease monotonically with halo mass while others decline toward the group/cluster boundary then rise again at the highest masses.  

The amount of gas and its thermal energy content are two key bulk measures accessible by X-ray and SZ observations.  Figure~\ref{fig:MPQzero} shows that both simulations find $\YSZ$ to be the best mass proxy at the group and low-mass cluster scale, with values starting at $\sim 13\%$ declining to $6\%$ near $10^{14.5} \msol$.  Above this halo mass scale, $\Mgas$ is a slightly better proxy, with $\sigmamus$ declining steadily to a value of $4\%$ at $10^{15} \msol$.  This minimum value is consistent in both simulation samples. 

As explained below, the MPRs of $\Mgas$ and $\YSZ$ differ in subtle ways. Consistently in TNG and {\textsc{flamingo}}, both the slope and scatter in the $\Mgas$ MPR gently decline across the cluster mass range, in a manner that produces the declines seen in Figure~\ref{fig:MPQzero}.  Those of $\YSZ$, however, tend toward near constant values of $0.1$ in scatter and a nearly self-similar slope, $\alpha = 5/3$, yielding the nearly constant MPQ of roughly $5\%$ for cluster-scale halo with $\Mfivehc > 10^{14} \msol$.  

These bulk measures integrate over the internal thermal structure of the hot ICM plasma.  The phase space structure expressed as filling fractions in the plane of density and temperature/entropy is driven by a number of physical and numerical factors.  We now turn to examine two properties that are sensitive to this internal structure. 

\subsubsection{Non-monotonic MPQ behaviors}\label{sec:MPQnonMono}

At all mass scales and in both simulation samples, the two least effective mass proxies are X-ray luminosity, $\LX$, and temperature, $\Tsl$.  Specific behaviors emerge for the two simulation methods, particularly for $\LX$ at the highest halo masses. 

In {\textsc{flamingo}}, the MPQs of $\LX$ and $\Tsl$ start near a value of 0.2 and decline at different rates with increasing halo mass. In $\Tsl$, the mass scatter is minimized  near $10\%$ at $10^{14} \msol$ at which scale the value for $\LX$ is significantly higher, $\sim 15\%$.  For $\Tsl$ the mass scatter rises at larger halo masses while that of $\LX$ remains nearly flat. As a result, the MPQ of both properties track each other above $10^{14.5} \msol$, gently rising to a value $0.16$  at $10^{15} \msol$. 

For TNG, the MPQ of $\Tsl$ has a similar shape to that of {\textsc{flamingo}}, but its magnitude is shifted consistently high, by $\sim 5\%$, throughout the halo mass range shown.  The MPQ of $\LX$ is strongly mass-dependent, falling from $35\%$ to $16\%$ in the group mass regime before rising again to values above $50\%$ at the highest masses.

Differences in internal thermal structure, particularly near the site of strong AGN feedback within the core region, are likely to be driving the sample-based MPQ differences for $\LX$ and $\Tsl$ seen in Figure~\ref{fig:MPQzero}.  We show in Appendix~\ref{sec:LX-vs-LXce} that core-excision significantly lowers the mass scatter for X-ray luminosity in TNG at high halo masses, bringing it into better agreement with {\textsc{flamingo}} sample behavior. 

In Appendix~\ref{sec:Tmw}, we show that the MPQs for mass-weighted temperature, $\Tmw$, in the two simulations are more consistent in magnitude compared to the case of $\Tsl$. Again, the density weighting of the latter makes it more susceptible to thermal structure of the plasma.  Note that we excise the core gas contribution when determining $\Tsl$, so the phase structure contributions lie at radii beyond $0.15 \Rfivehc$.  We defer a detailed examination of thermal phase structure in these simulations to future work. 

\subsubsection{Thermal energy measures}\label{sec:MPQthermal}

At the group scale, $\YX$ and $\YSZ$ perform better than $\Mgas$ for both simulation samples.  The $\YX$ MPQ is nearly identical to that of $\Mgas$ in TNG for halos with masses $3 \times 10^{13}\msun < \Mfivehc < 10^{14}\msun$. A the cluster scale, $\YX$ and $\YSZ$ exhibit a higher mass scatter than $\Mgas$ in both simulations for halos larger than $3 \times 10^{14}\msun$, making $\Mgas$ the best mass proxy for the largest halos. The primary reason for this is the increased scatter of $\YX$ at the cluster scale and the mass-independent scatter of $\YSZ$, while the heightened correlation between $\Mgas$ and $\Tsl$ or $\Tmw$ increases the mass scatter as a secondary effect (see Section \ref{sec:correlations}).

The increase in $\YX$ mass scatter at the cluster scale is more pronounced in TNG compared to that in {\textsc{flamingo}}. The reason for this is that TNG-{\textsc{cluster}} halos contain more hot gas at high densities than their TNG300 counterparts, which drives extra variance in $\Tsl$ and, thereby, $\YX$.  The MPQ of $\YX$ in the TNG300 sample alone agrees more closely with values seen in {\textsc{flamingo}} at cluster masses.

\subsection{MPQ Evolution with Redshift}\label{sec:MPQzevol}

\begin{figure}
    \includegraphics[width = \linewidth]{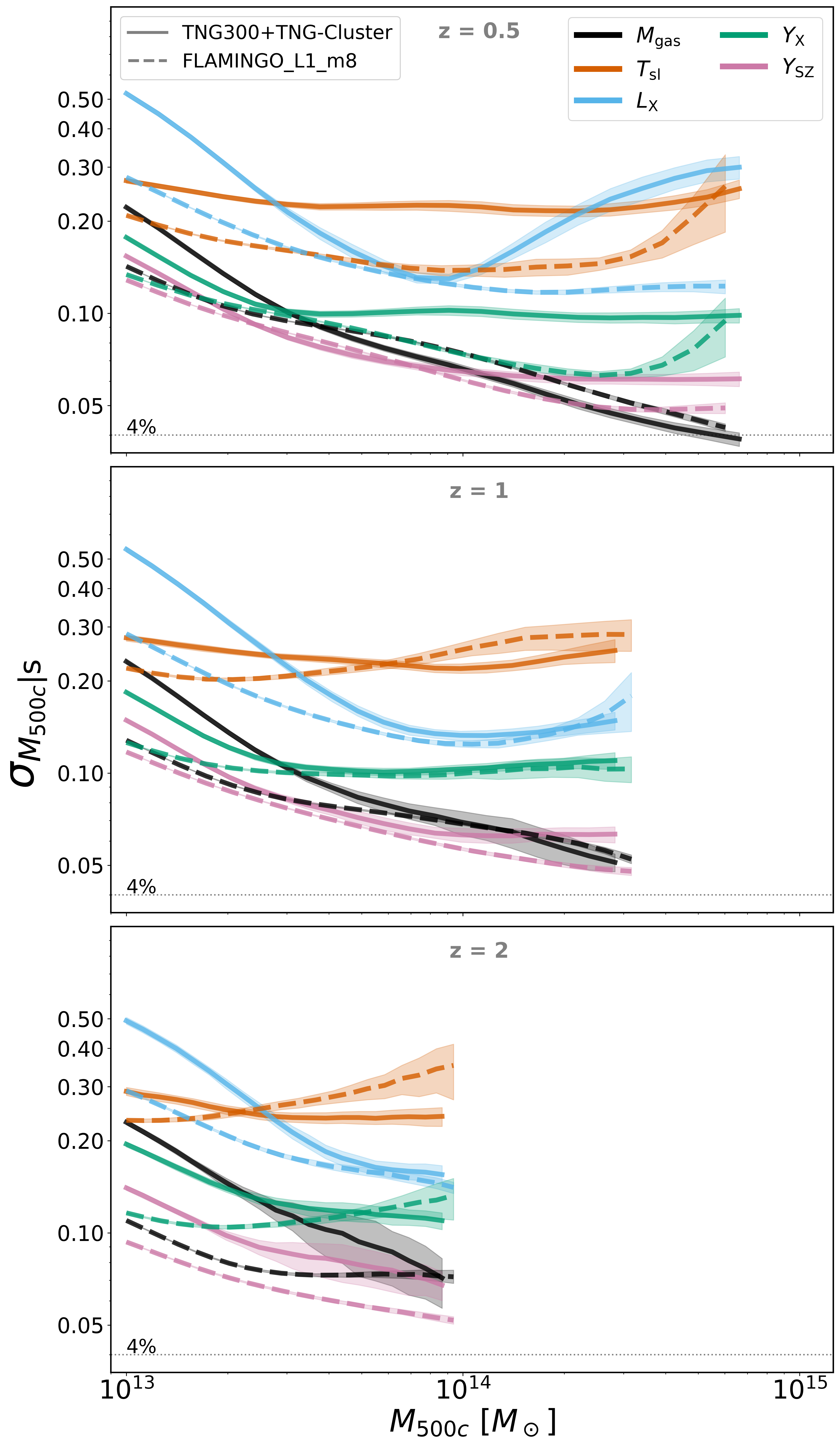}
    \caption{Halo mass and redshift dependence of the MPQs using the same format as in Figure~\ref{fig:MPQzero} for redshifts 
    $z = 0.5$ (top), $1$ (middle), and $2$ (bottom).  At each redshift, lines extend up to the mass of the $21^{\rm st}$ ranked halo in each simulation sample.}
    \label{fig:MPQ-redshifts}
\end{figure}

Figure \ref{fig:MPQ-redshifts} shows true halo mass scatter, $\sigmamus(\mu, z)$, at redshifts $z = 0.5,\ 1,\ 2$ (top to bottom).  Colors and scales are consistent in each panel and consistent with that of Figure~\ref{fig:MPQzero}.  

One way to consider this evolution is to examine the highest mass scale representing rare peaks in the initial density field able to have collapsed at each redshift \citep{Bardeen1986, Bond1991excursionSetPS}. Effectively the mass limit represents a fixed comoving number threshold of $20$ objects per cubic gigaparsec. The MPQs for $\YSZ$ and $\Mgas$ at this limit remain close to their $z=0$ values, lifting gently above the $4\%$ dotted line at $z \ge 1$.  

In general, the redshift dependence of MPQs is weaker than the mass dependence.  An exception is the case of $\LX$ mass scatter in TNG.  We anticipated that the large discrepancy with {\textsc{flamingo}} at $z=0$ is due to the inclusion of dense hot gas residing near large central galaxies.  This effect is sensitive to redshift.  At $z>1$ the MPQs for $\LX$ of the two simulations lie in agreement at the highest sampled masses.  The MPQ values diverge toward the lowest masses, as they do at $z=0$ but at higher absolute values.  

The redshift evolution of the hot gas mass MPQ behaves differently in the two simulation samples. While similar values are found at $z=0$ across the entire halo mass range, by $z=2$ a difference of a factor of two emerges at $10^{13}\msol$. We show in the following section that this is primarily driven by redshift-independent scatter in $\Mgas$ at this mass scale in the TNG halo sample.

At the mass scale of $10^{14}\msol$, halos are sufficiently numerous to be accessible to observations across the full redshift range, $0<z<2$, studied here. At that scale, both simulations provide a reasonably consistent picture of mass proxy quality.  The measures $\YSZ$, $\Mgas$, and $\YX$ provide 0.1 or better total mass scatter across all redshifts, while $\Tsl$ and $\LX$ have scatter greater than 0.1. 

The relatively poor performance of X-ray luminosity should not be interpreted as a mark against cluster samples selected by X-ray flux.  At $10^{14}\msol$, we note that there is a nearly redshift-independent value of the MPQ of $\sim 0.15$, consistent in both simulated samples.  As we show in Appendix~\ref{sec:LX-vs-LXce}, the divergence in behaviors at higher masses is due to excess scatter in core emission in TNG.  This is not true at the group scale, however.  

The MPQ of thermal SZ is also nearly redshift independent at $10^{14}\msol$, with values $6-8\%$. Projection effects will broaden this intrinsic measure, however, in a manner that will generally depend on cluster mass, redshift and angular scale of the measurement \citep[\eg][]{Shaw2008, Gupta2017}.  

\subsection{MPQ vs. Direct True Mass Scatter Measurement}\label{sec:directMassScatter}

We now examine how well the MPQ measure of equation \eqref{eq:mass-proxy-quality} reflects direct measurement of true mass scatter obtained by performing the KLLR fits to the MPR inverse, $\Pr(M|S,z)$. We perform this regression for each gas property with minimum values set by requiring the sample be mass complete above $\Mfivehc = 10^{13.25}\msol$.  A KLLR kernel width of $0.2$ dex is applied to all. The log-mean mass of each property is used to map each to a common halo mass axis.

\begin{figure}
    \includegraphics[width=\linewidth]{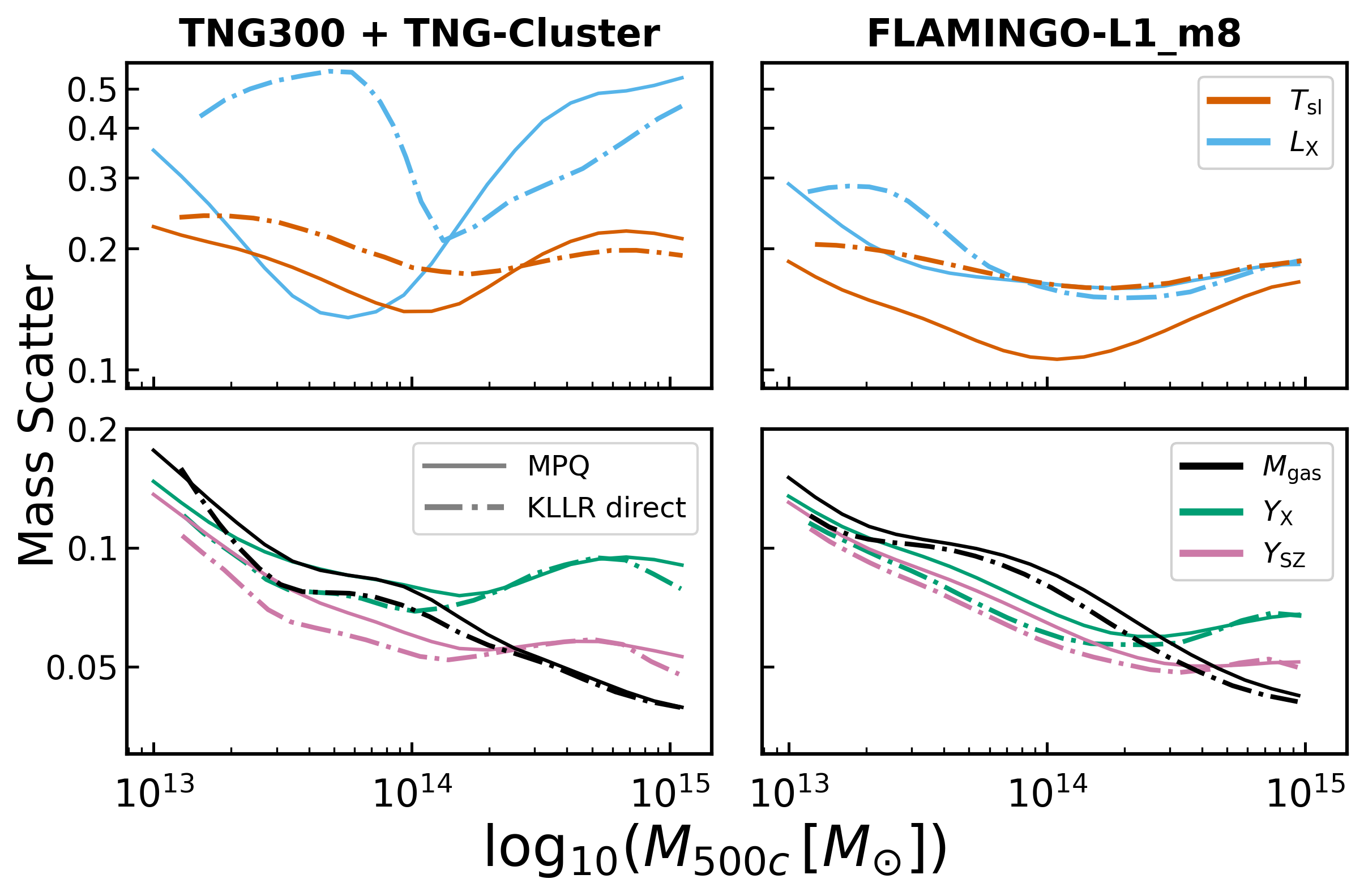}  
 \vspace{-12pt}
    \caption{Comparison of $z=0$ scatter in true halo mass inferred by the MPQ, equation~\eqref{eq:mass-proxy-quality} (solid lines), with values directly measured by KLLR applied to the conditional likelihood, $P(M | S)$, of each hot gas property, $S$ (dot-dashed).  The lower panels show good agreement for properties that adhere well to log-normality while the agreement degrades in the upper panels for properties with more complex likelihood shapes (see \S\ref{sec:kernel-shapes}).  
    }
    \label{fig:MPQ-bias}
\end{figure}

Figure \ref{fig:MPQ-bias} compares the $z=0$ MPQ estimates (solid lines) with the direct halo mass scatter measurements (dot-dashed) for TNG (left) and {\textsc{flamingo}} (right).  Lower panels show cases of properties which adhere closely to log-normal likeihood shapes ($\Mgas$, $\YSZ$ and $\YX$).  The MPQ values tend to lie somewhat higher than the direct measures, but the differences are consistently below $20\%$.  The upper panels show that larger discrepancies exist for $\LX$ and $\Tsl$.  Both of these properties have more complex conditional likelihood shapes, as discussed in \S\ref{sec:kernel-shapes}, so the larger disagreement is expected.  Importantly, the mass-dependent shapes and relative ordering of mass proxy quality are consistent in both measures, with the exception of the behavior of X-ray luminosity in the TNG sample.  We show in Appendix~\ref{sec:LX-vs-LXce} that core-excised X-ray emission adheres more closely to a log-normal shape in both simulation samples. 

We note that a similar assessment of the E14 model was performed by \citet{Farahi2018}, but that work examined the mean halo mass conditioned on either hot gas mass or total stellar mass. \citet{Farahi2018} demonstrate that the second-order HMF correction term for mean halo mass achieves a sub-percent level accuracy relative to direct KLLR measures (see their Figure 8) at the cluster scale. This work is the first to test E14 estimates of the second moment.


\section{Mass-Property Relation (MPR) Analysis}\label{sec:MPR-parameters}

We now turn to the mass and redshift dependence of the mass--property relations themselves. 
In Section \ref{sec:slope-scatter}, we present the slope and scatter of each property in light of the MPQ trends seen above. 
Property normalizations, framed as deviations from simple self-similar expectations, are  discussed in Section \ref{sec:normalization-redshift-evolution}. 

We selectively compare our results with those of past simulation studies in this section, offering a more extended review of MPR slope and scatter in Section \ref{sec:literature-comparison}. In future work, we plan to investigate the impact of different AGN and supernovae feedback models and to compare them with observations using the physics variations provided by the {\textsc{flamingo}} simulation suite \citep{Schaye2023Flamingo}.

\begin{figure*}
    \includegraphics[width = 0.8\textwidth]{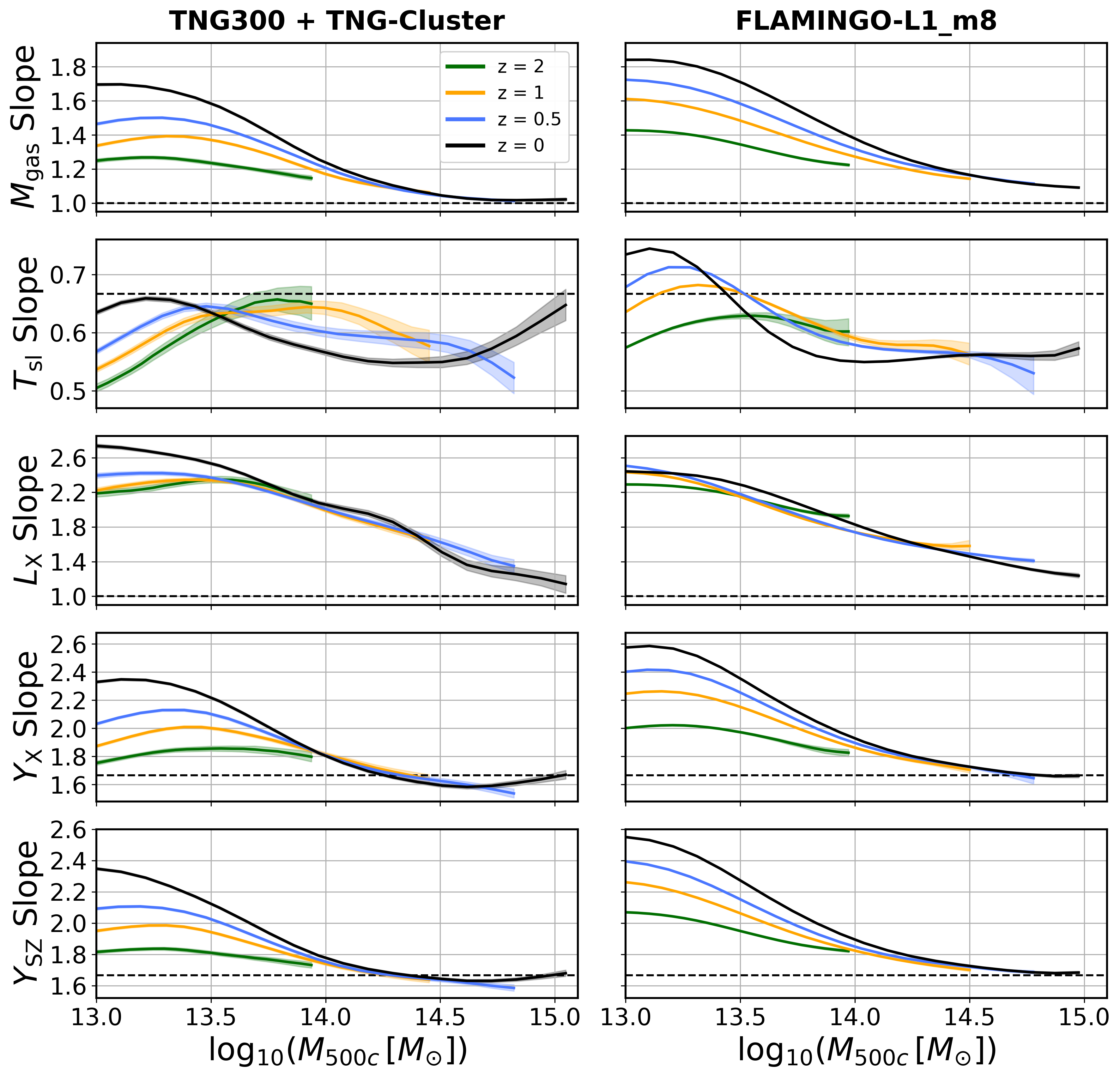}
    \caption{Halo mass and redshift dependence of the MPR slopes for (top to bottom) 
    $\Mgas$, $\Tsl$, $\LX$, $\YX$, $\YSZ$.  The 
    KLLR-derived values (Equation~\eqref{eq:KLLR-scaling}) are shown at redshifts $z = 0$ (black), $0.5$ (blue), $1.0$ (gold), and $2$ (green) for TNG (left) and {\textsc{flamingo}} (right) halo populations. Dotted lines indicate the self-similar slopes given in \S\ref{sec:self-similar-theory}. Note that, by definition, the slope in $\YX$ is the sum of the slopes in $\Mgas$ and $\Tsl$. }
    \label{fig:MPR-slope-plots}
\end{figure*}

\subsection{Slope, scatter, and their relationships to MPQ}\label{sec:slope-scatter}
In Figures \ref{fig:MPR-slope-plots} and \ref{fig:MPR-inrinsic-scatter-plots}, we present the mass and redshift dependent slopes and scatters --- resulting from Equation ~\eqref{eq:KLLR-scaling} for the KLLR slope and Equation ~\eqref{eq:KLLR-scatter} for the KLLR scatter---, respectively, for the five primary gas properties at four redshifts.
The column on the left presents results for the TNG halo population and the right column shows those of {\textsc{flamingo}}. Horizontal dotted lines in each panel of Figure~\ref{fig:MPR-slope-plots} display the self-similar model expectations for the property slopes (Section \ref{sec:self-similar-theory}).

\subsubsection{Hot gas mass}\label{sec:Mgas-M500c-slope-scatter}
For the hot gas mass-halo mass relation, the slope is steeper than the self-similar value of $1$ throughout the mass range across all redshifts. In TNG, the slope decreases from $1.7$ to $1.02$ as mass increases from $\Mfivehc  = 10^{13}\msol$ to $10^{15}\msol$ at $z = 0$, while in {\textsc{flamingo}} it is systematically steeper by $10-20\%$ for all redshifts, approaching a value of $1.09$ for the largest halos at $z = 0$. This significant deviation from the self-similar expectation can be attributed to two primary factors: (1) the conversion of gas content into stars through radiative cooling and star formation \citep{Nagai2007, Battaglia2013, Planelles2014, LeBrun2017, Truong2018}, and (2) the ejection of gas by AGN feedback events, which occurs more efficiently in lower mass halos due to their shallower potential wells \citep{Puchwein2008, Fabjan2010, McCarthy2011, Gaspari2014, LeBrun2017, Truong2018, Ayromlou2023}. Specifically, \citet{Ayromlou2023} show that the ratio of closure radius $R_{\rm c}$, the radius within which all baryons associated with a halo can be found, to $\Rtwohc$ for TNG groups is more than three times larger than that for clusters. The deeper potential well of clusters helps retain their gas content within $\Rtwohc$.

Both simulations show a $10-20\%$ decrease in slope with increasing redshift for halos with mass $\Mfivehc \lesssim 10^{14}\msun$, while the slope remains approximately redshift-independent for more massive halos. This trend aligns with previous numerical studies \citep{LeBrun2017, Farahi2018, Henden2019}, which link it to the increased binding energy at a fixed halo mass with increasing redshift, $E_{\rm bind}(z) \propto \Mfivehc^{5/3}E(z)^{2/3}$. At fixed mass, the larger binding energy with increased redshift suggests that it is more difficult to eject gas due to AGN feedback at larger redshifts, resulting in a shallower $\Mgas$ slope. Moreover, \citet{Ayromlou2023} show that $R_{\rm c}/\Rtwohc$ decreases with increasing redshift, suggesting that halos retain an increasing fraction of their hot gas at higher redshifts. \citet{Truong2018} observe a slight increase in the $\Mgas$-$\Mfivehc$ slope for $z > 1$ which they attribute to declining gas fractions in lower mass halos at those redshifts.

\begin{figure*}
    \includegraphics[width=0.8\textwidth]{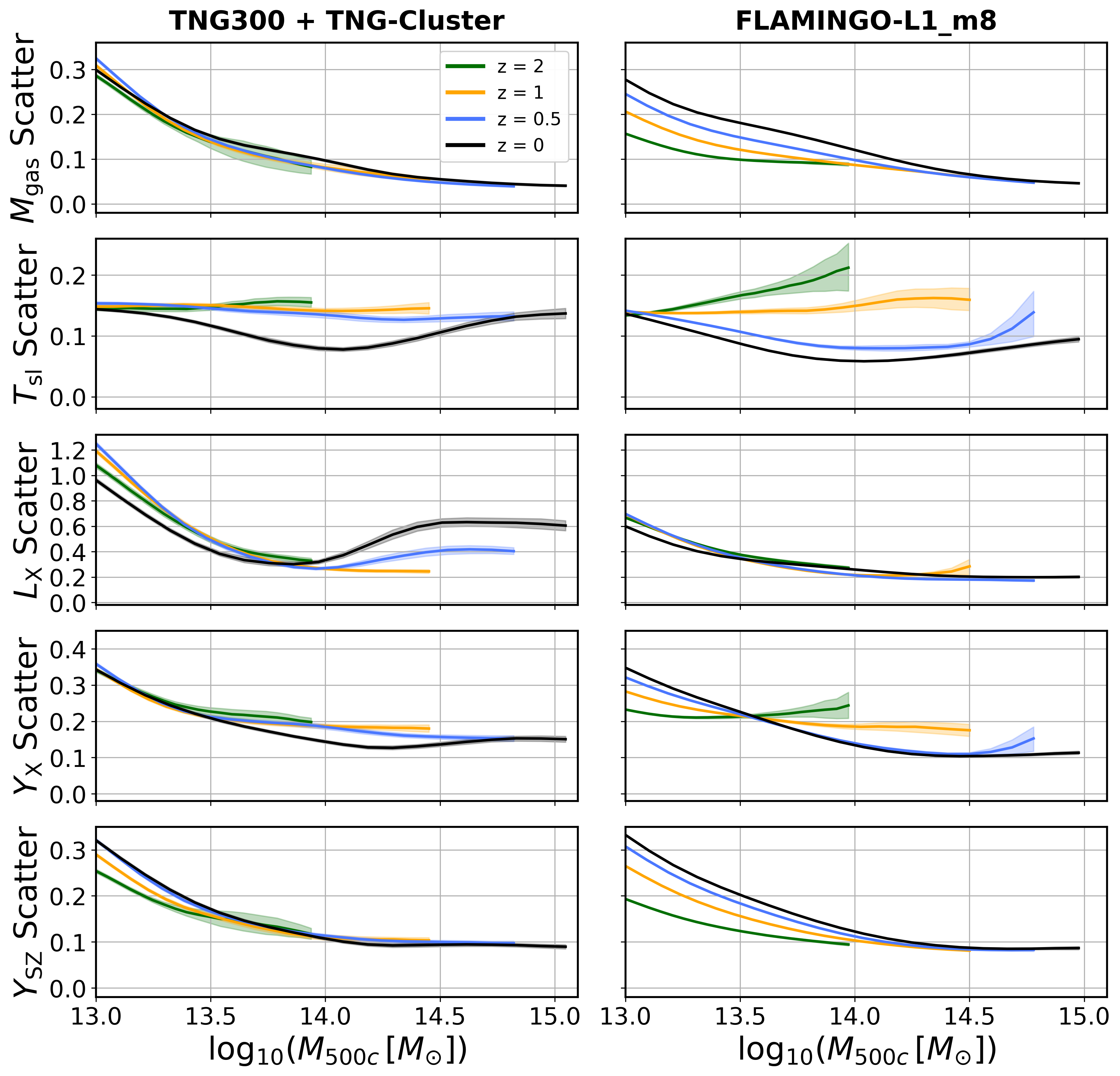}
    \caption{Halo mass and redshift dependence of the MPR intrinsic scatter values (Equation~\eqref{eq:KLLR-scatter}). Format is identical to that of Figure~\ref{fig:MPR-slope-plots}.}
    \label{fig:MPR-inrinsic-scatter-plots}
\end{figure*}

The intrinsic scatter in the $\Mgas$-$\Mfivehc$ relation, shown in the top panels of Figure~\ref{fig:MPR-inrinsic-scatter-plots}, decreases monotonically from $\sim 30\%$ for low-mass groups, reaching $\sim 4\%$ for the most massive halos.  Consistent with other simulation studies \citep{LeBrun2017, Farahi2018, Henden2019}, {\textsc{flamingo}} shows a $10\%$ reduction of the scatter only at the group scale with increasing redshift, likely due to the reduced influence of non-gravitational physics on galaxy formation as the binding energy rises at fixed mass. In contrast, the scatter is redshift-independent in TNG throughout the mass range corroborating the results by \citet{Barnes2017a} that show no redshift dependence in the scatter. 

The declining MPQ for $\Mgas$ at $z=0$ with increasing halo mass is primarily due to the falling $\Mgas$ scatter, with the mass-dependent slope having a softening effect on the magnitude range from groups to clusters. 
In {\textsc{flamingo}}, the decrease in slope with redshift compensates the decrease in scatter, rendering the MPQ to be approximately redshift-independent. Conversely, in TNG, the lower slope and constant scatter at the group scale with increasing redshift lead to a rise in mass scatter for $z > 0$. Though {\textsc{flamingo}}'s slope is larger than that of TNG, the scatter at most masses is also slightly larger, leading to an agreement in the $\Mgas$ MPQ between the two simulations.

Overall, the finding of very small, redshift-independent scatter in $\Mgas$ at high cluster masses supports the use of X-ray gas mass fraction as a cosmological probe \citep{Mantz2022gasfractions}.

\subsubsection{Gas temperature}\label{sec:Tsl-M500c-slope-scatter}
The $\Tsl$--$\Mfivehc$ relation lacks a clear monotonic trend in both the slope and the scatter compared to the other properties, the range of both parameters is relatively modest.  At $z = 0$, the slope in TNG is slightly shallower than the self-similar expectation of $2/3$, but increases toward this value for halos with $\Mfivehc > 10^{14.5}\msun$, in agreement with \citet{Pop2022} who use a smoothly broken power-law and \citet{Kay2024} who implement KLLR on the lower resolution {\textsc{flamingo}}-L1\_m9. In contrast, in {\textsc{flamingo}}, the slope is slightly steeper than self-similar for $\Mfivehc \lesssim 10^{13.5}\msun$, becoming shallower for larger halos, approaching a value of $0.57$ for $\Mfivehc > 10^{14.5}\msun$.  As shown in Appendix~\ref{sec:Tmw}, many of these features are also apparent in the mass-weighted temperature, $\Tmw$, although the scatter is considerably smaller, consistent with \citet{Kay2024}.  

Departures from the self-similar expectation, where the temperature reflects only the depth of the potential well, can be physically linked to the steeper than self-similar slopes observed for $\Mgas$-$\Mfivehc$ seen above. Radiative cooling removes low entropy gas through the formation of stars, decreasing gas density and increasing temperature. Moreover, AGN feedback ejections remove low entropy gas and increase the average entropy, thereby increasing the temperature while lowering the gas fraction. This increase in temperature flattens the slope of the temperature-halo mass scaling relation.

The slopes of both the $\Tsl$--$\Mfivehc$ and $\Tmw$--$\Mfivehc$ relations slightly increase with redshift in the mass range $10^{13.5}\msun < \Mfivehc < 10^{14.5}\msun$ and decrease by $\sim 15\%$ for halos with $\Mfivehc < 10^{13.5}\msun$. This aligns with the modest increases reported in the slope of the $\Tmw$--$\Mfivehc$ relation by \citet{LeBrun2017} and the slope of the $\Tsl$--$\Mfivehc$ relation reported by \citet{Truong2018}. Other numerical studies \citep{Barnes2017a, Henden2019} report no significant redshift-induced changes in the temperature-halo mass relation. Because the effectiveness of AGN feedback in expelling gas from halos of a fixed mass diminishes with redshift, the departure of the temperature slopes from self-similarity become less pronounced with increasing redshift.

At $z = 0$, the scatter in $\Tsl$ exhibits weak mass dependence, decreasing from $14\%$ at the group scale to $\sim 8\%$ at $\Mfivehc = 10^{14}\msun$, then increasing for larger halos. These values are consistent with the intrinsic scatter of the temperature-halo mass relation reported by \citet{Nagai2007, LeBrun2017, Henden2019} who measure a scatter of $14\%$, $5\%$ and $9-12\%$, respectively.  

In TNG, the scatter in $\Tsl$ increases at $z = 0.5$ relative to $z = 0$ but then remains both mass- and redshift-independent for $z \geq 0.5$. In {\textsc{flamingo}}, the scatter closely follows that of TNG at $z=0$, but the  increase with redshift is delayed to $z > 0.5$.
In contrast, we show in Appendix~\ref{sec:Tmw} that the $\Tmw$ scatter is redshift-independent for $\Mfivehc \gtrsim 10^{13.5}\msun$.  Above $10^{14} \msol$, the scatter lies between 5 and 7 percent independent of redshift in both simulation samples.  From the $\Tsl$ panels of Figure~\ref{fig:MPR-inrinsic-scatter-plots}, we see that the two simulations produce remarkably similar scale-dependent scatter at $z=0$, with non-monotonic behavior minimized at $10^{14} \msol$.  In TNG, the scatter is roughly $40\%$ higher at the upper mass limit of $10^{15} \msol$. 
At larger redshifts, the scatter at $10^{13}\msol$ remains consistent for both simulations, with amplitude close to $0.15$, but TNG halos show no mass or redshift dependence about this value whereas the {\textsc{flamingo}} samples show mild dependence on both.

Overall, the mild scale behavior of the slope and scatter are reflected in the  approximately flat $\Tsl$ MPQ seen in Figures~\ref{fig:MPQzero} and \ref{fig:MPQ-redshifts}. For $z <  1$, the lower intrinsic scatter of $\Tsl$ in {\textsc{flamingo}} leads to a reduced mass scatter compared to TNG values.  This difference is not apparent at $z \ge 1$.  

\subsubsection{X-ray luminosity}
The slope of the $\LX$-$\Mfivehc$ relation is significantly steeper than the self-similar expectation of $1$, particularly at the group scale, consistent with the bolometric luminosity-total mass relation slopes reported previously \citep{LeBrun2017, Barnes2017a, Truong2018, Henden2019, Pop2022}. At $z = 0$, the slope decreases approximately linearly from the mid-group scale to massive clusters, from $\sim 2.3$ to $\sim 1.2$, approaching the self-similar expectation of one for the most massive halos. The nearly redshift-independent, linearly declining slope for halo mass scales $> 10^{13.5} \msol$ indicates that a quadratic form for the $\LX-M$ relation would sufficiently capture the scale dependence.  

Since the X-ray luminosity is a volume integral of the square of the gas density, the departure from self-similarity in $\Mgas$,
is reflected in the steep $\LX$ scaling with halo mass. In Appendix~\ref{sec:LX-vs-LXce}, we show that, unlike the scatter, the slopes are largely unchanged when cores are excised.  The scatter is dramatically reduced for the TNG sample but less so for {\textsc{flamingo}} (see Figure \ref{fig:LXce-slope-scatter}).  

Strikingly, $\LX$ parameters exhibit some of the weakest redshift evolution among all gas properties, particularly in {\textsc{flamingo}}.  We note that the exclusion of recently heated gas particles in that simulation will reduce upward excursions in $\LX$. For TNG-{\textsc{cluster}} halos at $z=0$, there is a sub-population with X-ray luminosities  nearly one dex higher than those observed \citep{Nelson2024}, which can be attributed to the abundance of cool cores in TNG-{\textsc{cluster}} halos \citep{Lehle2024}: $24\%$ and $60\%$ of the $352$ TNG-{\textsc{cluster}} halos are strong and weak cool cores, respectively. This tail drives the TNG scatter in Figure~\ref{fig:MPR-inrinsic-scatter-plots} upward to a value of $0.6$ at the massive cluster scale, significantly higher than the $0.2$ value for the {\textsc{flamingo}} sample. The scatter in TNG-{\textsc{cluster}} for high halo masses is significantly larger ($\sim 50\%$) than that in TNG300 at the same mass scale. In Figure \ref{fig:LXce-slope-scatter}, we observe that $\LXce$ does not exhibit the same upturn in intrinsic scatter as $\LX$ for high mass halos which are dominated by the TNG-{\textsc{cluster}} sample, providing further evidence that the increased scatter is a result of the abundance of luminous cool-cores.

Figure~\ref{fig:MPR-slope-plots} shows that TNG slopes lie slightly above those of  {\textsc{flamingo}}, consistent with the slight elevation in $\Mgas$ slopes. For poor groups, the TNG population shows a steepening slope over time that is more pronounced than that seen for {\textsc{flamingo}}. The lack of redshift evolution in slope at scales $> 10^{13.5} \msol$, noted above, contrasts with prior findings \citep{LeBrun2017, Truong2018, Henden2019} of slopes that increase with redshift.  We note that fixed-slope linear regression will be probing lower halo mass ranges at larger redshift, so the drift reported may simply reflect population evolution in the HMF.  Consistent with our findings, \cite{Barnes2017a} report no change in the slope of the core-excised bolometric luminosity with redshift.

Despite similar slopes, the intrinsic scatter in Figure~\ref{fig:MPR-inrinsic-scatter-plots} differs between the two simulations.  They are closest at the group-cluster transition scale of $10^{14} \msol$, where values $0.20-0.25$ are seen at $z \ge 0.5$. The TNG population scatter is consistently larger at the group scale  at all redshifts, reaching values above one at $10^{13} \msol$.

As noted in Section \ref{sec:MPQ}, the late-time ($z<1$) increase in TNG scatter at high halo masses degrades the MPQ of $\LX$ on those scales. The scatter in core-excised luminosity does not exhibit this upturn and 
aligns better with the scatter in {\textsc{flamingo}} halos.  For {\textsc{flamingo}} clusters, the scatter is essentially flat at $\sim 20\%$ above $\Mfivehc \gtrsim 10^{14}\msun$. In Figure~\ref{fig:MPQzero}, the MPQ of $\LX$ in {\textsc{flamingo}} gently rises with mass at these scales because the slope gently falls.


\subsubsection{Gas thermal energy}

The total thermal energy measure, $\YSZ$, is a fundamental measure of halo gas since  pressure gradients strive to balance gravity as a halo dynamically relaxes \citep{Lau2009}.  As the product of gas mass and mass-weighted temperature, the scatter in $\YSZ$ is sensitive to the correlation between $\Tmw$ and $\Mgas$ conditioned on halo mass. Similarly, the scatter in $\YX$ depends on the correlation between $\Tsl$ and $\Mgas$. We discuss property correlations in \S~\ref{sec:correlations}. 

At group-scale masses, the $\YX$-$\Mfivehc$ and $\YSZ$-$\Mfivehc$ relations display slope and scatter behaviors similar to those of the $\Mgas$-$\Mfivehc$ relation. Both relations have slopes that deviate increasingly from self-similar over time. At $z = 0$, the maximum slopes of $\YX$ ($\YSZ$) in TNG and {\textsc{flamingo}} are $2.33$ ($2.34$) and $2.58$ ($2.55$), respectively. 
Similar to $\Mgas$, slopes at all redshifts are consistently larger by $\sim 10-20\%$ in {\textsc{flamingo}} relative to TNG.

The scatter in $\YX$ and $\YSZ$ decreases from $\sim 30\%$ at $\Mfivehc = 10^{13}\msun$ to $\sim 10\%$ and $12\%$ at $\Mfivehc = 10^{14}\msun$, respectively. While this scatter is larger than the value of $0.053$ found in \cite{Nagai2007}, it aligns with \cite{LeBrun2017, Barnes2017a, Henden2019} who measure values of $10\%$, $12\%$ and $14\%$, respectively. For more massive halos, the scatter in $\YX$ plateaus at $\sim 10-15\%$ for halos with $\Mfivehc > 10^{14}\msun$. In {\textsc{flamingo}}, the redshift dependence of the $\YX$ scatter is confined to groups, decreasing with increasing redshift.  The larger scatter in $\YX$ at the massive cluster scale allows $\Mgas$ to be a superior mass proxy for halos with $\Mfivehc > 10^{14.5}\msun$.

Unlike the previous properties, Figure~\ref{fig:MPR-slope-plots} shows that the SZ signal, $\YSZ$, scales in a nearly self-similar manner at high halo masses.  
Above $10^{14.5} \msol$ in both simulations, the slopes of both gas thermal energy measures lie within $0.05$ of the $5/3$ self-similar value.  This finding is consistent with several existing results \citep{LeBrun2017, Barnes2017a, Truong2018, Henden2019, Pop2022}. 

At high masses, Figure~\ref{fig:MPR-inrinsic-scatter-plots} shows that the scatter in $\YSZ$ in both simulated samples are remarkably mass and redshift independent with values close to $0.1$.  Hence, the MPQ of the SZ thermal energy  at $z=0$ (Figure~\ref{fig:MPQzero}) is nearly constant, taking values of $6-8\%$ for high mass halos at all redshifts in both populations.   

This finding reinforces observational efforts to identify clusters using the SZ effect \citep{Sehgal2011ACTclustercosmo, deHaan2016SPTclustercosmo, PlanckXXIV2016clustercosmo, Bocquet2019SPTclustercosmo, SPT2024}.  A recent Atacama Cosmology Telescope (ACT-DR6) sample of nearly 10,000 clusters is nearly complete for halos with $10^{14.3} \msol$  for $z\le 1$ (Bolliet \textit{et al.}, in prep.).  We find that the variance in $\YSZ$ above this mass scale is $(0.098 \pm 0.004)^2$ and $(0.082 \pm 0.002)^2$ for TNG and {\textsc{flamingo}}, respectively.  Based on older simulations, the value assumed in the ACT analysis was $(0.2)^2$. We anticipate that future SZ survey analyses could benefit from more precise estimates of this variance, particularly when verified by multiple simulations.  


\subsection{MPR Mean Behaviors}\label{sec:normalization-redshift-evolution}

Over the two orders of magnitude in halo mass we study, the log-mean values of most properties varies by several orders of magnitude (see Appendix~\ref{sec:scaling-relations}). To enhance mass- and redshift-dependent features, we reduce the dynamic range by presenting in \S\ref{sec:normalization-redshift} each property's mass-dependent mean relative to the simple self-similar power-law form.  We use reference normalizations from KLLR at $10^{15} \msol$ and $z=0$ derived separately for each simulated halo sample.  

In \S\ref{sec:normalization-simulations} we present absolute $z=0$ normalizations for each simulated sample, finding agreement at the few percent level for thermal energies at the highest masses despite $\sim 15-30\%$ shifts in hot gas mass. 

\subsubsection{Deviations from self-similarity}\label{sec:normalization-redshift}

For each property $S_a$, let $\langle S_a(\mu,z) \rangle$ be the exponential of the log-mean MPR derived from KLLR analysis of a given halo sample.  We define the $z = 0$ self-similar relation, $S_{a, \rm SS}(\mu,0)$, as the single power-law scaling that matches the KLLR normalization for each population found at $\Mfivehc = 10^{15}\msun$ (because the highest mass halo used in the {\textsc{flamingo}} analysis is just below $10^{15}\msun$, we linearly extrapolate the normalization to $10^{15}\msun$) but with self-similar slope given in Section \ref{sec:self-similar-theory}. We then define deviations from self-similar scaling at any redshift by 

\begin{equation}\label{eq:deviation-evolution}
    \DeltaSS(\mu, z) = \frac{\langle S_a(\mu, z)\rangle}{E^{\beta}(z) S_{a, \rm SS}(\mu, z = 0)}
\end{equation}
where $\beta$ denotes the self-similar redshift evolution exponent described in Section \ref{sec:self-similar-theory}. We choose the $10^{15}\msun$ reference values at $z=0$ since the hot gas scalings are tending toward self-similar slopes at the highest masses.

\begin{figure}
    \includegraphics[width=\linewidth]{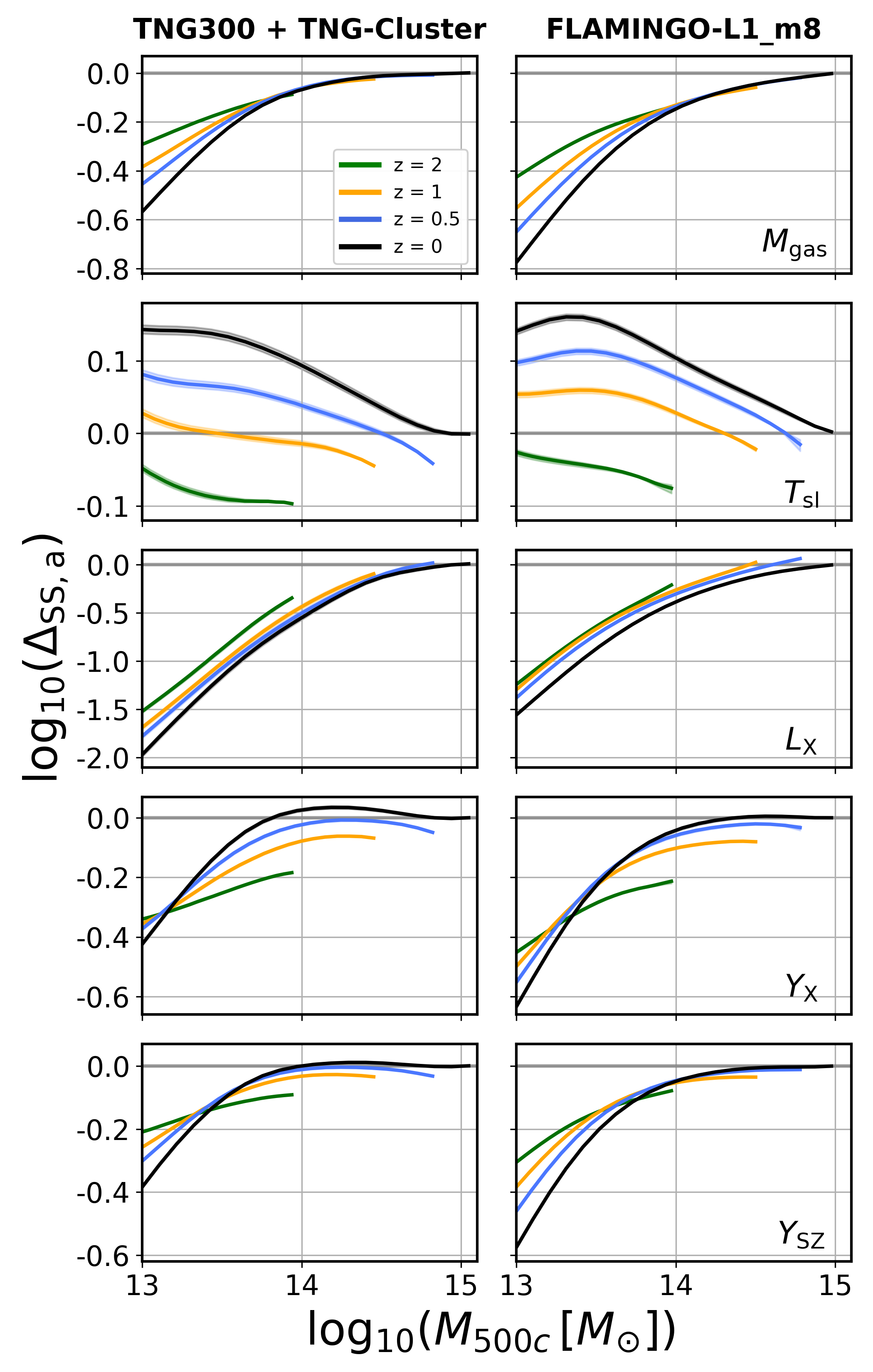}
\vspace{-8truept}
    \caption{Hot gas property mean values quantified in terms of differences from the self-similar behavior $\DeltaSS(\mu, z)$, equation~\eqref{eq:deviation-evolution}.  Note the logarithmic shifts are presented in dex.  For a non-scaled version, see Figure~\ref{fig:normalizations-self-similar}.  }
    \label{fig:deviations-of-evolution}
\end{figure}

Figure \ref{fig:deviations-of-evolution} presents the mass and redshift dependence of $\DeltaSS(\mu, z)$ for the five gas properties in each simulation (Figure~\ref{fig:normalizations-self-similar} offers the original KLLR fits).  In magnitude, the shifts vary least for $\Tsl$ ($-0.1$ to $0.15$ dex) and most for $\LX$ ($-2$ to $0$ dex) with gas mass and thermal energy being intermediate.   

Across a majority of the mass and redshift ranges shown there is striking similarity in the behaviors exhibited by the two simulation samples. This is not surprising given that the same astrophysical ingredients underlie both numerical methods. Star and SMBH formation removes hot gas and heats the remainder.  For halos of low mass, the actions develop over time to both reduce the gas mass and elevate the temperature.  Above $10^{14} \msol$, reduced efficiency of star and SMBH formation produces a higher gas mass and a progressively lower temperature. These trends are apparent in the slopes shown in Figure~\ref{fig:MPR-slope-plots}.

Above $\Mfivehc > 10^{14}\msun$, the $\Mgas$ mean values are independent of redshift, whereas below this mass scale the gas mass increases toward larger redshifts.  At the group scale, the higher binding energy at earlier times allows for more gas to be retained, in agreement with previous works \citep{LeBrun2017, Barnes2017a}.

In contrast, $\Tsl$ evolves at all mass scales such that the ICM plasma in halos of fixed mass is cooler, by up to 0.2 dex at $z=2$ compared to the scaled $z=0$ expectations.  While some of this effect may be driven by increased merger frequency and lack of complete virialization at high redshift \citep{LeBrun2017}, much is likely due to the heating effects of AGN feedback over time.

The X-ray luminosity shows large deviations in mass, with low mass halos depressed by an order of magnitude relative to self-similar expectations.  The redshift evolution, however, adheres much more closely to self-similar expectations, in that the mass-dependent shape is consistently reproduced, especially for TNG halos at $z \le 1$.  

For both $\YX$-$\Mfivehc$ and $\YSZ$-$\Mfivehc$ relations, there is a interesting sign change in the evolution at group scales, occurring at slightly different masses for the two simulations. At low masses, the evolution with redshift is positive, with higher thermal energies relative to the self-similar scaled $z=0$ values, while for larger halos with $\Mfivehc \gtrsim 10^{13.5}\msun$, it becomes negative.  Similar behavior is seen by \citet{LeBrun2017}.  This sign change is a consequence of the mass-dependent evolution of $\Mgas$ and the negative evolution of $\Tsl$ and $\Tmw$. At low masses, the positive evolution of $\Mgas$ more than compensates for the negative evolution in $\Tsl$ (or $\Tmw$ for $\YSZ$), and at larger masses, $\Mgas$ evolves self-similarly while $\Tsl$ still evolves negatively. 

Observational constraints regarding redshift evolution remain mixed and are often limited by small sample sizes, particularly at high redshift. In a joint analysis of $14$ cluster samples focusing on high mass groups and clusters and extending to $z \sim 1.5$, \citet{Reichert2011} find no deviation from self-similar evolution in the $M$-$T$ relation but report a significant deviation for the $M-L_{\rm X}^{\rm bol}$ relation ($\propto E(z)^{-0.81\pm0.12}$ compared to the SS value of $-7/4$). Additionally, \citet{Sereno2015} observe that the evolution of the $\YSZ$--$\Mfivehc$ relation is compatible with the SS model, while that of $\LX$--$\Mfivehc$ exhibits a steeper evolution than the theoretical prediction. A more recent analysis of a large sample of groups and clusters by \citet{Chiu2022} reports that the $\Mgas$--$\Mfivehc$, $\LX$--$\Mfivehc$, and $\YX$--$\Mfivehc$ relations are consistent with SS evolution, while the $\TX$--$\Mfivehc$ relation showed a shallower than SS behavior in redshift at $\sim 2\sigma$ significance.

\subsubsection{MPR absolute normalizations}\label{sec:normalization-simulations}

Figure~\ref{fig:normalizations-self-similar} normalizes mean behaviors separately for each simulated halo sample.  In 
Table \ref{tab:normalization-z-0}, we present absolute log-mean values of the gas properties at $z = 0$ for TNG and {\textsc{flamingo}} (black lines in Figure~\ref{fig:normalizations-self-similar}) at five mass scales.  

For the lowest-mass groups, TNG halos contain $60\%$ (0.27 dex) more gas than those in {\textsc{flamingo}}, and this difference decreases with scale to $16\%$ for the largest halos. 
In contrast, the spectroscopic-like temperatures in {\textsc{flamingo}} are approximately $15\%$ higher than TNG independent of halo mass.  The thermal energy product, $\YX$, thus begins higher in TNG at the group scale but the difference drops to $ < 0.01$ dex at $10^{15} \msol$.  
The direct thermal energy measure, $\YSZ$, also begins higher in TNG, by $45\%$, but the difference at cluster scales declines to $7\%$ at $10^{15} \msol$.  

Surprisingly, the X-ray luminosity differences deviate strongly from the pattern seen in $\Mgas$.  At the group scale, TNG halos are only $4\%$ more luminous than those of {\textsc{flamingo}}, suggesting significant structural differences in the hot gas interiors of low-mass halos as well as differences in metallicity which is shown to be too high in {\textsc{flamingo}} halos \citep{Braspenning2024}.  At the high mass end, this difference grows dramatically, to a magnitude of $0.42$ dex (a factor of 2.6) at $10^{15} \msol$.  This upward pull in the TNG mean $\LX$ is driven by a significant increase in positive skewness of the mass-conditioned distribution, $\Pr(\LX | \mfiveh)$ \citep{Nelson2024}, a topic we discuss further in \S\ref{sec:kernel-shapes} below. Excising emission from the core, the inner $0.15 \Rfivehc$ region, cuts the gap at high halo masses roughly in half, to $0.24$ dex, which lies closer to the simple expectation of twice the gap in $\Mgas$, or $0.14$ dex.


\begin{table*}
\centering
\caption{Normalizations, in decimal exponent format, of the scaling relations at five halo mass scales 
for TNG and {\textsc{flamingo}} (FLAM) at $z=0$.  Values at $10^{15} \msol$ are used to set the zero-points for the deviations from self-similarity shown in Figure~\ref{fig:deviations-of-evolution}. 
}\label{tab:normalization-z-0}
\begin{tabular}{ccccccccccc}
\hline
$\log_{10}(\Mfivehc/[\msun])$ & \multicolumn{2}{c}{$\log_{10}(\Mgas/[\msun])$} & \multicolumn{2}{c}{$\log_{10}(\Tsl/[{\rm K}])$} & \multicolumn{2}{c}{$\log_{10}(\LX/[{\rm erg/s}])$} & \multicolumn{2}{c}{$\log_{10}(\YX/[\msun \cdot {\rm K}])$} & \multicolumn{2}{c}{$\log_{10}(\YSZ/[\msun \cdot {\rm K}])$} \vspace{5pt} \\ 
 & \textbf{TNG} & \textbf{FLAM} & \textbf{TNG} & \textbf{FLAM} & \textbf{TNG } & \textbf{FLAM }& \textbf{TNG} & \textbf{FLAM} & \textbf{TNG} & \textbf{FLAM} \\ \hline
$\boldsymbol{13.0}$ & $11.58$ & $11.31$ & $6.65$ & $6.72$ & $40.97$ & $40.96$ & $18.24$ & $18.03$ & $18.29$ & $18.07$ \\ 
$\boldsymbol{13.5}$ & $12.40$ & $12.19$ & $6.98$ & $7.07$ & $42.28$ & $42.14$ & $19.38$ & $19.27$ & $19.40$ & $19.27$ \\ 
$\boldsymbol{14}$ & $13.08$ & $12.93$ & $7.27$ & $7.35$ & $43.39$ & $43.13$ & $20.36$ & $20.28$ & $20.34$ & $20.27$ \\ 
$\boldsymbol{14.5}$ & $13.64$ & $13.54$ & $7.55$ & $7.63$ & $44.31$ & $43.90$ & $21.18$ & $21.17$ & $21.19$ & $21.15$ \\
$\boldsymbol{15.0}$ & $14.15$ & $14.08$ & $7.84$ & $7.91$ & $44.94$ & $44.52$ & $21.99$ & $21.99$ & $22.01$ & $21.98$ \\ \hline
\end{tabular}
\end{table*}

\subsection{Conditional likelihood shapes}\label{sec:kernel-shapes}

Modeling the count of dark matter halos as a function of an observable property and utilizing it for cosmological inference requires careful handling of the mass-observable relations and selection effects \citep{Angulo2012, Mantz2019, Kugel2025}. Both are sensitive to the form of the mass-conditioned likelihood, $\Pr(S|M,z)$  \citep[\emph{e.g.},][]{Shaw2010, Erickson2011, Costanzi2019}, which is often assumed to be log-normal.  Indeed,  log-normal covariate behavior is an implicit assumption of the E14 multi-property population model. 

Although some observations support the log-normality of conditional statistics \citep[\emph{e.g.},][]{Pratt2009, Mantz2010XrayScaling, Czakon2015, Mantz2016a}, sample size tend to be too small to make sensitive tests.  Simulations of large cosmic volumes, however, create samples with substantial statistical power.  

Log-normality in halo mass-conditioned deviations was found in both hot gas and total stellar mass within $\Rfivehc$ for 22,000 halos above $10^{13} \msol$ in the MACSIS and BAHAMAS simulations \citep{Farahi2018}.  Log-normality in total stellar mass was also confirmed using samples of thousands of halos with $\Mtwohc > 10^{13.5} \msol$ in the TNG300 and Magneticum simulations \citep{Anbajagane2020}.  
However, that work also identified a form with modest skewness, common to all three simulations sets mentioned above, for the likelihood of satellite galaxy number conditioned on halo mass, with low satellite galaxy number associated with early forming halos.  

\begin{figure}
    \includegraphics[width=0.95\linewidth]{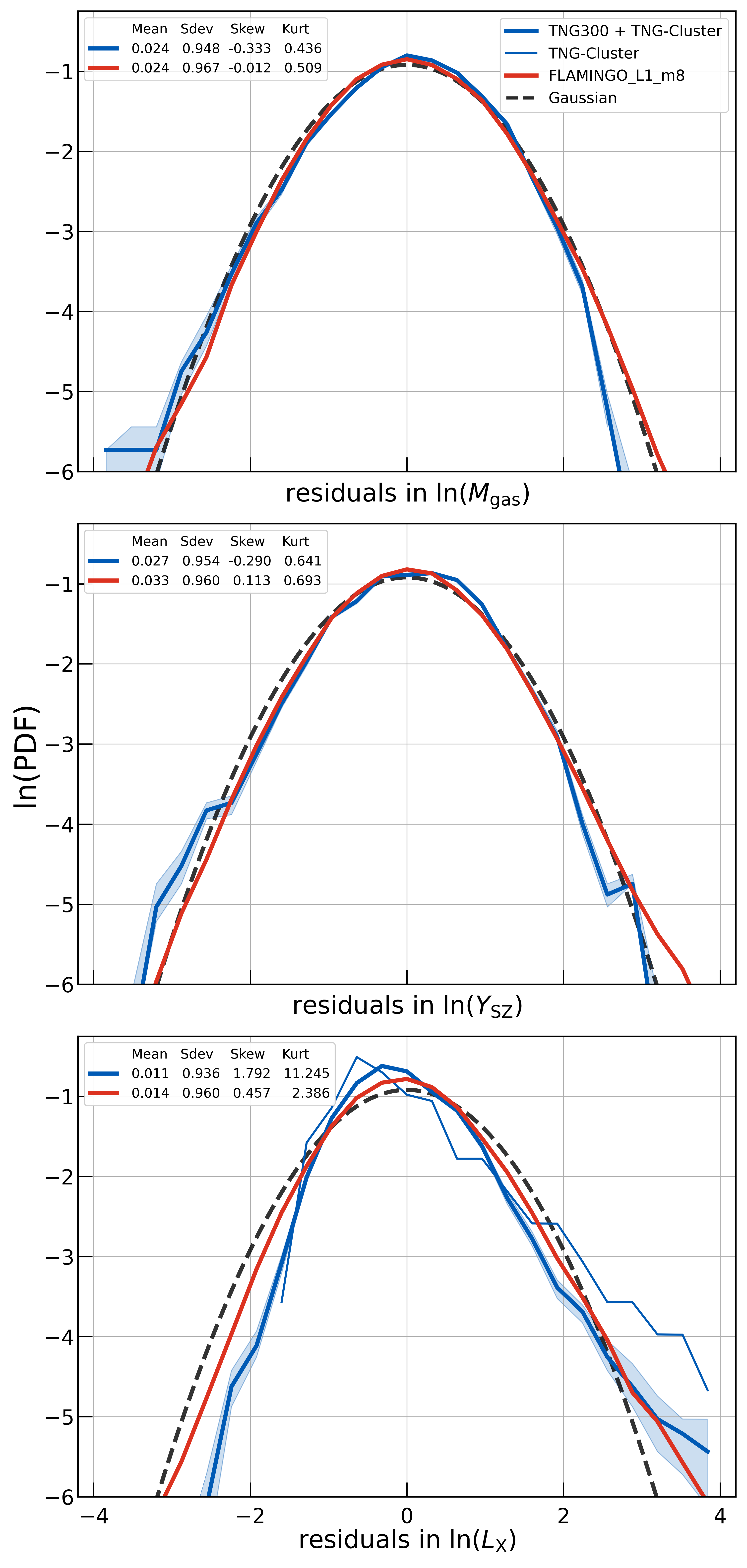}
    \caption{Probability density functions of normalized deviations from mean KLLR behavior at $z=0$ formed by marginalizing across halo mass for $\Mgas$ (top panel), $\YSZ$ (middle) and $\LX$ (bottom).  Deviations are expressed in units of the KLLR standard deviation at each halo mass.  A KLLR kernel width of $0.1$ dex is used to reduce numerical effects on low-order moment estimates, values for which are given in the panel insets.  }
    \label{fig:Mgas-YSZ-LX-kernels}
\end{figure}

In Figure~\ref{fig:Mgas-YSZ-LX-kernels} we present normalized likelihood  distributions of residuals in three properties at $z=0$. Of the three, two adhere closely to log-normality while the third does not.  Additional property likelihood forms are presented in Figure \ref{fig:LXce-YX-Tsl-Tmw-kernels}. The residuals are integrated across the full mass range of halo mass using equal weight per halo.  Each halo's contribution is centered at the KLLR mean and normalized by the KLLR standard deviation at the halo's total mass value.  In the presence of curvature in the slope or scatter as a function of $\mu$, and also in the presence of a declining HMF with increasing mass, the KLLR estimates of the first two moments can be biased by small amounts.  To minimize this bias, we employ a narrower KLLR kernel width of $0.1$ dex to produce these likelihood shapes.

Each panel in Figure~\ref{fig:Mgas-YSZ-LX-kernels} lists the first four moments of the measured residual distribution and dashed lines show a Gaussian reference.  
Means and standard deviations of the normalized residuals are very close to the expected values of $0$ and $1$, respectively.
The skewness and the kurtosis are the key unconstrained moments shown here that measure deviations from log-normality---our definition of kurtosis is shiften so that a log-normal distribution has a kurtosis of $0$. 

\begin{figure*}
    \includegraphics[width= 0.8\linewidth]{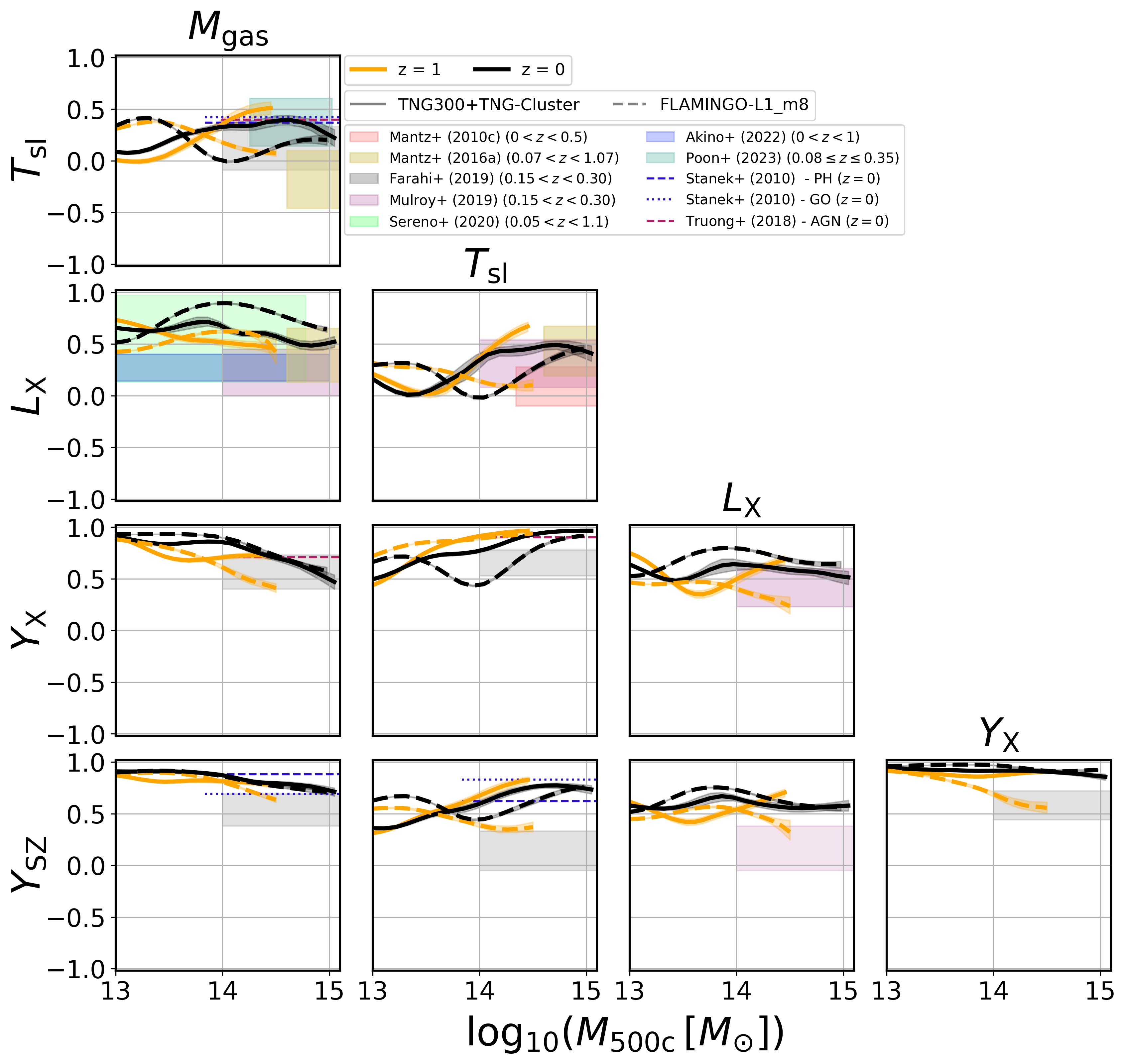}
    \caption{KLLR hot gas property correlations at redshifts zero (black) and one (orange) in TNG (solid) and {\textsc{flamingo}} (dashed) simulations. Observational estimates are shown with shading with vertical extent given by the $1\sigma$ reported error and horizontal extent given by the mass range used for the analyses. Correlation measures in past numerical studies, with no errors reported, are given by the dashed and dotted lines, with horizontal extent given by mass ranges used. See text for discussion.}
    \label{fig:correlations-corner-plot}
\end{figure*} 

The residuals in both $\ln \Mgas$ and $\ln \YSZ$ are well-fit by a Gaussian, particularly in {\textsc{flamingo}}, while in TNG, there is slightly reduced support at high values. The skewness is therefore negative in TNG with value close to $-0.3$ in both properties.  The kurtosis is slightly larger than zero in both populations. 

The choice of radial scale is important in these measures, as $\Rfivehc$ is well within the zone where virial and hydrostatic equilibrium are expected to generally hold. For {\textsc{flamingo}} halos, \citet{Kugel2024} observe a significant tail in the Compton-Y signal at a much larger scale of $5\times \Rfivehc$. The larger aperture reaches out into the infall regime where surrounding massive halos can be encountered.  The extra thermal energy from this 2-halo term will boost the positive tail of residuals around the mean.  \citet{Kugel2024} show that the significance of this tail decreases when the aperture is reduced to $\Rfivehc$, consistent with our findings.  

In contrast to the gas mass and thermal energy, the overall shape of the X-ray luminosity likelihood has substantial skewness and kurtosis.  
The distributions feature a high-value tail in both TNG and {\textsc{flamingo}}, the latter previously reported by \citet{Kugel2024}, with less support at low values.  The non-guassianity is more extreme in TNG halos.  The thin line in the bottom panel of Figure~\ref{fig:Mgas-YSZ-LX-kernels} shows the likelihood shape for the massive TNG-{\textsc{cluster}} sample only. This sub-population features a more extreme tail, with $3\sigma$ upward fluctuations roughly ten times more likely than for the full TNG300 sample.  

We note in Appendix~\ref{sec:additionalKernelShapes} that core-excision does not strongly alter these shapes.  While the $\LX$ variance in TNG-{\textsc{cluster}} halos is significantly reduced by core excision (see Figure~\ref{fig:LXce-slope-scatter}) this is not the case for the much more numerous TNG300 halos.  This finding suggests that other factors that affect the  phase structure outside the core are responsible for driving the non-normal shape of the X-ray luminosity likelihood. 


\section{Mass-conditioned property correlations}\label{sec:correlations}

Property correlations at fixed halo mass are higher-order statistical features of the group and cluster population that can have important consequences.  
For example, \citet{Nord2008} illustrate that, if covariance is not properly taken into account, selection effects can mimic redshift evolution in the $\LX-T$ scaling relation derived from flux-limited samples. For massive clusters, \citet{Wu2015} and \citet{Farahi2018} show that total baryon mass serves as a superior mass proxy compared to both stellar mass and gas mass alone, due to strong anti-correlation between the stellar and gas masses at a given halo mass. 

Additionally, \citet{Anbajagane2020} find an anti-correlation between the number of satellite galaxies and the BCG stellar mass across multiple simulations, with early forming systems preferring higher BCG mass and fewer satellite galaxies. Relatedly, \citet{Farahi2020} find that the total stellar mass-BCG stellar mass and total stellar mass-satellite stellar mass correlations in simulated halo samples become stronger when conditioned on both the magnitude gap -- the difference between the fourth and brightest cluster member -- in addition to halo mass.  The jointly conditioned properties show reduced intrinsic scatter relative to halo mass conditioning alone.

Measuring these correlations can be challenging due to the need for high signal-to-noise measurements, trustworthy estimates of observational errors, and careful modeling of survey selection and other sources of systematic uncertainty.
Nonetheless, recent studies have successfully measured the covariance matrix of several X-ray properties \citep{Mantz2010XrayScaling, Mantz2016a, Mantz2016b, Andreon2017, Sereno2020, Akino2022, poon2023} as well as multi-wavelength properties \citep{Farahi2019, Mulroy2019}.  
The latter work used $41$ clusters from the LoCuSS X-ray selected sample and analyzed up to 11 optical and hot gas properties conditioned on weak lensing mass.

Lines in Figure \ref{fig:correlations-corner-plot} present KLLR mass-conditioned correlations (Equation~\eqref{eq:KLLR-correlation}) for the five principal hot gas properties at redshifts of zero and one (we omit other redshifts for clarity).  Shaded regions show observational constraints, with the caveat that projection effects may drive deviations from the intrinsic, spherical values we measure in the simulations.  

Strongly positive correlations exist between $\Mgas$ and its direct derivative quantities, $\LX$, $\YSZ$ and $\YX$.  The SZ thermal energy, $\YSZ$, and $\Mgas$ have correlation coefficients in the range $0.6-0.9$, with only weak dependence on halo mass and redshift.  

The $\YX$ and $\Mgas$ correlation has scale-dependence that reflects the mass-dependent scatter of gas mass and temperature. At group scales, the MPR scatter in gas mass dominates that of temperature, which drives the correlation to high values.  For cluster-scale halos above $10^{14} \msun$, the MPR scatter in in $\Tsl$ is the dominant contributor to $\YX$ scatter.  The relatively weak correlation of $\Tsl$ with $\Mgas$ thus results in a decline in the $\YX - \Mgas$ correlation, while the $\YX-\Tsl$ grows, at these mass scales.  While qualitatively similar, there are differences in detail between TNG and {\textsc{flamingo}} behaviors.  

These differences also emerge in the $\YX$ and $\YSZ$ correlation (lower right of the triangle plot) which is large, $r \sim 0.9$ independent of halo mass.  A weakening trend is apparent in {\textsc{flamingo}} halos at $z=1$ that mimics the weakening in the $\YX -\Mgas$ correlation. 

The $\Mgas$-$\LX$ correlation is positive and remains above $0.4$ at all mass scales and for all redshifts. At $z = 0$, both simulations predict a mild decrease with system mass in the correlation for clusters.  At the group scale, the correlation increases from $\sim 0.5$ to $\sim 0.9$ in {\textsc{flamingo}} and from  $0.65$ to $0.78$ in TNG. Both simulation samples show a slight decline in correlation at $z=1$. 
The $\Tsl$-$\LX$ correlation is mildly positive across the mass range at all redshifts, with  dips toward zero at different halo mass scales.  This correlation rises to $\sim 0.4$ at the cluster scale. These correlation values agree with the observational constraints of \citet{Mantz2016a} and  \citet{Sereno2020} and are only slightly larger than the results reported in \citet{Mulroy2019} and \citet{Akino2022}.

Within this lower part of the triangle, observational estimates, shown in shaded colors with $1\sigma$ uncertainties, are largely in agreement with the simulation values.  The LoCuSS sample \citep{Mulroy2019, Farahi2019} finds correlation coefficients of $0.33^{+0.21}_{-0.25}$ for $\TXce$-$L_{\rm X, RASS}$, and  $0.13^{+0.20}_{-0.22}$ for $\TXce$-$\Mgas$.  Using 238 clusters drawn from the ROSAT All-Sky Survey (RASS), \citet{Mantz2010XrayScaling} measure a correlation of $0.09 \pm 0.19$ between the scatter of the core-excised temperature and the X-ray luminosity in the ROSAT broad band. In contrast, using a sample 19 X-ray selected clusters, \citep{poon2023} measure a larger value, $0.43^{+0.172}_{-0.293}$, that is consistent within the uncertainties.

Finally, the observed $\YSZ$ correlations show mixed agreement, both internally and with the simulation expectations. The Planck measurements, both from the same LoCuSS study \citep{Mulroy2019, Farahi2019}, are consistent at the $2\sigma$ level with the simulation values, but some values are also consistent with zero. Projection effects, particularly in SZ measurements, may be affecting the observational estimates. More accurate estimates of these correlations using  larger, more sensitive surveys are clearly needed.

\subsection{MPQ from five properties combined}\label{sec:MPQcombined} 

\begin{figure}
    \includegraphics[width = \linewidth]{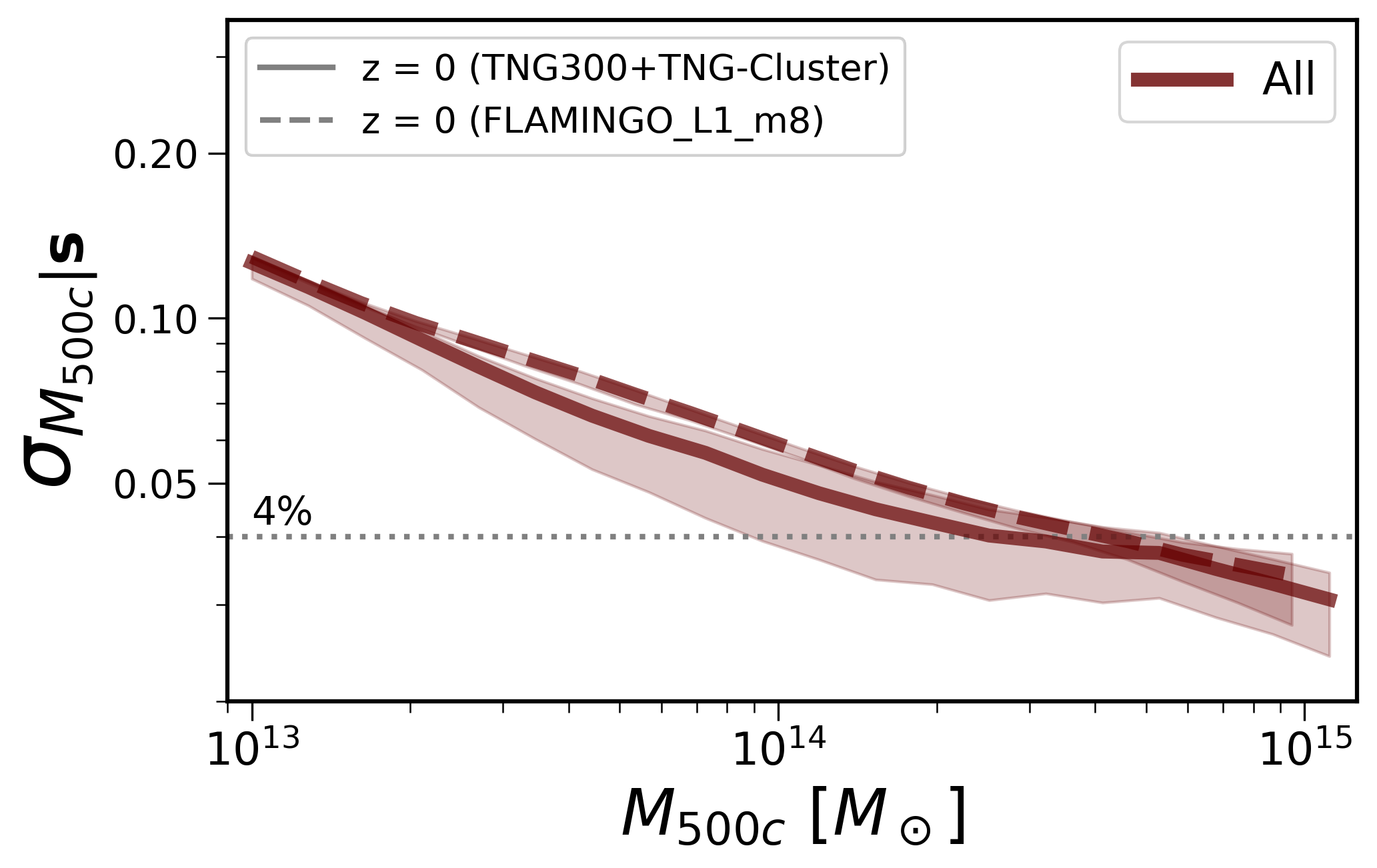}
        \vspace{-12 truept}
    \caption{Halo mass proxy quality at $z=0$ implied by combining the five principal properties listed in Table~\ref{tab:prop-definitions}.  Because correlations among parameters are mostly positive, the MPQ of the combined properties improves only modestly upon the best single measures of Figure~\ref{fig:MPQzero}.   
}
    \label{fig:MPQ-all-properties-zero}
\end{figure}

We close this section on correlations by using the full covariance and set of slopes to determine the mass proxy quality achieved by combining all five properties, Equation~\eqref{eq:implied-variance}, comparing it to scatter in mass estimates produced by novel machine learning methods in the following section. The result at $z=0$ is shown in Figure~\ref{fig:MPQ-all-properties-zero}.  Strongly negative correlations can greatly reduce the effective halo mass variance while strongly positive correlations are less helpful.  This makes qualitative sense, in that the limit of correlation coefficient $r \rightarrow 1$ for any pair of properties implies having a pair of redundant measurements.  

There is a remarkable level of consistency in the combined MPQs for TNG and {\textsc{flamingo}} halo samples, with values dropping gently from $12\%$ for $10^{13} \msol$ low mass groups to $3\%$ for $10^{15} \msol$ high mass clusters.  While improvements in halos mass scatter may appear relatively modest, the reduction in variance brought about by combining all properties is more substantial, a $43\%$ drop from $(0.04)^2$ to $(0.03)^2$ for high mass clusters, for example. 

\subsection{Comparison to Machine Learning mass estimates}

Recent campaigns for cluster mass estimation have employed various machine learning (ML) models, such as convolutional neural networks (CNNs) and random forests (RFs), trained on X-ray \citep{Ntampaka2019MLXraymasses, GreenMLXraymasses2019, Ho2023MLmasses, Krippendorf2023ERositaMLmasses} and SZ \citep{deAndres2022DLYSZ, WadekarMLSZmasses2023, WadekarMLSymbSZmasses2023} mock observations from hydrodynamical simulations.

\citet{Ntampaka2019MLXraymasses} utilized a CNN trained and X-ray mock observations from 329 {\textsc{illustris}}TNG clusters to reproduce cluster masses with a low bias of $-0.02$ dex and average intrinsic scatter of $12\%$. Using a similar architecture, \citet{Ho2023MLmasses} demonstrated that a CNN trained on bolometric X-ray photon maps derived from the Magneticum hydrodynamic simulation achieves a predictive mass scatter of $17.8\%$, while a CNN trained on a multi-channel maps split into low, medium, and high energy bands reduces the scatter to $16.2\%$. In a study using mock \textit{eROSITA} observations for a cluster sample within a mass range of $10^{13}\msun < \Mfivehc < 10^{15}\msun$ in the Final Equatorial Depth Survey (eFEDS), \citet{Krippendorf2023ERositaMLmasses} reduced the scatter of $45\%$ resulting from the luminosity-mass scaling relations \citep{Chiu2022} to $43\%$ by using a CNN augmented with redshift information. They also showed that a CNN can properly handle major contaminators such as AGNs, without a priori filtering. In contrast to CNNs, \citet{GreenMLXraymasses2019} used RF regression trained on \textit{Chandra} and \textit{eROSITA} mock X-ray observations of the core-excised luminosity for 1615 clusters from the Magneticum simlations with additional set of morphological parameters, showing a scatter of $15\%$ on predicted masses, a $20\%$ reduction relative to the scatter in the mass-luminosity relation. This level of scatter for the core-excitsed luminosity is similar to what we find for TNG and {\textsc{flamingo}} (see Figure \ref{fig:LX-LXce-MPQ}). Beyond X-ray images, \citet{deAndres2022DLYSZ} used a CNN trained mock Planck compton-y observations from the Three Hundered simulations to estimate the cluster masses, achieving a scatter of $17\%$.

The levels of scatter in the predicted cluster mass from the ML studies above are large compared to the scatter achieved by the combined set of idealized mass proxies reported in this work. Therefore, we hope that future ML estimates of cluster masses will provide smaller levels of scatter, closer to the value we report here.


\section{Comparison to Previous Works}\label{sec:literature-comparison}
The literature on observed scaling relations of cluster properties is now in its fifth decade \citep{Bahcall1974, Kaiser1986, Evrard1991, Markevitch1998, Arnaud1999, McKay2001, McKay2002, Novicki2002, Eckmiller2011, Maughan2012, Schellenberger2015, Lovisari2020}, and simulation campaigns to theoretically address them began in the 1990s   \citep{Navarro1995, EMN1996, Bryan1998}.  
In this section, we place our findings into the context of prior studies of mass proxy quality and property scaling relations.  

It is important to note that any sample has specific selection criteria that yields varying coverage in mass and redshift.  Simulation studies are often complete in a certain volume above some minimum mass, but some approaches, such as zoom-in re-simulations, will trade completeness for uniform coverage across a mass range.  
In addition, definitions of halo properties, including true mass, can differ by construction. Here we focus on studies using $\Mfivehc$.  
Simulation works differ in their physical assumptions for gas evolution, and we include examples of: shock heating under gravity only (GO); preheating (PH), the simple  assumption of a rapid rise in entropy at some fixed epoch to mimic galaxy feedback; cooling, star formation and feedback from supernovae (CSF); and SMBH formation and feedback from active galactic nuclei (AGN).  All AGN simulation models also include a CSF treatment.  

Observational measures will include projection effects and other real-world features that are often absent from simulation studies. Caution must be thereby be exercised when making comparisons, particularly between computational and observational samples.  We make no attempt to explicitly include redshift evolution; we quote $z=0$ values for simulations while observed samples have varying depths but tend to be near field, $z<0.5$.  
Despite these caveats, the literature review we present here offers insights into the state of our understanding of the astrophysics operating within the massive halo population.  

In the figures below, we show values obtained in this study, using dark shading for values found for high mass halos, $10^{14.5} - 10^{15} \msol$, and light shading for values across the full halo mass range $10^{13} - 10^{15} \msol$.  The former is indicative of the typical median mass scale for the samples in the literature that we include. We note that the lack of error bars in any figures or tables reflects the fact that they were not reported in the original source.

This analysis is not intended to be complete with respect to all published scaling relations for clusters or massive halos.  This exercise simply aims to compare the MPQ and MPR parameters of the current halo samples with those of previous studies.  We leave a more complete review of these topics to future work.  Indeed, a combined KLLR statistical analysis of all available data from extant simulations and observations would be a first step toward a global ``particle data book'' \citep{ParticleDataBook2020} approach to massive halo population statistics. 


Figure~\ref{fig:MPQ-literature} presents MPQ values for the five principal hot gas properties studied in this work.  For simulations, the true 3D total mass is used while for observations total masses are derived using hydrostatic X-ray or weak lensing methods. 
The shaded bands shown for each property (left is TNG, right is {\textsc{flamingo}}) provide extremal values for the MPQs shown in Figure~\ref{fig:MPQzero} for high mass systems ($10^{14.5} - 10^{15} \msol$, dark shading) and for all halo masses (light).  

For the most massive simulated halos at $z = 0$, the finding that hot gas mass, $\mgas$, is a slightly better mass proxy than either thermal energy measures, $\YX$ or $\YSZ$, is supported by the studies of \citet{Fabjan2011} and \citet{Truong2018}.  The former work employed CSF and a sample of roughly 140 halos while the latter used an AGN treatment on 58 resimulated halos. The midpoints of halo mass ranges in both works lie close to our dark shaded region.  Both studies find $\YX$ to be a slightly worse halo mass proxy compared to $\mgas$.  

This finding contradicts the original work of \citet{Nagai2007} based on a sample 16 halos simulated with a CSF approach.  That study found $\YX$ to be superior, with $7\%$ scatter in true mass relative to $11\%$ for gas mass and $20\%$ for gas temperature.  While that sample extended into the mass range of groups, no evidence for power-law slope deviation or increased scatter was found, presumably because of the small overall number of systems.  

An observational study of ten relaxed clusters by \citet{Arnaud2007} also found that the intrinsic scatter in hydrostatic masses derived from X-ray observations was smallest for $\YX$, with slightly larger values for temperature and gas mass.  The statistical significance of this comparative result is unclear, and the result reflects values for a small sample selected on criteria that indicate dynamical relaxation.

\subsection{Mass Proxy Quality}
\begin{figure}
    \includegraphics[width=0.95\linewidth]{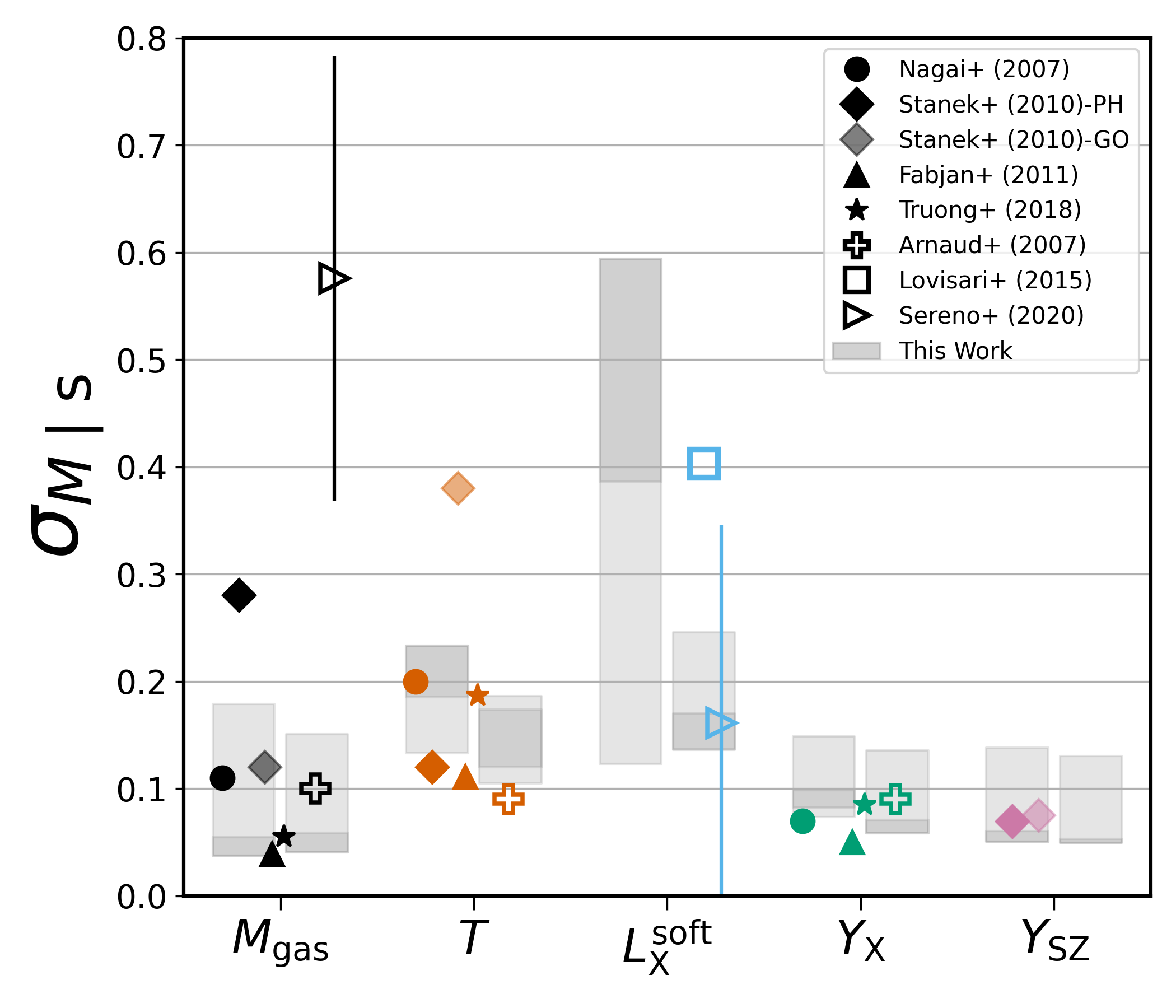}
    \vspace{-8 truept}
    \caption{Comparison of MPQ values (the scatter in total halo mass conditioned on each of the listed properties) from prior works (symbols) with those presented here (shaded, left=TNG, right={\textsc{flamingo}}). Dark shading shows the MPQ range for high mass systems ($\mfiveh \in 10^{14.5} - 10^{15} \msol$) while light shading shows the range of values for all halos in this study,  $10^{13} - 10^{15} \msol$.  Filled symbols show values from computational halos samples at $z=0$ (typically mass-complete) while open symbols are results from observational cluster samples in the cosmic near-field. For simulations, specific temperature definitions vary (see Table~\ref{tab:ScatterLiterature}).}
    \label{fig:MPQ-literature}
\end{figure}

In simulations, the implied mass scatter for $\Tsl$ at $z = 0$ ranges between $10\%$ and $20\%$, with the exception of Millennium Gas simulation halos under GO treatment of \citet{Stanek2010}. That work used the same $\Tsl$ definition used here (and in all other cited studies except \citet{Nagai2007}, who derived temperatures from fits to synthetic X-ray spectra) and found a mass scatter near 0.4 for the GO halo sample.  Under purely gravitational evolution, the internal phase space structure of halos includes low entropy core gas that would be consumed by star and SMBH formation or heated and ejected by AGN feedback events under more advanced treatments \citep{Voit2005}. The existence of variable amounts of such high density and low temperature gas in the cores of halos increases the variance in $\Tsl$ relative to models for which this gas phase is consumed by compact object formation.  Indeed, the preheated (PH) halo sample of \citet{Stanek2010}, in which a floor is imposed on core entropy, has a much reduced MPQ of $0.12$.  A value of $0.1$ was found for the scatter in hydrostatic mass at fixed X-ray temperature in the small sample of \citet{Arnaud2007}.

Fewer studies have explicitly reported the scatter in halo mass at fixed thermal energy, but the values shown in Figure~\ref{fig:MPQ-literature} for $\YX$ and $\YSZ$ lie consistently below $10\%$.  In our study, we find good consistency between MPQ derived from TNG and {\textsc{flamingo}} samples for $\YSZ$,  with values falling from $\sim \! 15\%$ at the group scale to $\sim \! 5\%$ for high mass clusters.  As previously described, MPQs for $\YX$ in TNG are higher than those in {\textsc{flamingo}} in the high mass cluster regime. 
For $\YSZ$, the PH and GO treatments of \citet{Stanek2010} produce nearly identical MPQs, and both values lie within the range seem in the full physics simulations used here.  This suggests that the MPQ of gas thermal energy may be insensitive to cluster astrophysics, lending further support to the use of SZ selection for cosmological studies. 

\begin{figure}
    \includegraphics[width=\linewidth]{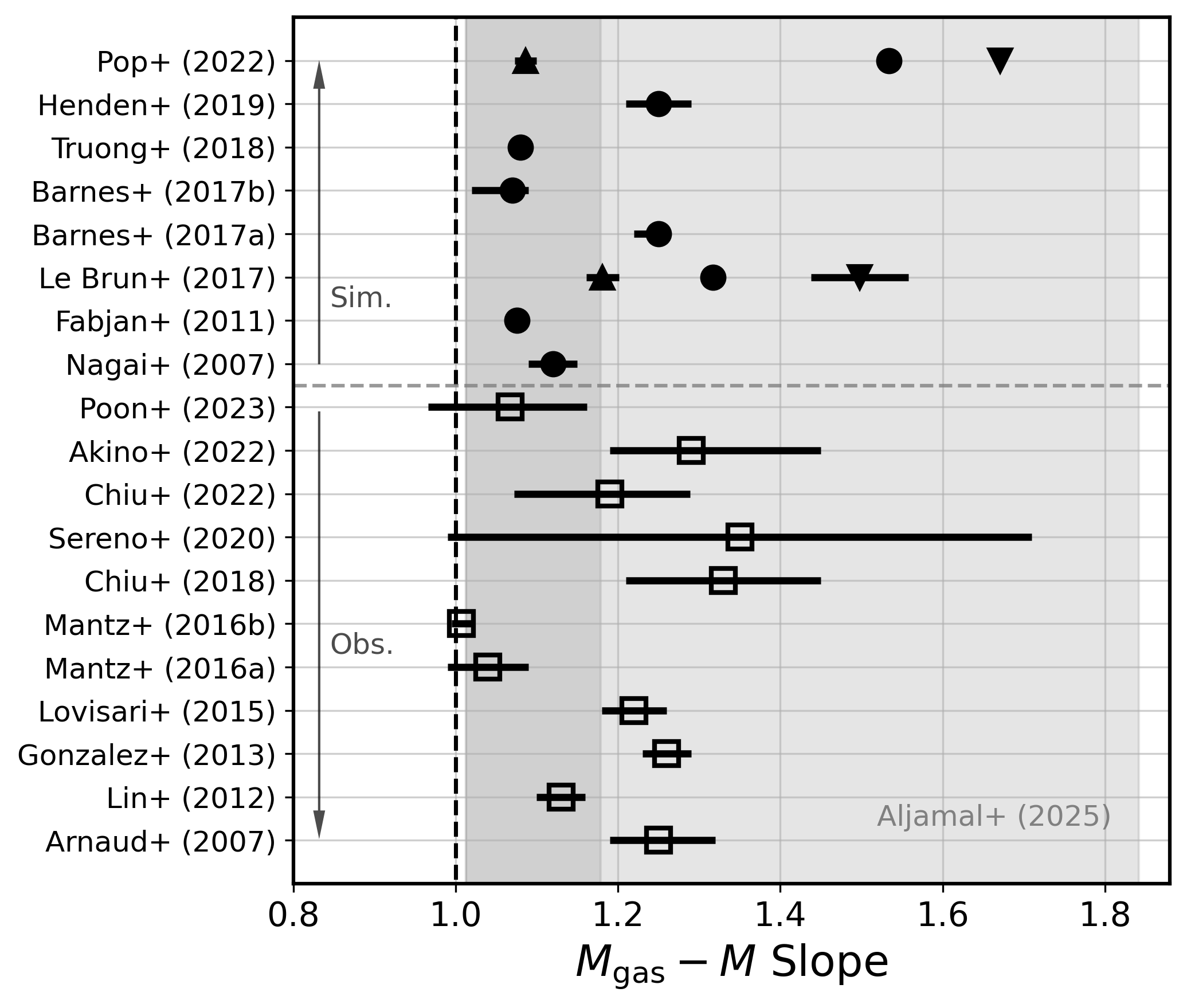}  
     \vspace{-12 truept}
    \caption{Hot gas scaling relation slopes reported for $z=0$ simulated halo samples (top) and from local observational cluster samples (bottom). The vertical dashed line represents the self-similar slope of one. The light gray shaded area indicates the extremal range of KLLR slopes from TNG and {\textsc{flamingo}} samples across the full mass range studied here ($10^{13} - 10^{15}\msun$), while the dark shading highlights the slopes observed in the mass range $10^{14.5}\msun \leq \Mfivehc \leq 10^{15}\msun$. For \citet{LeBrun2017} and \citet{Pop2022}, we report the simple power law (SPL) slopes (circle) as well as broken power law slopes for the low (downward-pointing triangle) and high (upward-pointing triangle) halos.
    }
    \label{fig:Mgas-slope-literature}
\end{figure} 

Because of the sensitivity of X-ray emission measure to internal cluster structure, the MPQ of soft X-ray luminosity is somewhat poorer relative to the other measures shown in Figure~\ref{fig:MPQ-literature}.  In a study of 23 groups of galaxies using hydrostatic mass estimates, \citet{Lovisari2015} find a mass scatter of 40\%, noting that hydrostatic mass measurements can introduce significant scatter \citep{Braspenning2024}, consistent with the range of values seen in the TNG halo sample but not in {\textsc{flamingo}}.

A far larger number of studies report scaling relation slope and scatter values from which MPQs could be derived.  We quote in this section only studies that explicitly measure variance in total system mass conditioned on an observed property.  In the next section we turn to examining past MPR measures of slope and intrinsic scatter. 

\subsection{MPR Slopes and Standard Deviations}

\begin{table*}
\centering
\caption{Intrinsic scatter estimates for hot gas properties studied here. Simulation results are present epoch ($z=0$) while observational results are for various median redshifts. Our scale-dependent results at $z=0$ are shown as the black lines in Figure~\ref{fig:MPR-inrinsic-scatter-plots}.}
\label{tab:ScatterLiterature}
\begin{threeparttable}
\begin{tabular}{cccccccc}
\hline
Source & $N_{\rm sample}$ & \textbf{$\sigma_{\ln \Mgas}$} & \textbf{$\sigma_{\ln T}$} & \textbf{$\sigma_{\ln \LX}$} & \textbf{$\sigma_{\ln \YX}$} & \textbf{$\sigma_{\ln \YSZ}$} \vspace{1pt}  & Notes \\ \hline
\multicolumn{8}{c}{\textbf{Simulations}} \\ \hline & \\[-8pt]
\citet{Henden2019} & $27$ & $0.276^{+0.069}_{-0.046}$ & --- & $0.576^{+0.092}_{-0.069}\tnote{e}$ & $0.322 \pm 0.046$ & $0.23^{+0.046}_{-0.023}$ & CSF + AGN \\
\citet{Truong2018} & $58$ & $0.060$ & $0.111\tnote{a}$ & $0.516\tnote{e}$ & $0.140$ & --- & CSF + AGN\\
\citet{Barnes2017b} & $30$ & $0.299$ & $0.138\tnote{b}$ & $0.691$ & $0.230$ & --- & CSF + AGN\\
\citet{Barnes2017a} & $1294$ & $0.161 \pm 0.023$ & $0.111 \pm 0.007\tnote{c}$ & $0.345^{+0.023}_{-0.046}\tnote{f}$ & $0.276 \pm 0.023$ & $0.230 \pm 0.023$ & CSF + AGN\\
\citet{Planelles2017} & $\sim100$ & --- & --- & --- & --- & $0.154$ & CSF + AGN\\
\citet{LeBrun2017} & $\sim10\,000$ & $0.10$ & $0.10\tnote{c}$ & $0.25\tnote{e}$ & $0.10$ & $0.10$ & CSF + AGN\\
\citet{Pike2014} & $30$ & --- & $0.048 \pm 0.005\tnote{a}$ & $0.283 \pm 0.04$ & --- & $0.078 \pm 0.009$ & CSF + AGN\\
\citet{Planelles2014} & $160$ & --- & $0.092\tnote{a}$ & --- & $0.184$ & $0.161$ & CSF + AGN\\
\citet{Biffi2014} & $179$ & --- & $0.11\tnote{b}$ & $0.25\tnote{e}$ & --- & --- & CSF\\
\citet{Fabjan2011} & $\sim140$ & $0.042$ & $0.069\tnote{d}$ & --- & --- & $0.084$ & CSF\\
\citet{Stanek2010} & $\sim10\,000$ & $0.086 \pm 0.001$ & $0.069 \pm 0.001\tnote{a}$ & $0.193 \pm 0.002\tnote{e}$ & --- & $0.125 \pm 0.002$ & PH\\
\citet{Short2010} & $187$ & --- & $0.071\tnote{a}$ & $0.226\tnote{e}$ & $0.111$ & --- & CSF + AGN\\\hline
\multicolumn{8}{c}{\textbf{Observations}} \\ \hline & \\[-8pt]
\citet[SPT][]{SPT2024} & $727$ & --- & --- & --- & --- & $0.20 \pm 0.05$ & SZ selection\\
\citet{poon2023} & $19$ & $0.207^{+0.045}_{-0.033}$ & $0.137^{+0.032}_{-0.025}$ & $0.219^{+0.062\dagger}_{-0.051}$ & $0.412^{+0.100}_{-0.073}$ & --- & X-ray selection\\ & \\[-8pt]
\citet{Akino2022} & $136$ & $0.39 \pm 0.08$ & --- & 
 $0.73^{+0.12}_{-0.14}$ & --- & --- & X-ray selection\\ & \\[-8pt]
\citet{Chiu2022} & $434$ & $0.074^{+0.063}_{-0.019}$ & $0.069^{+0.061}_{-0.014}$ & 
 $0.120^{+0.138}_{-0.060}$ & $0.106^{+0.171}_{-0.047}$ & --- & X-ray selection\\ & \\[-8pt]
\citet{Sereno2020} & $118$ & $0.253 \pm 0.23$ & --- & 
$1.266 \pm 0.299$ & --- & --- & X-ray selection\\
\citet{Chiu2018} & $91$ & $0.11 \pm 0.05$ & --- & --- & --- & --- & SZ selection\\
\citet{Mantz2016b} & $139$ & --- & $0.16 \pm 0.017$ & $0.43 \pm 0.03$ & $0.185 \pm 0.016$ & --- & X-ray selected\\
\citet{Mantz2016a} & $40$ & $0.09 \pm 0.02$ & $0.13 \pm 0.02$ & $0.24 \pm 0.05$ & --- & --- & combination\\
\citet{Lovisari2015} & $23$ & --- & --- & $0.564$ & --- & --- & X-ray selection\\
\citetalias{Planck2014} & $71$ & --- & --- & --- & --- & $0.145 \pm 0.025$ & SZ selection\\
\citet{Lin2012} & $94$ & $0.08$ & --- & --- & --- & --- & combination\\
\citet{Andersson2011} & $15$ & --- & --- & $0.253 \pm 0.092$ & --- & $0.207 \pm 0.115$ & SZ selection\\
\citet{Mantz2010XrayScaling} & $238$ & --- & $0.126 \pm 0.018$ & $0.414 \pm 0.044$ & --- & --- & X-ray selection\\
\citet{Vikhlinin2009} & $88$ & --- & --- & $0.396 \pm 0.039$ & --- & --- & X-ray selection\\ \hline
\end{tabular}
\begin{tablenotes}
    \item[a] Spectroscopic-like temperature
    \item[b] X-ray spectroscopic temperature
    \item[c] Core-excised X-ray spectroscopic temperature
    \item[d] Mass-weighted average temperature
    \item[e] bolometric X-ray luminosity
    \item[f] core-excised bolometric X-ray luminosity
\end{tablenotes}
\end{threeparttable}
\end{table*}

The majority of sources used in this section are listed in Table~\ref{tab:ScatterLiterature} where we also give the MPR standard deviation estimate.  We will refer to this table in each subsection below.   
Sample sizes from previous simulations vary from a few dozen halos to more than 10,000. The size of the {\textsc{flamingo}} sample used here is considerably larger than any previously reported.  

The observational samples we include vary in size from 15 to 434, with a median value near 70. Many are X-ray selected while three are SZ-selected samples.  The selection is sometimes complex, based on multiple characteristics, so caution must be employed when inter-comparing results.

Once again we emphasize that the values from the numerical literature includes samples with different selected mass ranges.  We report simulation slopes at $z = 0$, while for observational studies this is not the case.  The TNG and {\textsc{flamingo}} halos show little redshift dependence for slopes at cluster mass scales (see Figure~\ref{fig:MPR-slope-plots}), while low mass groups typically display slopes that decrease with increasing redshift. 

Simulations use different codes, physical models, and a variety of fitting methods. Most works fit the scaling relations using a single power law (SPL) model while both \citet{LeBrun2017} and \citet{Pop2022} employ both SPL and broken power law (BPL) models. The former uses a fixed break scale of $\Mfivehc = 10^{14}\msun$ while the latter fits for this scale. For these sources we report the SPL circle value along with the low and high mass slopes of the BPL fits; these are denoted by a circle, downward-pointing triangle, and upward-pointing triangle, respectively, in the figures below. 

Subsections that follow present results for individual hot gas properties.   
Slope estimates are summarized in 
Figures \ref{fig:Mgas-slope-literature} through \ref{fig:Y-slope-literature}.  
In each figure, the vertical dashed line represents the self-similar expected slope. The light shaded area indicates the range of KLLR slopes observed across the full mass range studied in this work, while the darker gray shade highlights the slope at high masses, $\log_{10} (\mfiveh/\msun ) \in 14.5-15$.  Since TNG and {\textsc{flamingo}} slopes shown in Figure~\ref{fig:MPR-slope-plots} are similar, we base these ranges on the minimum and maximum values from the combined set.  Note that some works that present slopes but not intrinsic scatter are not listed in Table~\ref{tab:ScatterLiterature}.   

\subsubsection{Hot gas mass}

Figure \ref{fig:Mgas-slope-literature} shows slopes of the hot gas mass scaling relation reported in prior literature compared to the ranges we find within the high and full mass ranges applying KLLR to TNG and {\textsc{flamingo}} samples. 

All existing evidence supports a scaling of gas mass with total mass that is steeper than self-similar, with magnitudes ranging from $1$ to $1.3$  The high mass portion of broken power law fits lie in the range we find here, as do most previous simulation works that focus on cluster-scale halos.  

While \citet{Barnes2017a} report a slope of $1.25^{+0.01}_{-0.03}$ in their full BAHAMAS + MACSIS sample, they find a slope of $1.02 \pm 0.03$ in their hot sample (clusters with core-excised temperature $\kB T \geq 5\kev$), aligning more closely with the trend of decreasing slope with halo mass scale seen in Figure~\ref{fig:MPR-slope-plots}.  The 27 halos of the FABLE simulation \citep{Henden2019} are uniformly selected in halo mass across the two orders of magnitude we study here.  That study finds a slope $1.25$ with no evidence for steepening at the group scale. 

Supporting the trend of steeper $\Mgas$ slope at the group scale, \cite{LeBrun2017} find a slope $1.50 \pm 0.06$ for halos with $\Mfivehc \leq 10^{14}\msun$. Similarly, \cite{Pop2022} find a break in the $\Mgas$ scaling relation at $\Mfivehc \sim 5.7\times 10^{13}\msun$, with slopes of $1.671 \pm 0.002$ for smaller halos and $1.086^{+0.014}_{-0.013}$ for larger halos.

On the observational side, estimates are mostly consistent with a value near 1.2 with the exceptions of the Weighing the Giants (WtG) study of 139 clusters \citep{Mantz2016b} and an independent analysis of 40 relaxed, massive clusters \citep{Mantz2016a} using X-ray hydrostatic masses.  In the WtG study, weak lensing  masses were available for only 27 systems.  For the remaining 112 clusters, total mass was inferred from gas mass using a fixed gas fraction overdensity model (see equation (1) of \citet{Mantz2016b}).  
Given the prevalence of this mass definition, a slope of one is to be expected by construction. The focus by \citet{Mantz2016a} on high mass, relaxed systems may reflect our finding that slopes approach one as masses approach $10^{15} \msol$.  

Reports of intrinsic scatter in hot gas mass are listed in Table~\ref{tab:ScatterLiterature}, seven from simulation campaigns and five from observational samples.  The values reported in these studies occupy the full range we find in TNG and {\textsc{flamingo}} samples, from a high value of $30\%$ at the group scale to a low of $4\%$ for the highest mass clusters.  Previous simulations have noted a similar scale dependence to the scatter in hot gas mass \citep{Barnes2017a, LeBrun2017}.

For simulations, the high value of 0.30 from \citet{Barnes2017b} was derived using both gas and total hydrostatic masses from synthetic X-ray observations.  A similar value was found by \citet{Henden2019} for the FABLE simulations that uniformly sample halo masses from group to cluster scales.  The studies of \citet{Fabjan2011} and \citet{Truong2018} find values of $0.04$ and $0.06$, respectively, close  to the minima we find here. 

On the observational side, values lie in the range 0.1 to 0.2.  The high value of $0.207^{+0.045}_{-0.033}$ found by \citet{poon2023} is based on X-ray hydrostatic masses for 19 X-ray selected clusters.  In that work, a relaxed subset of 9 clusters displays smaller scatter of $0.120^{+0.045}_{-0.028}$.  This finding is consistent with the result of \citet{Mantz2016a}, who find a hot gas scatter of $0.09 \pm 0.02$ for 40 relaxed clusters. 


\subsubsection{Hot gas temperature}

\begin{figure}
    \includegraphics[width=\linewidth]{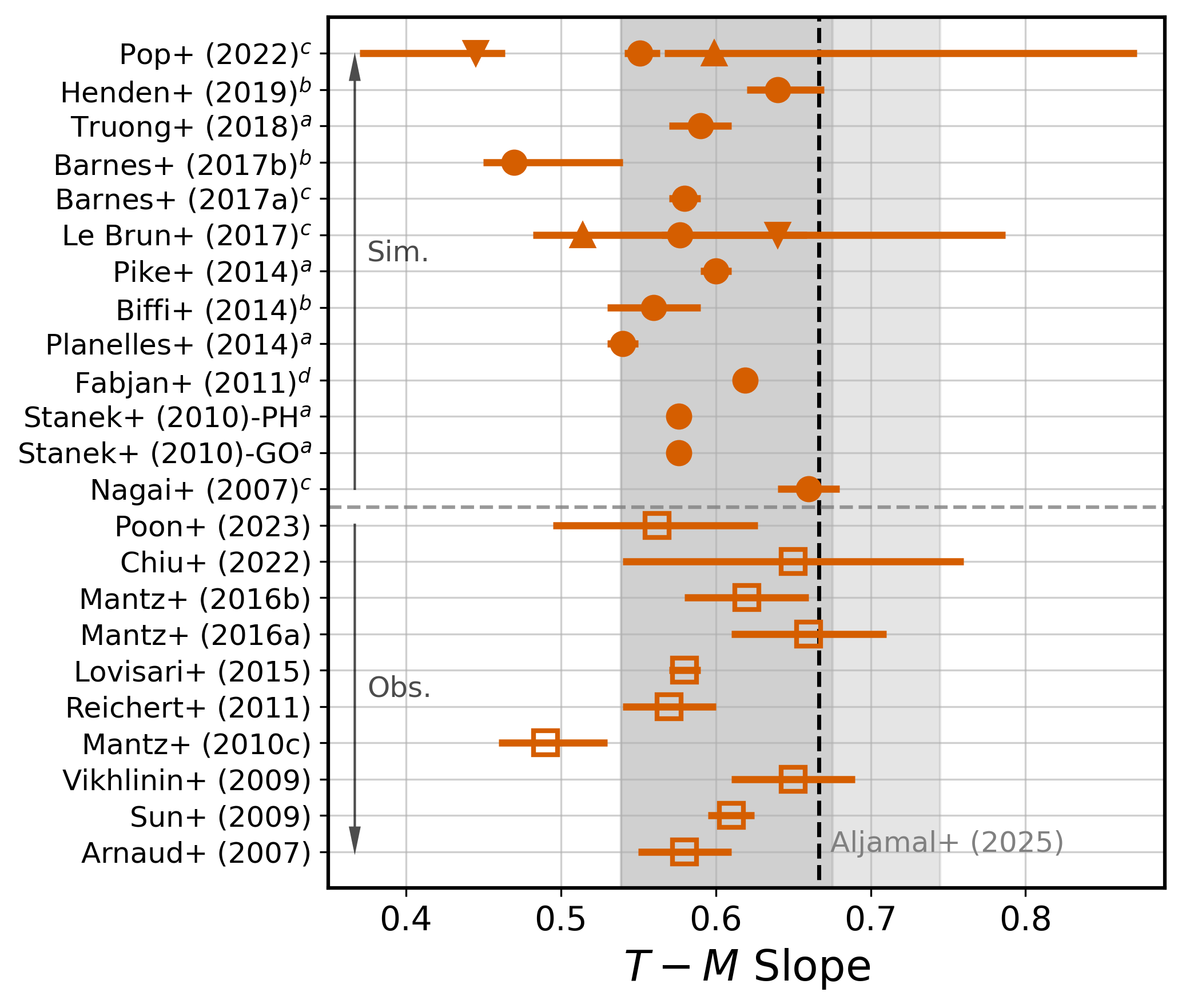}  
 \vspace{-12 truept}
 \caption{Temperature-mass scaling relation slopes from simulations and observations. The vertical dashed line represents the slope of 2/3 predicted by the self-similar model. Superscripts for simulations identify the specific temperature definition used in each study; key is given in Table \ref{tab:ScatterLiterature}. Format is the same as that used in Figure~\ref{fig:Mgas-slope-literature}.
    }
    \label{fig:T-slope-literature}
\end{figure}

Figure \ref{fig:T-slope-literature} illustrates a trend among observations and simulations indicating that the slope of the temperature-total mass scaling relation is shallower than predicted by the self-similar model.  While some works report slopes that are consistent with 2/3 \citep{Nagai2007, Vikhlinin2009, Henden2019}, most measurements agree with our results over both the full and reduced mass ranges, with the exception of \citet{Barnes2017b}, who find a much lower slope, $0.47^{+0.07}_{-0.02}$, with large uncertainty.   

Our findings in Section \ref{sec:Tsl-M500c-slope-scatter} show no clear mass dependence for the slope, which is supported previous works  \citep{Barnes2017a, Pop2022}. 
The method employed by \citet{Pop2022} identify a break scale in the temperature-total mass relation at $0.83^{+4.17}_{-0.24} \times 10^{14}\msun$ (note the large error bar), with slopes of $0.445^{0.019}_{-0.075}$ below and $0.599^{+0.273}_{0.032}$ above. 
In contrast, \cite{LeBrun2017} observe a significant decrease in the slope from $0.64 \pm 0.03$ for halos with $\Mfivehc \leq 10^{14}\msun$ to $0.514 \pm 0.009$ for more massive halos.

\begin{figure}
    \includegraphics[width=\linewidth]{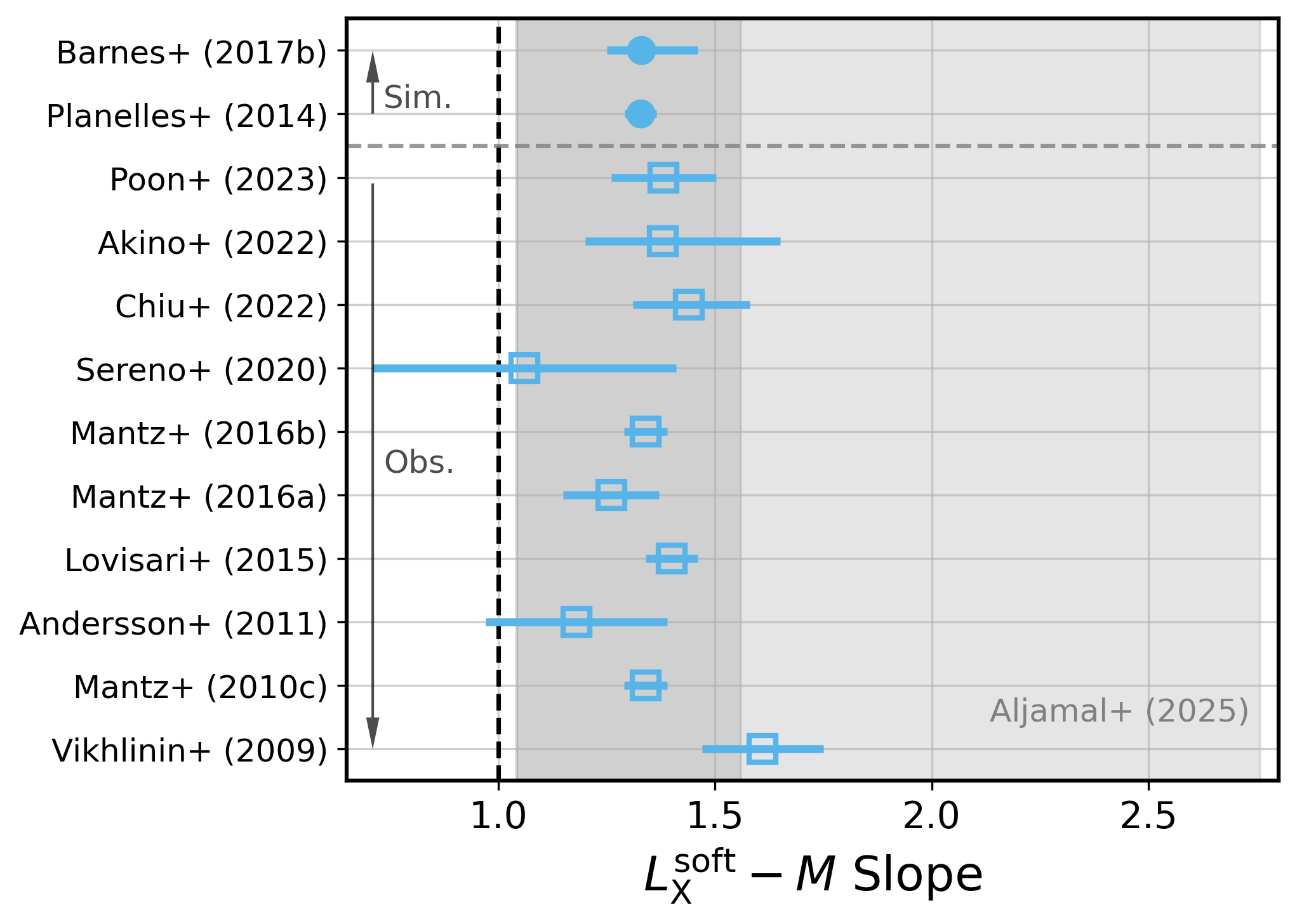}  
     \vspace{-12 truept}
    \caption{Soft-band luminosity-mass scaling relation slopes from simulations and observations. The vertical dashed line represents the self-similar slope of one.  Format is the same as that used in Figure~\ref{fig:Mgas-slope-literature}.
}
    \label{fig:LX-slope-literature}
\end{figure} 

For intrinsic scatter, our results in Figure~\ref{fig:MPR-inrinsic-scatter-plots} show that the intrinsic scatter in $\Tsl$ varies only moderately with mass at $z=0$, with values ranging from 0.06 to 0.15.  
Simulation values listed in Table~\ref{tab:ScatterLiterature} are mainly  consistent within this range.  Remarkably, multiple observational studies find consistency with a value of approximately 13\%.  


\subsubsection{Soft X-ray luminosity}

Figure \ref{fig:LX-slope-literature} presents four observational and two numerical measurements of the slope for the soft-band luminosity-total mass scaling relation. We found only a small number of simulation studies reporting the scaling for soft-band X-ray luminosity.  Both works find values near 1.3, but note that we inferred the $1.33 \pm 0.05$ value from \citet{Planelles2014} by combining their reported slopes for the luminosity-temperature ($2.46 \pm 0.05$) and temperature-total mass ($0.54 \pm 0.01$) relations.  Seven other simulations, listed in Table~\ref{tab:ScatterLiterature}, employed bolometric X-ray luminosity. We discuss their estimates of intrinsic scatter below, as this feature is less dependent on passband than is the scaling relation slope.  

\begin{figure*}
    \includegraphics[width=0.45\linewidth]{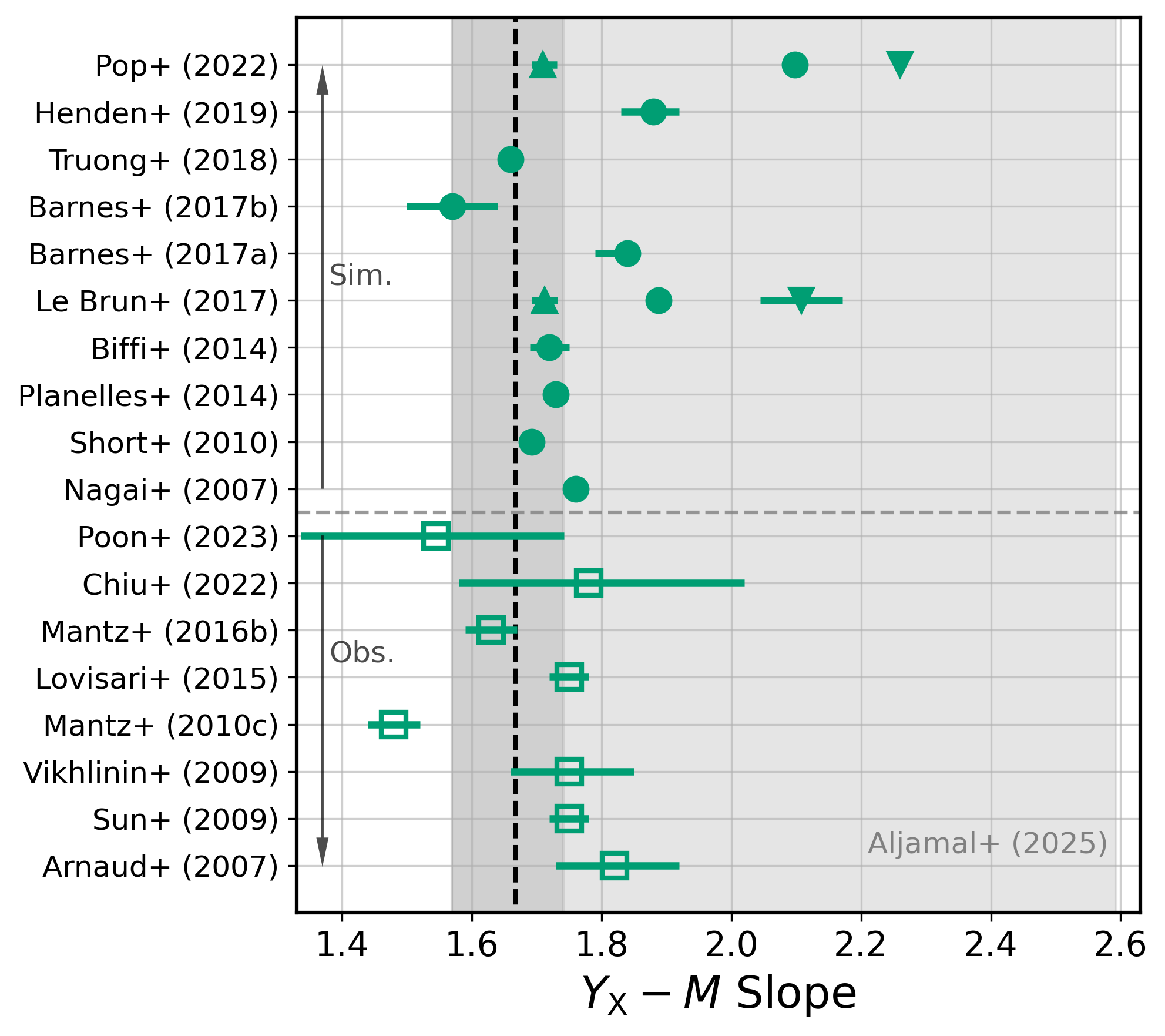}  
     \includegraphics[width=0.505\linewidth]{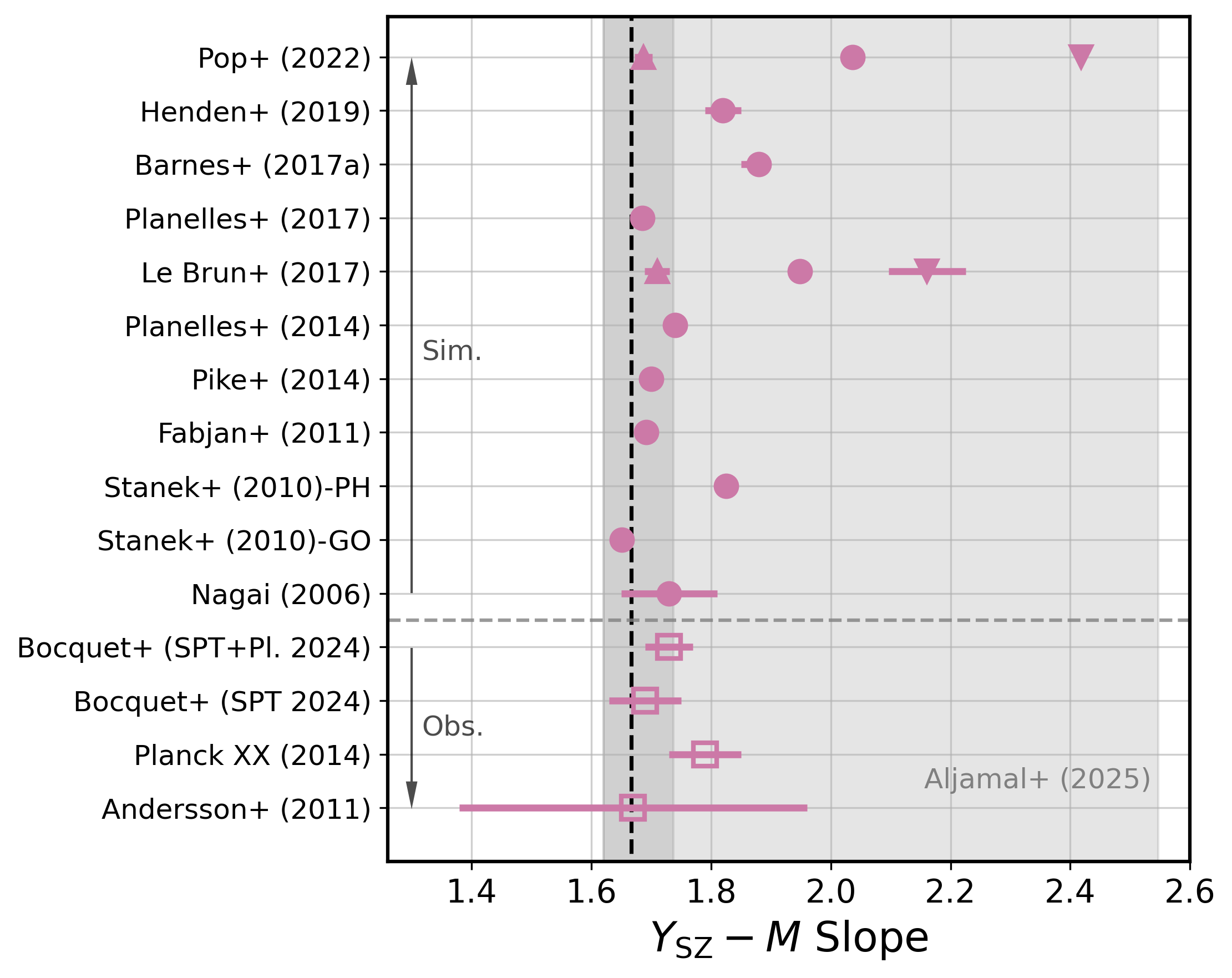}  
      \vspace{-4 truept}
    \caption{Hot gas thermal energy-mass scaling relation slopes from simulations and observations for $\YX$ (left) and $\YSZ$ (right). Vertical dashed lines represent the self-similar slope of 5/3.  Format is the same as that used in Figure~\ref{fig:Mgas-slope-literature}.}
    \label{fig:Y-slope-literature}
\end{figure*}

Observational measurements also find steeper than self-similar slopes, with values that agree with those found in our study using the high mass range of $10^{14.5}-10^{15} \msol$.  All studies are consistent with a central value of 1.3, though the work of \citet{Vikhlinin2009} lies in $2\sigma$ tension with this value.  

The intrinsic scatter estimates are generally consistent with a range of 0.2 to 0.5, with come simulation estimates lying even higher.  The presence of a cool core has long been known to increase variance in the $\LX-T$ relation \citep{Fabian1994} yet \citet{Barnes2017a} still find a scatter of $0.345^{+0.023}_{-0.046}$ for core-excised $\LX$ in the MACSIS halo sample. 


\subsubsection{Hot gas thermal energy}

The findings that hot gas mass scaling is steeper than self-similar, while gas temperature is shallower, combine to result in thermal energy scaling relations having slopes that, for high mass clusters, lie close to the 5/3 self-similar value. Figure \ref{fig:Y-slope-literature} shows slopes of thermal energy scaling relations for both X-ray and SZ variants. 

In most simulations, the direct SZ measure, $\YSZ$, has slopes that lie within 0.05 of the self-similar value.  In the cases using broken power-laws \citep{LeBrun2017, Pop2022}, the high mass values lie in this regime, with lower mass halos having much steeper values  The FABLE simulations \citep{Henden2019} find a steeper slope, which is not surprising given their uniform sampling of a wide range in halo mass and our finding of a scale-dependent slope.  The BAHAMAS+MACSIS study of \citet{Barnes2017a} also find a steeper slope.  That work uses a synthetic X-ray hydrostatic mass in place of true mass.  Finally the preheating treatment of \citet{Stanek2010} also finds a steeper than self-similar slope.  In general, these patterns are also reproduced in the X-ray thermal energy measure, $\YX$.  For poor galaxy groups, the steepening of the mean hot gas mass slope pushes the thermal energy slope well above self-similarity, toward values near 2.6.

The small number of observational studies to date for thermal SZ signal, $\YSZ$, also find slopes near 5/3.  Values for the X-ray equivalent also tend to lie near or slightly above this value.  The result of a shallower slope by \citet{Mantz2010XrayScaling} can be traced to the use of $\Mgas$ as a direct proxy for total system mass.  Under this approach the $\YX$ slope lies low because the $\TX$ slope is also shallower than self-similar.  

In terms of intrinsic scatter at $z=0$, our simulation measurements show a decrease with total mass at the group scale that flattens above $10^{14} \msol$ to values near 0.1.  This value lies near the low end of simulation results shown in Table~\ref{tab:ScatterLiterature}, although \citet{Pike2014} find an even smaller scatter of $0.078 \pm 0.009$ from a small sample of 30 halos.  We note that the larger values seen in simulated samples arise in cases where the halo mass lower limit extends into the group range below $10^{14} \msol$ \citep{Henden2019, Barnes2017a}. 

Observational estimates shown in Table~\ref{tab:ScatterLiterature} lie in the range of 0.1 to 0.2, with the exception of the 19-cluster sample of \citet{poon2023}, who find a larger value, $0.41^{+0.10}_{-0.07}$, lying $3\sigma$ above that range. 


\section{Conclusions}\label{sec:conclusion}

Motivated by the need to assess the quality of various hot gas properties in tracing the total mass of dark matter halos across group and cluster scales, 
we perform localized linear regression on five key properties using halo populations produced by two large-volume cosmological simulations. These simulations, containing thousands of halos in the combined TNG300 and TNG-{\textsc{cluster}} samples and close to a hundred thousand in {\textsc{flamingo}} (L1\_m8), enable detailed examination of the mass and redshift dependence of the mass--property scaling relations, the slope and intrinsic scatter of which set the limiting precision on inferred total mass.  
We focus on five aggregate gas properties: the hot gas mass $\Mgas$, the spectroscopic-like temperature $\Tsl$, the soft $[0.5-2.0]\kev$ X-ray luminosity $\LX$, the X-ray derived thermal energy $\YX$, and its thermal SZ equivalent, $\YSZ$.

Using Kernel Localized Linear Regression (KLLR) with a kernel resolving $0.2$ dex in halo mass, we extract mass-dependent log-means, slopes and covariance of the above properties across $10^{13} \lesssim \Mfivehc/\msol \lesssim 10^{15}$ at four redshifts, $z \in 0, 0.5, 1, 2$.  We compute mass proxy quality (MPQ) as the ratio of property's local scatter divided by its slope, which illuminates the \emph{precision} of that property's use as a total mass proxy.  The \emph{accuracy} of each property involves understanding the absolute normalization of the scaling relation, a more difficult task (see item vii below).  

Our main results are as follows:
\begin{enumerate}
    \item At all redshifts, slopes and standard deviations of hot gas properties are sensitive to total halo mass,  
    leading to scale-dependent MPQ values displayed in Figure \ref{fig:MPQzero}. As halo mass increases, intrinsic property variance decreases while slopes converge toward self-similar expectations. These features, common to both simulation samples, reflect the decreasing importance of star formation and AGN feedback in shaping hot gas properties in the deepest gravitational potentials.   
    
    \item At $z = 0$, the thermal energy measures, $\YX$ and $\YSZ$, as well as hot gas mass, $\Mgas$, compete for best halo mass proxy.  Property-conditioned mass scatter decreases from $\sim 18\%$ at $\Mfivehc = 10^{13}\msol$ to $4-5\%$ at $\Mfivehc = 10^{15}\msol$. With the exception of $\YX$ in TNG at high masses, the MPQ values and mass dependence are nearly identical similar in both simulation samples.  

    \item At cluster scales, $\mfiveh > 10^{14}\msol$, the slope and intrinsic scatter in $\Mgas$ and $\YSZ$ are nearly redshift independent, 
    and log-mean normalizations of these properties scale self-similarly with redshift. 
    These properties, observable via their X-ray and millimeter-wave observable signatures, are thus ideal proxies for identifying massive halos and characterizing the population in the near cosmic field. 
    
    \item Residuals about scale-dependent means in $\Mgas$ and $\YSZ$ are very consistent with a log-normal form. 
    The MPQ estimate of the E14 model thus agrees well with directly-measured mass variance for these properties.

    \item For galaxy groups, $\Mfivehc < 10^{14}\msol$, the slopes and normalizations of all properties are strongly redshift dependent in a manner that, at fixed total mass, reflects a progressive heating and loss of hot gas over time. The scale-dependent scatter of gas properties is nearly redshift independent in TNG while {\textsc{flamingo}} halos show significantly lower scatter at high redshift for $\Mgas$ and $\YSZ$.

   \item  Correlations among hot gas properties are generally positive, with some (\eg \, $\LX-\Tsl$) exhibiting complex scale-dependent behavior. Combining the entire set of properties yields a minimum halo mass scatter of $3\%$ for high mass clusters.  

    \item Mass-dependent means of hot gas properties at $z=0$ can differ significantly between the two astrophysical treatments, 
    challenging the role of such simulations in setting absolute expectations for observed group and cluster properties. However, good agreement is found for gas thermal energy, $\YSZ$, for high-mass clusters. Respect for virial equilibrium may be at play, as gas masses and temperatures in the two simulations differ in directions that cancel when combined. Cosmological analysis using thermal SZ cluster samples may benefit from these findings by employing more informative,  simulation-based priors on astrophysical scaling relation priors.
    
\end{enumerate}

These theoretical expectations should be verified by other hydrodynamical simulations and their robustness to different galaxy formation and feedback models examined. We have verified that the results summarized above are robust to numerical resolution, 
but similar analysis on higher resolution simulations with different gas physics and AGN feedback models remains imperative.  Scaling relation mean values are a particular area deserving more careful attention in both simulated and observed cluster samples.  

Validation studies should also be pursued.  Extraction of  population-level statistical features from multi-wavelength observational studies is an essential opportunity for samples of galaxy groups and clusters already in hand, as well as those to be defined by eROSITA, Euclid, LSST, the Simons Observatory, and other next-generation facilities.  As group and cluster samples with uniform multi-wavelength coverage grow to encompass thousands of systems, scale-sensitive studies of \emph{observable-conditioned} properties offer key pathways to precisely reveal the baryonic components of massive halos and, by comparison to simulated group and cluster population expectations, to understand their astrophysical evolution.

\section*{Acknowledgements}
EA and AEE are grateful for support from the National Aeronautics and Space Administration Data Analysis Program (NASA-80NSSC22K0476). AF acknowledges support from the National Science Foundation under Cooperative Agreement 2421782 and the Simons Foundation award MPS-AI-00010515. We thank Megan Donahue, Ian McCarthy, Daisuke Nagai, Elena Rasia, Volker Springel and Mark Voit for useful discussions. This work used the DiRAC@Durham facility managed by the Institute for Computational Cosmology on behalf of the STFC DiRAC HPC Facility (\url{www.dirac.ac.uk}). The equipment was funded by BEIS capital funding via STFC capital grants ST/K00042X/1, ST/P002293/1, ST/R002371/1 and ST/S002502/1, Durham University and STFC operations grant ST/R000832/1. DiRAC is part of the National e-Infrastructure as well as the Max Planck Computing and Data Facilitity (MPCDF) of the Max Planck Society (\url{https://www.mpcdf.mpg.de/}). 

All analysis in this work was enabled by the following {\sc{Python}} packages:{\sc{illustris\_python}} (\url{https://github.com/illustristng/illustris_python}), {\sc{numpy}} \citep{harris2020array}, {\sc{pandas}} \citep{Mckinney2011}, {\sc{scipy}} \citep{2020SciPy-NMeth}. {\sc{sci-kit Learn}} \citep{scikit-learn}, {\sc{pymc}} \citep{pymc2023}, {\sc{h5py}} 
\citep{collette_python_hdf5_2014}, {\sc{joblib}} (\url{https://github.com/joblib/joblib}), {\sc{swiftsimio}} \citep{Borrow2020}, and {\sc{matplotlib}} \citep{Hunter:2007}.


\section*{Data Availability}
The halo catalogues used in this work, the code used to analyze it, and KLLR analysis data, are available on EA's github: \texttt{https://github.com/ealjamal/MPQScaling} \href{https://github.com/ealjamal/MPQScaling}{\faGitSquare}.



\bibliographystyle{mnras}
\bibliography{MPQ-gas-properties} 



\appendix

\section{Kernel Localized Linear Regression (KLLR)}\label{sec:KLLR}

Simple Linear Regression using the least squares algorithm has been a staple 
approach to cluster scaling relations, providing a normalization, slope, and variance defining the relationship between two integral log-space properties. 
However, such an approach is too inflexible as it attempts to summarize a population of thousands of halos by only the three aforementioned quantities. Furthermore, as mentioned in Section \ref{sec:simulations}, the steeply falling nature of the HMF means that parameters recovered via linear regression are strongly biased by the behavior of low-mass halos. Kernel Localized Linear Regression \citep[KLLR][]{Farahi2022KLLR} allows for a more sensitive analysis of the scaling relations by conditioning the normalization, slope, and variance on halo mass.

\subsection{Scaling relations using KLLR}
We follow the treatment in E14 and in \cite{Farahi2018} by considering a set of halo properties, $\mathbf{S}$, associated with halos with total mass $\Mfivehc$. Under the assumption of log-normal distribution around the scaling relation, we define $\mathbf{s} = \ln \mathbf{S/S_{\rm fid}}$ and $\mu = \ln (\Mfivehc/\Mfid)$ where $S_{\rm fid}$ and $\Mfid$ are chosen fiducial scales. We write the mean scaling relation of property $a$ at a fixed redshift $z$ as:
\begin{equation}\label{eq:KLLR-scaling}
    \langle s_a \ \vert \  \mu, z \rangle = \pi_a(\mu, z) + \alpha_a(\mu, z)\mu,
\end{equation}
where $\pi_a(\mu, z)$ and $\alpha_a(\mu, z)$ are the mass- and redshift-dependent normalization and slope for the property of interest, respectively. At a given redshift, we estimate the local normalization and slope around some chosen mass scale $\mu_c$ by minimizing the weighted residual sum of squares :
\begin{equation}
    \epsilon_a^2(\mu_c) = \sum_{i = 1}^n w^2(\mu_i-\mu_c)[s_{a, i} - \pi_a(\mu_c) - \alpha_a(\mu_c)(\mu_i-\mu_c)]^2,
\end{equation}
where the sum $i$ is over all halos. This procedure is carried out for $20$ halo mass centers, $\mu_c$, equally spaced in $\mu$ while recording the normalization, slope, and intrinsic variance for each. Weights are Gaussian,
\begin{equation}
    w(\mu_i - \mu_c) = \frac{1}{\sqrt{2\pi}\sigma_{\text{KLLR}}} \exp\bigg[-\frac{(\mu_i-\mu_c)^2}{2\sigma^2_{{\rm KLLR}}} \bigg],
\end{equation}
where we choose $\sigma_{\rm KLLR} = 0.2 \ln(10)$ . The kernel width controls the size of the local features that the algorithm tracks, but like any hyperparameter it also controls the degree of overfitting to the noise in the data. Thus, a small kernel width would be ideal for more local analysis of the scaling behavior but that would result in a noisy fit, and if the kernel width is too large, any local features would be washed out and we would approach a simple linear regression fit.

\subsection{Scale-dependent scatter and covariance}
To calculate the mass scatter implied by a single property, we only need to use the slope and intrinsic scatter for that property, but to calculate the mass scatter implied by multiple properties, we also need to calculate the pair-wise covariance between the properties (Section \ref{sec:correlations}). Using the local slope and normalization, we can calculate the covariance between $s_a$ and $s_b$ by using the unbiased estimator \citep{Gough2009}:

\begin{equation}\label{eq:KLLR-covariance}
    C_{a, b} = A\sum_{i = 1}^n w_i \delta s_{a, i} \delta s_{b, i},
\end{equation}
where the residual for the $i$th data point is given by $\delta s_{a, i} = s_{a, i} - \alpha_a \mu_i - \pi_a$, the pre-factor is
\begin{equation}
    A = \frac{\sum_{i = 1}^n w_i}{\bigg(\sum_{i = 1}^n w_i\bigg)^2 - \sum_{i = 1}^n w_i^2},
\end{equation}
and the correlation coefficient normalizes this covariance to be in the range $[-1, 1]$
\begin{equation}\label{eq:KLLR-correlation}
    r_{a, b} = \frac{C_{a, b}}{\sqrt{C_{aa}} \sqrt{{C_{bb}}}}.
\end{equation}
The scale-dependent scatter is given by the square root of the diagonal entries of the covariance:
\begin{equation}\label{eq:KLLR-scatter}
    \sigma_{a} = \sqrt{C_{a, a}} =  \sqrt{A\sum_{i = 1}^n w_i \delta s_{a, i} \delta s_{a, i}}.
\end{equation}
For the mass-conditioned residual kernels, we focus on the \textit{normalized residuals}, 
\begin{equation}\label{eq:KLLR-normalized-res}
    \widehat{\delta}s_{a, i} \equiv \frac{\delta s_{a, i}}{\sigma_a} = \frac{ s_{a, i} - \alpha_a \mu_i - \pi_a}{\sigma_a}  .
\end{equation}
As discussed in the text, we employ a narrower kernel width of 0.1 dex to compute the distribution of residuals.  

\section{Core-excised luminosity}\label{sec:LX-vs-LXce}

\begin{figure}
    \centering
    \includegraphics[width=0.85\linewidth]{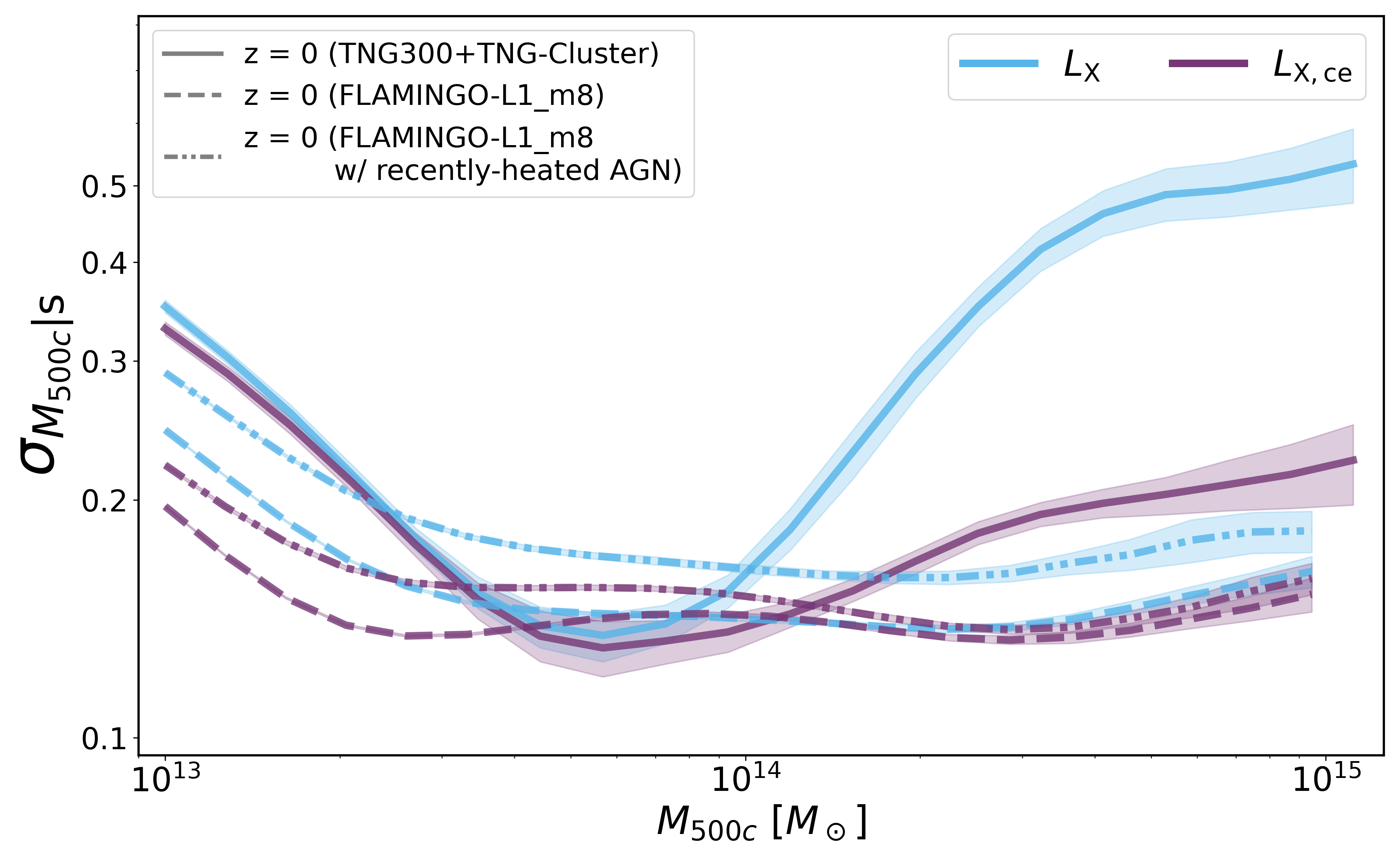}  
     \vspace{-8 truept}
    \caption{MPQ for X-ray luminosities without ($\LX$, cyan lines) and wih ($\LXce$, purple) core-excision at $z = 0$ for {\TNGcombined} (solid line) and {\Flam} excluding cells recently-heated by AGN feedback (dashed line) and including those cells (dash-dotted line).}
    \label{fig:LX-LXce-MPQ}
\end{figure}

\begin{figure}
    \includegraphics[width=\linewidth]{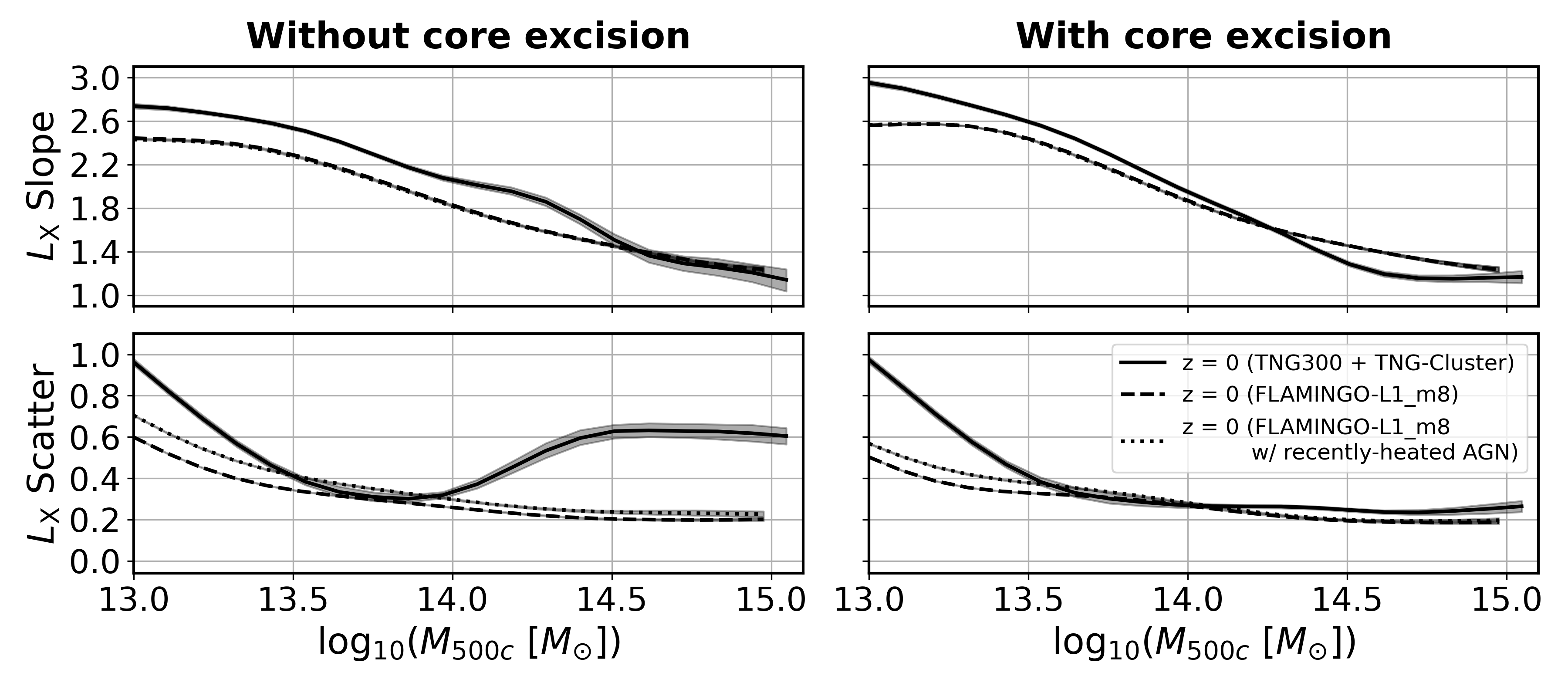}  
     \vspace{-6 truept}
    \caption{Scale-dependent MPR slope (top) and intrinsic scatter (bottom) of the $\LX$-$\Mfivehc$ (without core excision, left) and $\LXce$-$\Mfivehc$ (with core excision, right) at $z = 0$ in {\TNGcombined} (solid line), in {\Flam} excluding recently-heated AGN cells (dashed line) and including those cells (dotted line). The inclusion of recently-heated AGN cells does not affect the slope, but increases the scatter of $\LX$ and $\LXce$, specifically at the group scale.}
    \label{fig:LXce-slope-scatter}
\end{figure} 

Because cluster cores contain high density gas with complex thermal structure, excess X-ray emission can often be produced within this region, typically taken to be radii interior to $0.15\Rfivehc$.
Here we illustrate how core excision affects population statistics examined for non-core excised quantities in the main text. 
The core-excised luminosity, $\LXce$, is calculated in the same way as $\LX$ (see \S\ref{sec:defintions-of-properties}) but with all gas cells within $0.15\Rfivehc$ being excluded. 

Figure \ref{fig:LX-LXce-MPQ} present the MPQ for both measures in both simulated halo samples at $z=0$. In TNG, we observe no change in the MPQ for group-scale halos with $\Mfivehc < 10^{14}\msun$. However, for more massive halos the $\LXce$ MPQ is nearly mass-independent exhibiting a mass scatter of $\sim 20\%$, significantly smaller than that of $\LX$. Conversely, in the {\textsc{flamingo}} sample, the MPQ of $\LXce$ for low-mass groups is slightly smaller than that of $\LX$ while the two measures are nearly identical for larger halos. We also measure the luminosity MPQs in {\textsc{flamingo}} with the inclusion of gas cells recently-heated by AGN feedback. We observe that these cells introduce an added mass scatter of $< 5\%$ throughout the mass range for $\LX$ while for $\Mfivehc > 10^{14}\msol$, there is little to no added scatter for $\LXce$. The added scatter can be attributed to the increased heating of dense gas which results in extremely luminous gas cells--most of which located near the core. These cells contribute to the added intrinsic scatter of $\LX$ and $\LXce$ without affecting their MPR slopes (see Figure \ref{fig:LXce-slope-scatter}).

In Figure~\ref{fig:LXce-slope-scatter}, we display KLLR slope and scatter values for $\LX$-$\Mfivehc$ (left) and $\LXce$-$\Mfivehc$ (right) relations at $z=0$. We observe modest changes in slopes after core excision, with a slight increase in slope for halos with $\Mfivehc < 10^{13.5}\msun$. At these mass scales the intrinsic scatter is also unaffected, implying that cool core activity is not important in low-mass groups.  

At cluster scales, core excision reduces the scatter by nearly a factor of three for TNG halos, whereas there is no difference in scatter between $\LX$ and $\LXce$ at this scale in the {\textsc{flamingo}} sample. The differing behaviors indicate that the gas phase structure of cores in TNG-{\textsc{cluster}} and high-mass TNG300 halos contain very luminous gas cells near the core, which are likely part of the abundant cool cores.

\section{Mass-weighted temperature}\label{sec:Tmw}

\begin{figure}
    \centering
    \includegraphics[width=0.85\linewidth]{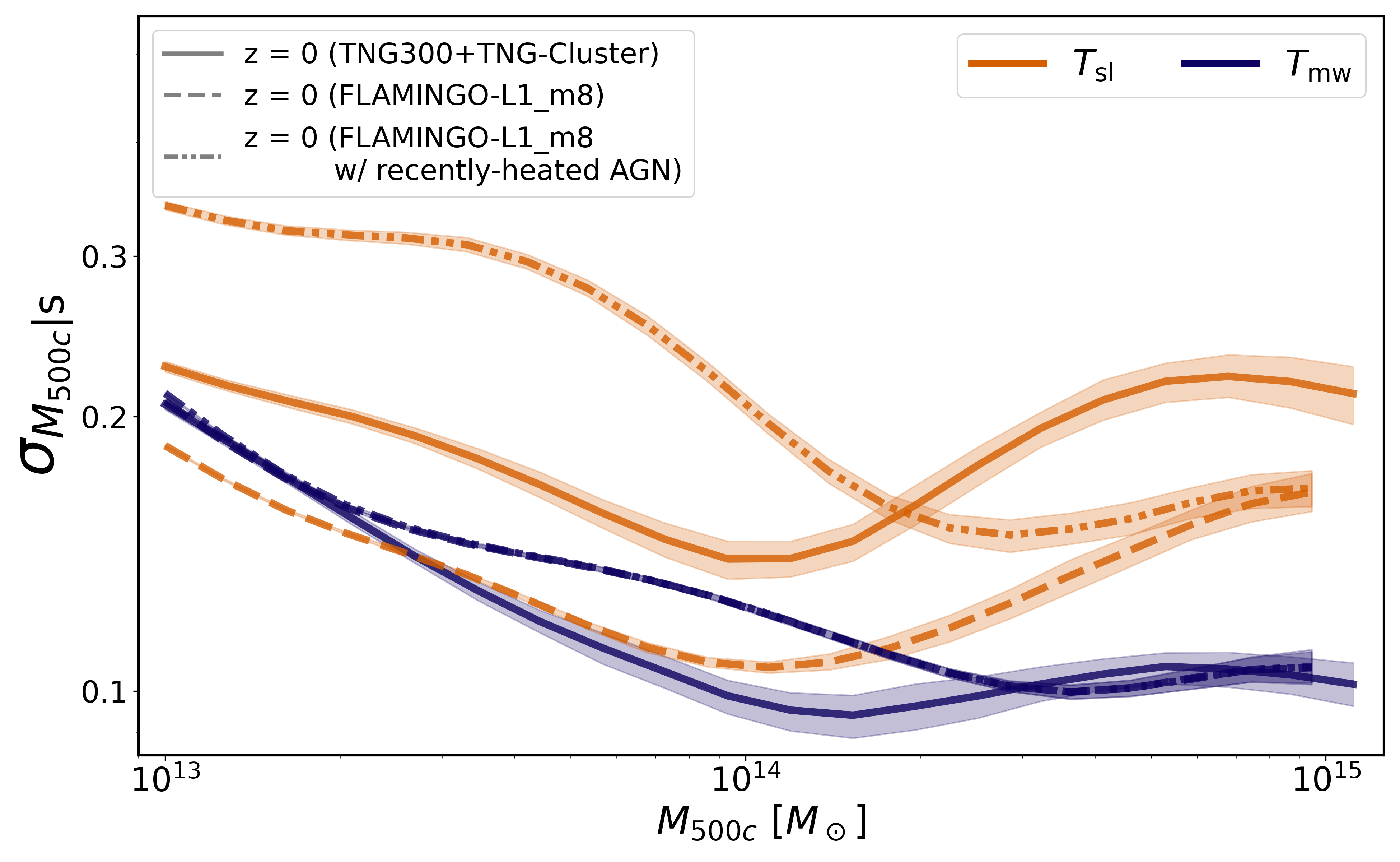}  
     \vspace{-6 truept}
    \caption{MPQs of core-excised spectroscopic-like temperature $\Tsl$ (orange) and mass-weighted average temperature $\Tmw$ (dark blue) at $z=0$ for {\TNGcombined} (solid line) and {\Flam} excluding cells recently-heated by AGN feedback (dashed line) and including those cells (dash-dotted line). For $\Tmw$, there is no difference in MPQ when adding the gas that is recently-heated by AGN feedback.}
    \label{fig:MPQ-temps-z0}
\end{figure}

\begin{figure}
    \includegraphics[width=\linewidth]{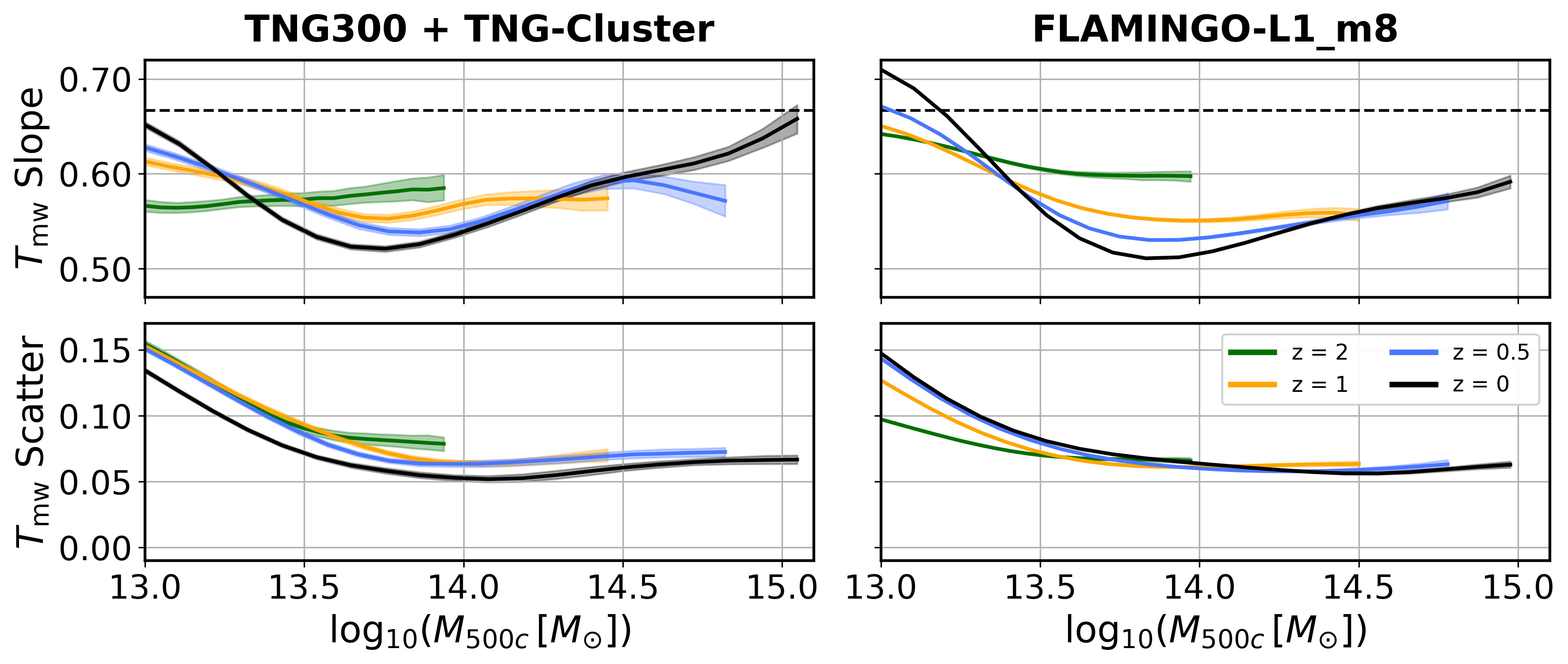}  
     \vspace{-12 truept}
    \caption{Scale-dependent MPR slopes (top) and intrinsic scatter values (bottom) of the mass-weighted temperature in {\TNGcombined} (left) and {\Flam} (right) samples at redshifts indicated in the legend.}
    \label{fig:Tmw-slopes-scatters}
\end{figure} 

While the spectroscopic-like temperature is sensitive to gas phase structure the simpler mass-weighted temperature is not.  Here we present behaviors for $\Tmw$.

\begin{figure*}
    \centering
    \includegraphics[width=0.7\textwidth]{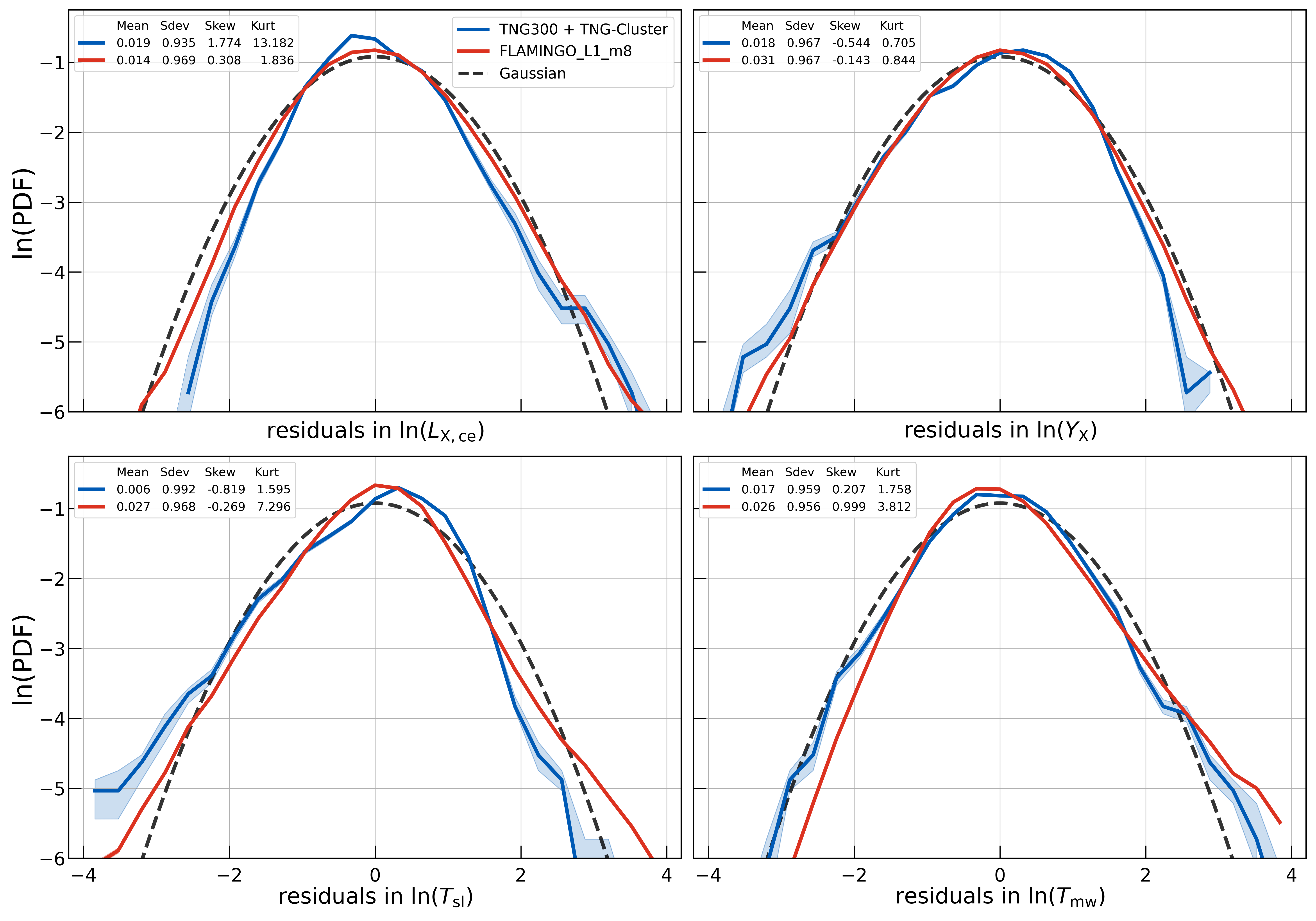}
         \vspace{-8 truept}
    \caption{Mass-conditioned likelihood forms at $z=0$ of core-excised X-ray luminosity $\LXce$ (top left), X-ray thermal energy $\YX$ (top right), core-excised spectroscopic-like temperature $\Tsl$ (bottom left), and mass-weighted average temperature $\Tmw$ (bottom right). 
    }
    \label{fig:LXce-YX-Tsl-Tmw-kernels}
\end{figure*}

Figure~\ref{fig:MPQ-temps-z0} compares the MPQs of $\Tsl$ and $\Tmw$ at $z = 0$. At the group scale, the mass scatter of $\Tmw$ is modestly reduced compared to that of $\Tsl$ for TNG halos, while remaining nearly unchanged for {\textsc{flamingo}}.   
However, at the cluster scale, $\Tmw$ has reduced scatter in both simulated samples, with the reduction being nearly a factor of two for TNG-{\textsc{cluster}} and high-mass TNG300 halos. 

The effect of including gas that is recently-heated by AGN feedback is minimal for $\Tmw$, while for $\Tsl$, there is an added mass scatter of $\sim 17\%$ at the group scale that decreases to $< 1\%$ at the cluster scale. The recently-heated AGN gas cells boosts the temperature of dense gas cells which increases the intrinsic scatter in the $\Tsl$-$\Mfivehc$ relation while slightly decreasing the slope, leading to a larger mass scatter.

Paralleling the time-dependent MPRs shown in Figures~\ref{fig:MPR-slope-plots} and \ref{fig:MPR-inrinsic-scatter-plots},  Figure~\ref{fig:Tmw-slopes-scatters} presents the slope and intrinsic scatter of the $\Tmw$-$\Mfivehc$ scaling relation over time.  The scale-dependence of the slope at $z=0$ is similar to that of $\Tsl$, with minimum slope values in both simulations occurring near $10^{13.8} \msol$.  In both simulations, the sensitivity to redshift is more modest in $\Tmw$ at nearly all mass scales. 

Intrinsic scatter values are substantially smaller for $\Tmw$ than $\Tsl$, and there is a much less pronounced scale-dependence, particularly at $z=0$.  
For {\textsc{flamingo}}, the scatter in $\Tmw$ is larger than that of $\Tsl$ 
for halos with mass $10^{13.7} \msun < \Mfivehc < 10^{14.2}$ which, when combined with smaller slopes, results in the halo mass scatter of $\Tmw$ shown in Figure~\ref{fig:MPQ-temps-z0} being slightly larger for this mass range. In contrast, for more massive halos, the scatter in $\Tmw$ is smaller by approximately $25\%$ which, when combined with a slightly larger slope, yields a smaller mass scatter.

\section{Additional Likelihood Shapes}\label{sec:additionalKernelShapes}

Expanding upon the three property kernels shown in Figure~\ref{fig:Mgas-YSZ-LX-kernels}, we show in Figure \ref{fig:LXce-YX-Tsl-Tmw-kernels} mass-conditioned likelihoods of four additional properties at $z=0$: core-excised luminosity $\LXce$; X-ray thermal energy $\YX$; core-excised spectroscopic-like temperature $\Tsl$; and mass-weighted average temperature $\Tmw$. 

The shapes are approximately Gaussian but deviations are apparent.  
Compared to the case of $\LX$ in Figure \ref{fig:Mgas-YSZ-LX-kernels}, the distribution of $\LXce$ residuals is similar in form, with a tail to high values particularly pronounces in TNG.  Residuals in both $\Tsl$ and $\YX$ are mildly skewed negative in both simulated samples, and the {\textsc{flamingo}} distribution of $\Tsl$ has  tails at both low and high values.  The $\Tmw$ distribution has a tail to high values and is skew positive in both simulation samples.

\section{Scaling Relation Absolute Means}\label{sec:scaling-relations}

\begin{figure*}
    \includegraphics[width=0.7\textwidth]{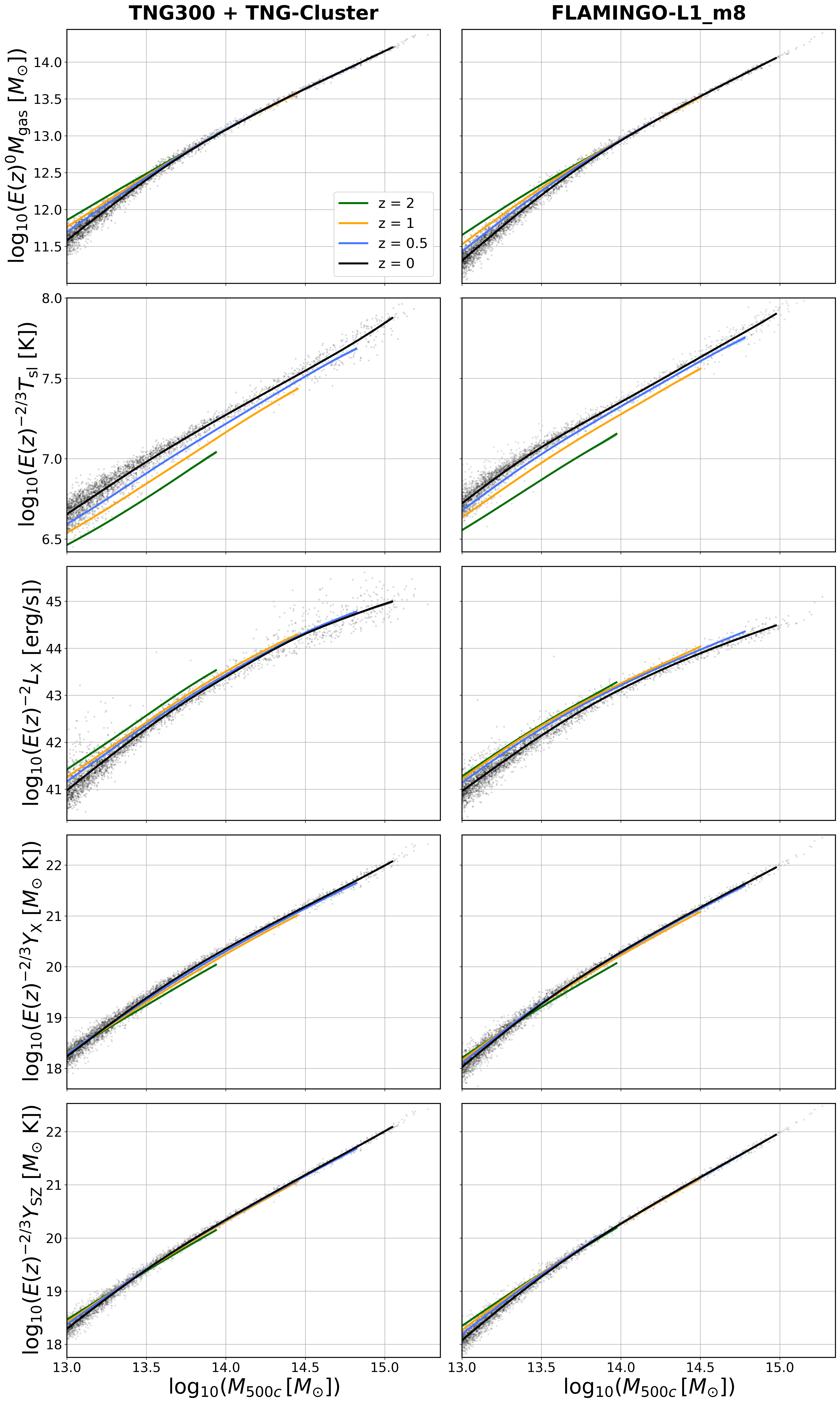}
    \caption{Logarithmic-mean gas property values as a function of halo mass, including self-similar redshift scaling for each property, using the same format as that of Figure~\ref{fig:deviations-of-evolution}.   
    The resampled bootstrap $1\sigma$ uncertainties are smaller than the width of the lines. We plot the individual halo properties at $z = 0$, omitting other redshifts for clarity, while sub-sampling {\textsc{flamingo}} halos to produce a comparable sample size to that of TNG in each bin.
    }
    \label{fig:normalizations-self-similar}
\end{figure*}

In Figure~\ref{fig:deviations-of-evolution} of the main text, the log-mean behaviors of the hot gas properties are shown as deviations relative to self-similar scaling in both halo mass and redshift.  
In Figure \ref{fig:normalizations-self-similar}, we present these scaling relations in a manner that retains the expected self-similar redshift scaling while removing the rescaling with halo mass scale. 
The $z=0$ normalization differences listed in Table~\ref{tab:normalization-z-0} are now apparent by close reading of this figure.

At the cluster scale, $\Mfivehc > 10^{14}\msun$, hot gas mass $\Mgas$ and thermal energy $\YSZ$ adhere remarkably closely to self-similar evolution.  Small deviations are seen for other measures, with $\Tsl$ and $\YX$ being slightly lower, and $\LX$ slightly higher, at larger redshifts relative to the $z=0$ self-similarly scaled values.  These drifts reflect the cumulative heating of the ICM plasma, primarily from AGN feedback, over time. 

The effects of such feedback are more apparent at the group scale.  Low mass halos at high redshift have more gas, are cooler, and emit more X-ray luminosity than those at $z=0$.  Over time, moderate star formation in these halos removes hot gas and feedback from growing SMBHs raises ambient gas temperatures in these systems. These effects conspire to reduce X-ray emission at late times. As noted in the main text, gas thermal energy measures adhere more closely to self-similar redshift evolution due to opposing trends in gas mass (declining) and temperature (rising) over time.  


\bsp	
\label{lastpage}
\end{document}